\documentclass[%
reprint,
 amsmath,amssymb, aps,
]{revtex4-2}

\usepackage{graphicx}% Include figure files
\usepackage{tabularx}
\usepackage{dcolumn}% Align table columns on decimal point
\usepackage{bm}% bold math
\begin{document}

%\preprint{APS/123-QED}

\title{Entropy and chirality in sphinx tilings}% Force line breaks with \\
%%% GHuber - discuss title

\author{Greg Huber}
\email{gerghuber@gmail.com}
\affiliation{Chan Zuckerberg Biohub -- San Francisco, 499 Illinois Street, San Francisco, CA 94158, USA}
\author{Craig Knecht}
\email{craigknecht03@gmail.com}
\affiliation{691 Harris Lane, Gallatin, TN 37066, USA}
\author{Walter Trump}
\email{w@trump.de}
\affiliation{Department of Physics, Gymnasium Stein, Faber-Castell-Allee 10, 90547 Stein, Germany}
\author{Robert M. Ziff}
\email{rziff@umich.edu}
\affiliation{Center for the Study of Complex Systems and Department of Chemical Engineering, University of Michigan, Ann Arbor, MI 48109-2800, USA}

%%\collaboration{SPHINX Collaboration}% \noaffiliation

\date{\today}% It is always \today, today,

\begin{abstract}
As a toy model of chiral interactions in crowded spaces, we consider sphinx tilings in finite regions of the triangular lattice. The sphinx tiles, hexiamonds composed of six equilateral triangles in the shape of a stylized sphinx, come in left and right enantiomorphs. Regions scaled up from the unit sphinx by an integer factor ({\it Sphinx frames}) require tiles of both chiral forms to produce tilings, including crystalline, quasicrystalline, and fully disordered tilings. 

For frames up to order 13, we describe methods that permit exact enumeration and computation of partition functions using {\it accelerated backtracking}, {\it seam}, and {\it dangler} algorithms. For larger frames, we introduce a Monte Carlo (MC) method to sample typical  tilings.  Key to the latter is the identification of fundamental shapes ({\it polyads}) that admit multiple tilings and which allow a rejection-free MC simulation.
\end{abstract}

%\keywords{Suggested keywords}%Use showkeys class option if keyword display desired
\maketitle

\section{Introduction}

Chiral interactions in crowded spaces are important to a number of subjects \cite{VisheratinaKumarKotov22} but remain poorly understood. As an extreme case of a highly packed and geometrically frustrated system, one may consider tilings. 
Within statistical mechanics, the study of tilings has been a fundamental and fertile subfield since at least 1937 with the introduction of the dimer (or domino) tiling  \cite{FowlerRushbrooke37}, which was solved exactly for the square lattice in 1961 \cite{TemperleyFisher61,Kasteleyn61}.
Dimer tilings are related to Pfaffian solutions of the Ising model \cite{KacWard52} and have reappeared in many guises \cite{Lieb67,PriezzhevRuelle08,Loebl10,Pham20,Dhar90}.  When the region tiled is the shape of an Aztec diamond, vast simplifications in the enumeration occur, yet nontrivial phase ordering emerges, as evidenced by the arctic-circle theorem \cite{JockuschProppShor95,ElkiesKuperbergLarsenPropp92,FerrariSpohn06,KenyonWilson11}. ``Arctic boundaries" have since been discovered for a variety of tiling systems. Pauling's ice model from 1935 can also be mentioned in this context, since there the state of a neighboring site (tile) is similarly constrained by the state of the reference site (tile) \cite{Pauling35,TemperleyLieb71}. Introduced in 1961, Wang tiles \cite{HWang1961} connected satisfiability problems to tilings of the square lattice, and they continue to be relevant to the study of random-search algorithms and phase transitions in statistical computational-complexity theory \cite{Tersenghi01,Tersenghi02}.

\begin{figure}[htbpbp]
\centering
\includegraphics[width=1\linewidth]{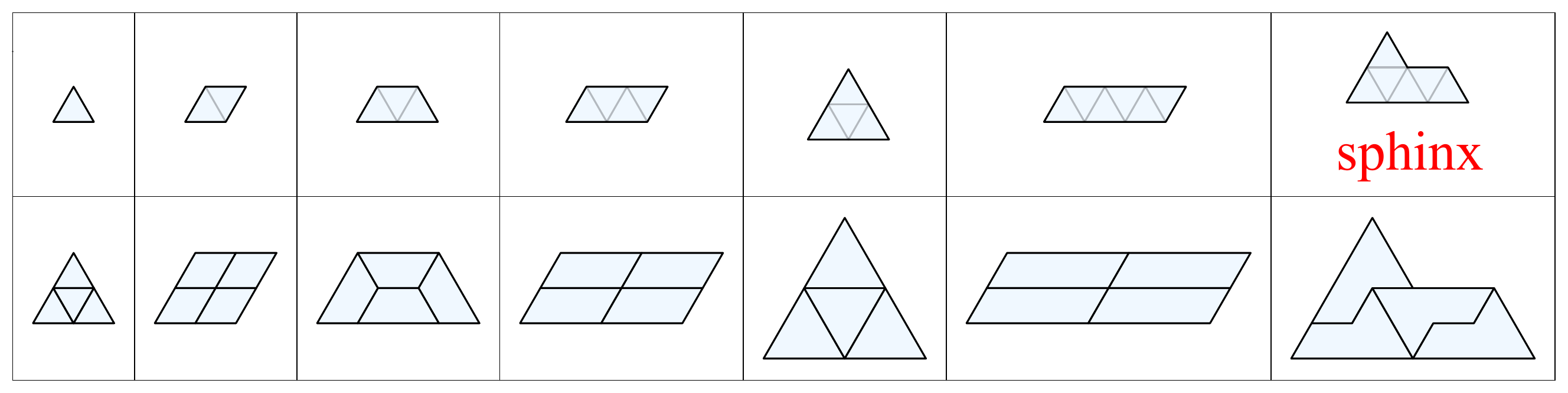}
\caption{The sphinx is the smallest asymmetric chiral rep-tile on the triangular lattice. Here we show the polyiamonds with area $\leq 7$ that are area-quadrupling rep-tiles (triangle, diamond, hemisphinx, tetriamond parallelogram, tetriamond triangle, hexiamond parallelogram, and sphinx). In the second row, area-quadrupling inflation rules are shown (the tetriamond parallelogram has more than one).}
\label{fig:Rep-tiles}
\end{figure}

\begin{figure}[htbp]
\centering
\includegraphics[width=0.85\linewidth]{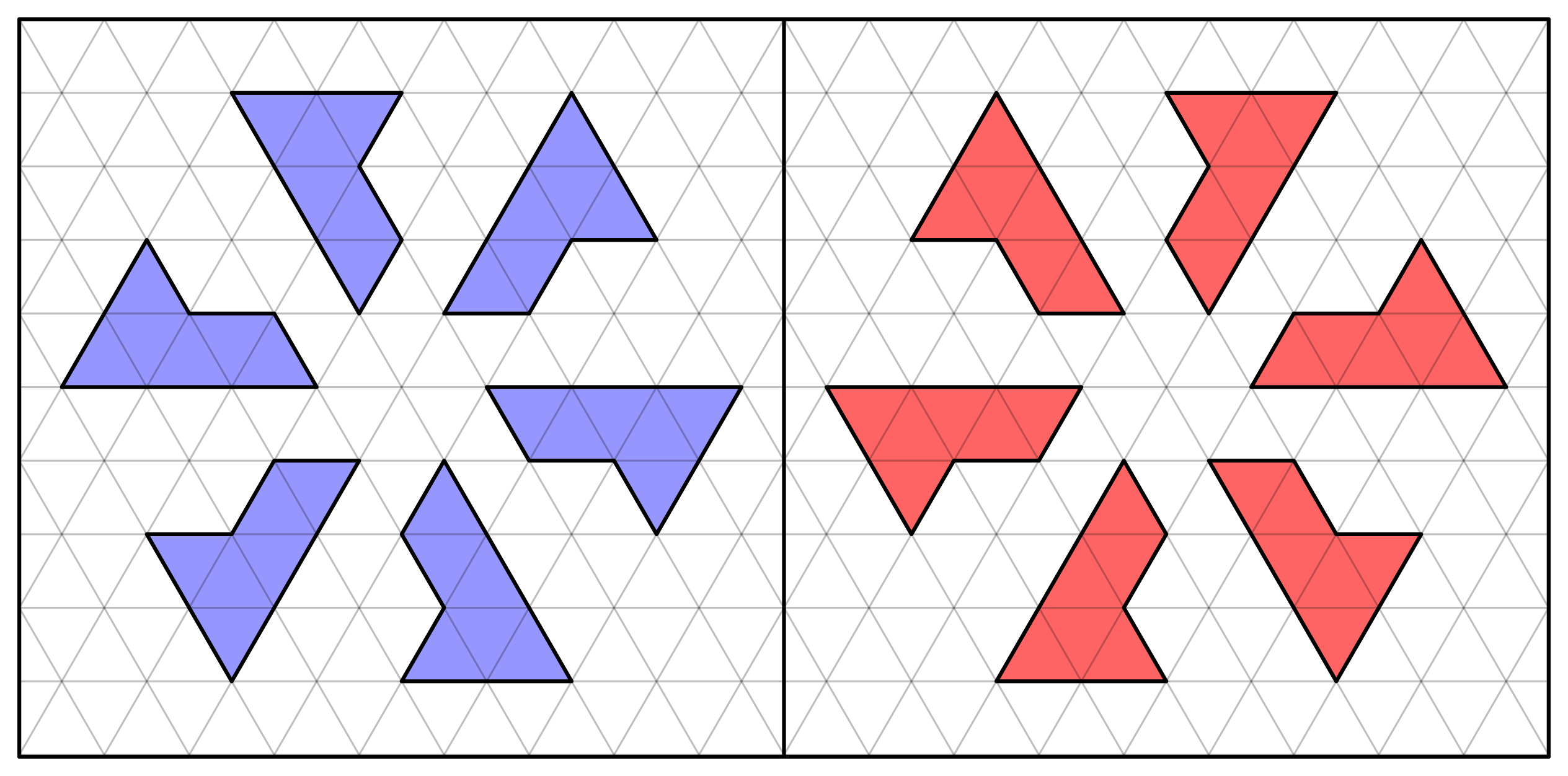}
\caption{L and R chiral tiles}
\label{fig:L-R-sphinxes-1}
\end{figure}

The study of more complex tiles formed by connected polygons was stimulated by Golomb \cite{Golomb54,Golomb94} and his polyominoes (generalizations of dominoes on the square lattice)---it was in this context that the sphinx tile and its unusual properties were discovered. The sphinx tile is composed of six connected equilateral triangles, as seen in the right side of Fig.\ \ref{fig:Rep-tiles}. Shapes made of connected equilateral triangles are termed {\it polyiamonds}. The sphinx makes an appearance, already with its name, along with those of 11 other shapes, in a catalog of {\it hexiamonds} (polyiamonds of size 6) due to O'Beirne \cite{OBeirne61,Weisstein}. The sphinx stands out from the other tiles in that it is both {\it asymmetric} (therefore, {\it chiral}) and {\it rep-tilian}, being the smallest such tile on the triangular lattice.
We discuss each of these properties in turn.
Sphinx tiles have an intrinsic handedness. The sphinx in Fig.\ \ref{fig:Rep-tiles} (upper right) we label left (``L"), and its mirror reflection we label right (``R"). Different orientations retain their chiral labels as seen in Fig.\ \ref{fig:L-R-sphinxes-1}. This and further properties of the sphinx tile are discussed in appendix\ \ref{sec:nomenclature}. The sphinx is a {\it rep-tile} (as coined by Golomb) because its shape can be dissected into repeated, smaller copies of itself; see the order-2 tiling by four unit sphinxes in Fig.\ \ref{fig:Rep-tiles} (lower right).
Here order refers to the integer scale factor of the sphinx frame over a basic sphinx tile.

\begin{figure}[htbp]
\centering
\includegraphics[width=1\linewidth]{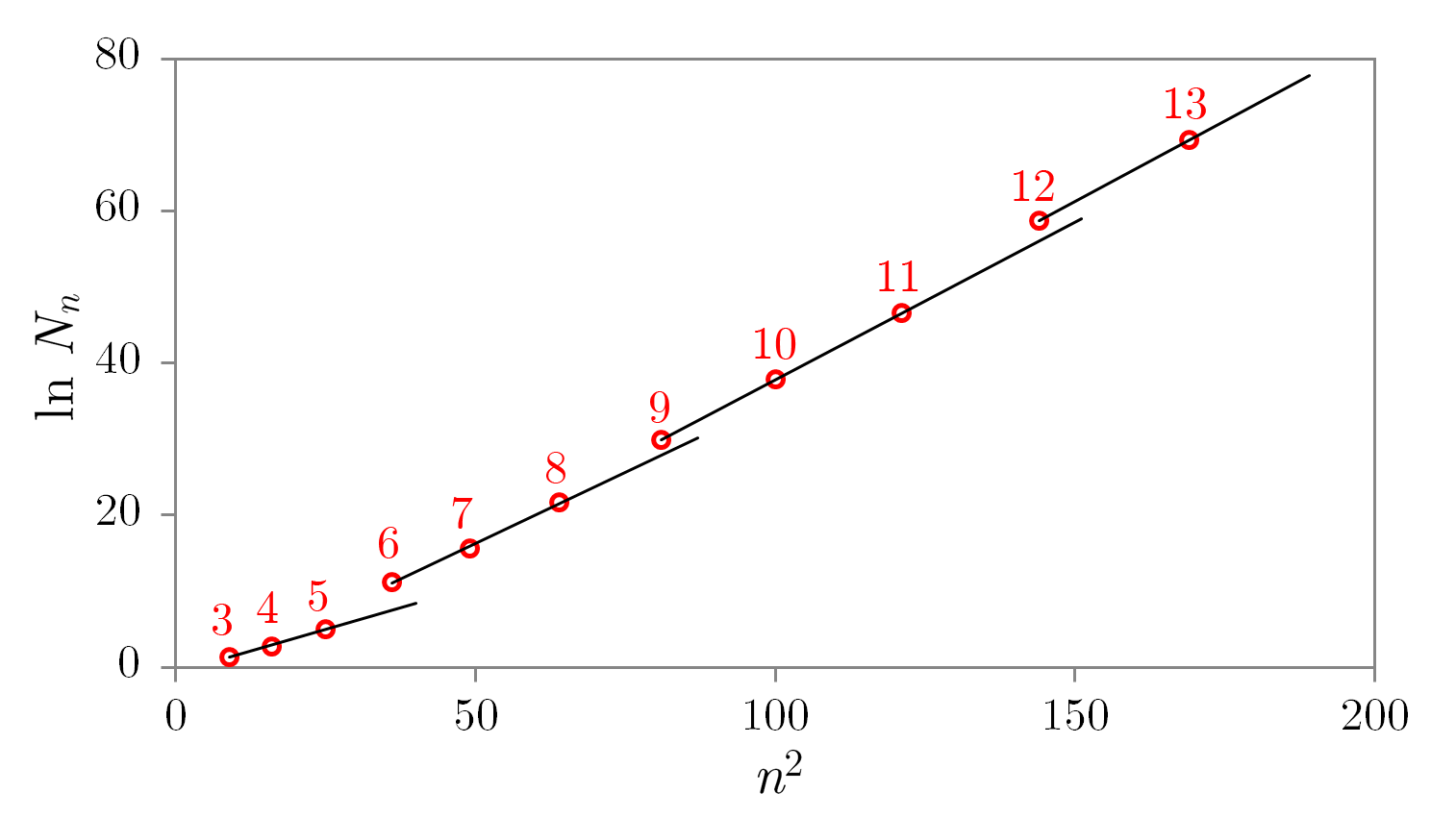}
\caption{$\ln N_n = S/k_B$ where $N_n$ is the number of tilings from table \ref{tab:tilings} vs.\ the square of the order $n$ ($n^2$ is the number of tiles).  Values of $n$ are displayed in red.}
\label{fig:ln-tilings}
\end{figure}

This allows one to recursively create large self-similar, aperiodic tilings
\cite{Godreche89,LeeMoody01,GoodmanStrauss16,GoodmanStrauss18}. 
Rep-tiles are examples of quasi-crystals, which have turned up in a number of contexts \cite{Penrose79,HajiAkbariEtAl09,Senechal95,SocolarTaylor12}, including the recently discovered tiles that force aperiodicity \cite{SmithEtAl1-23,SmithEtAl2-23} and whose physical properties, including chirality, have been studied \cite{SchirmannFrancaFlickerGrushin23,AguilarBarbosaDonangeloSouza23,JungChenGu23}.  Here we consider the ensemble of {\it all} possible tilings of a sphinx frame, not only those produced through a substitution or inflation rule.

\begin{figure}[htbp]
\centering
\includegraphics[width=0.35\linewidth]{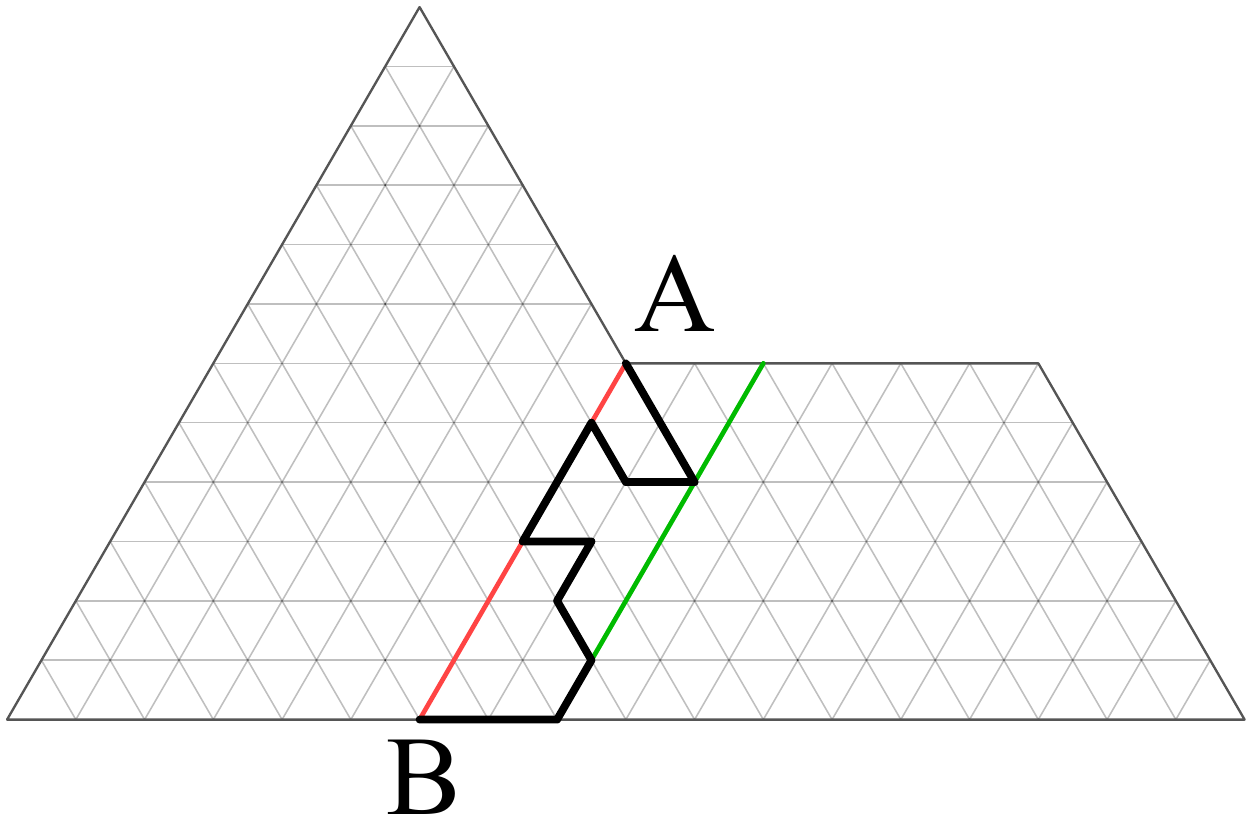}
\includegraphics[width=0.55\linewidth]{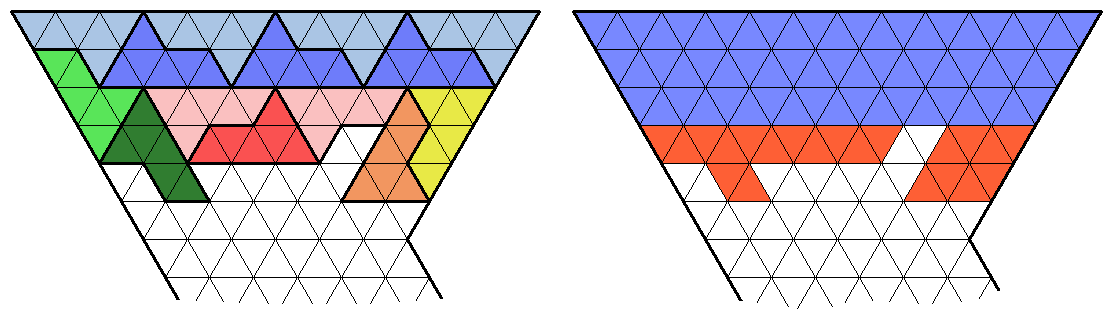}
\caption{Left: Illustration of the seam method, where the seam is the path in black. The tilings of the two regions on either side of the seam are displayed in Fig.\ \ref{fig:seammethod}.  Middle and right: A tiling of the first 3 rows using the dangler method, and its associated dangler shape (red triangular cells).}
\label{fig:dangler-small}
\end{figure}

\begin{figure}[htbp]
\centering
\includegraphics[width=0.6\linewidth]{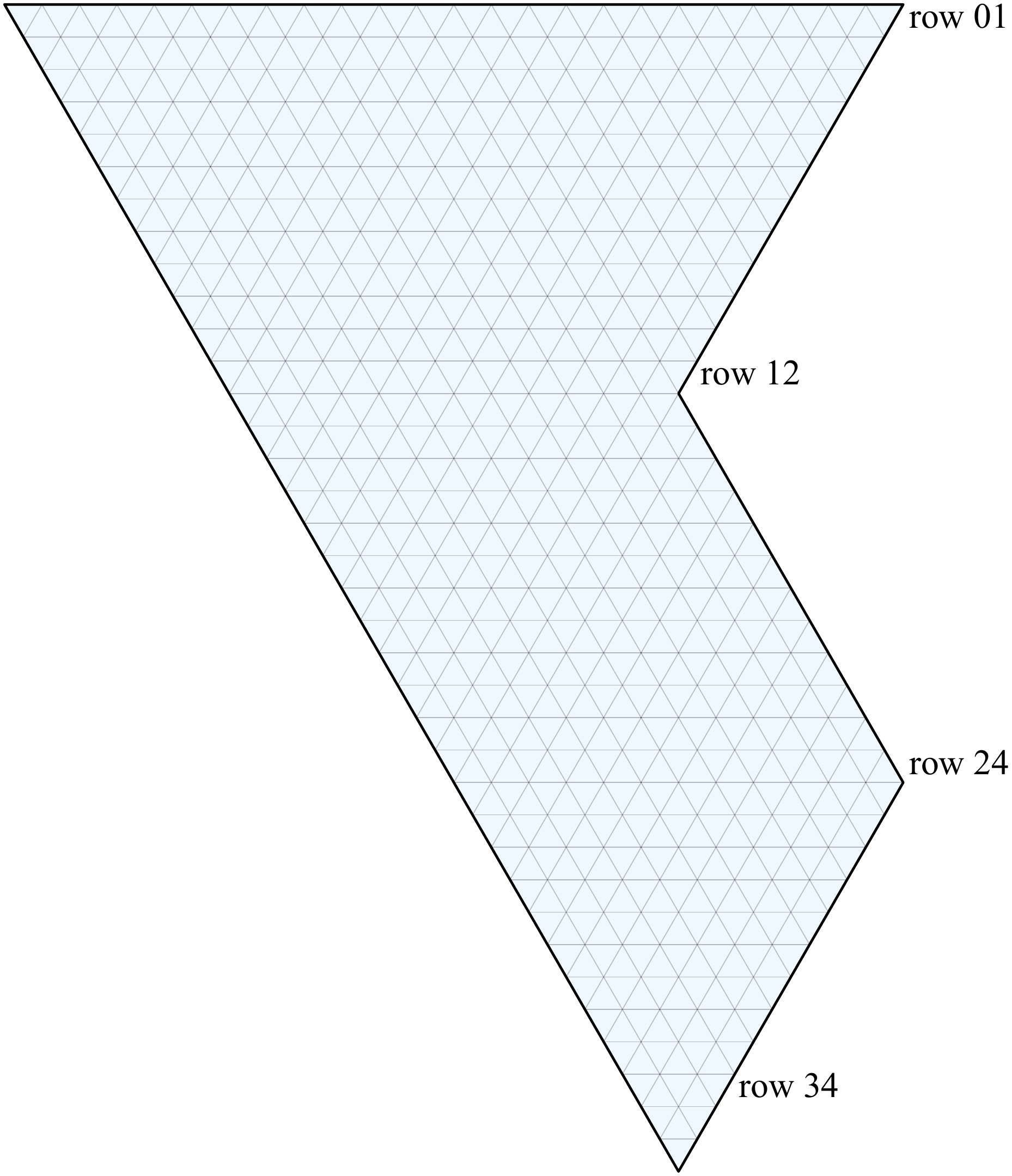}
\caption{An order-12 sphinx frame in orientational state 7 (see Fig.\ \ref{fig:standard-colors}) with rows from 1 to 36 indicated.}
\label{fig:dangler-state-7}
\end{figure}

\begin{figure}[htbp]
\centering
\includegraphics[width=0.9\linewidth]{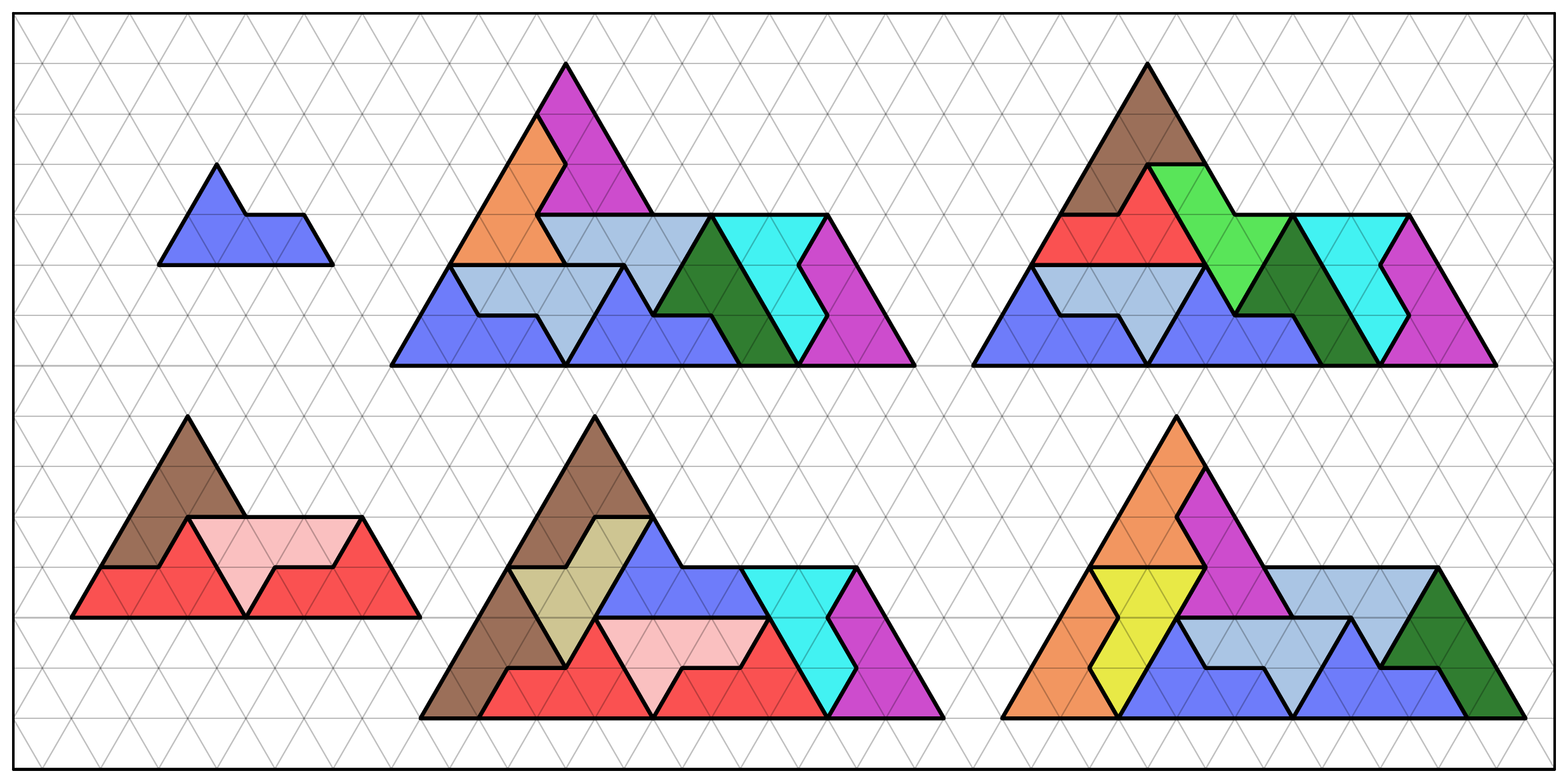}
\caption{All sphinx tilings of orders 1, 2 and 3. 
}
\label{fig:sphinx-1-2-3}
\end{figure}

\begin{figure*}[htbp]
\centering
\includegraphics[width=0.9\linewidth]{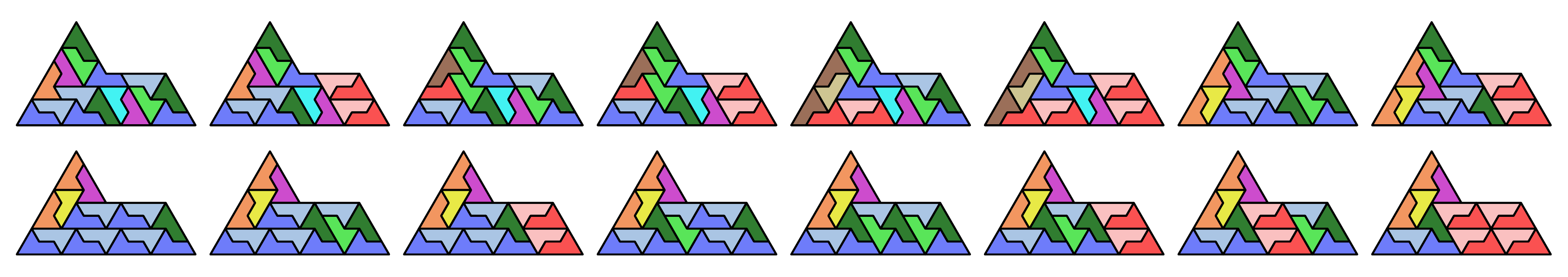}
\caption{The 16 sphinx tilings of an order-4 Sphinx frame.}
\label{fig:tilings-order-4}
\end{figure*}

\begin{figure*}[htbp]
\centering
\includegraphics[width=1.0\linewidth]{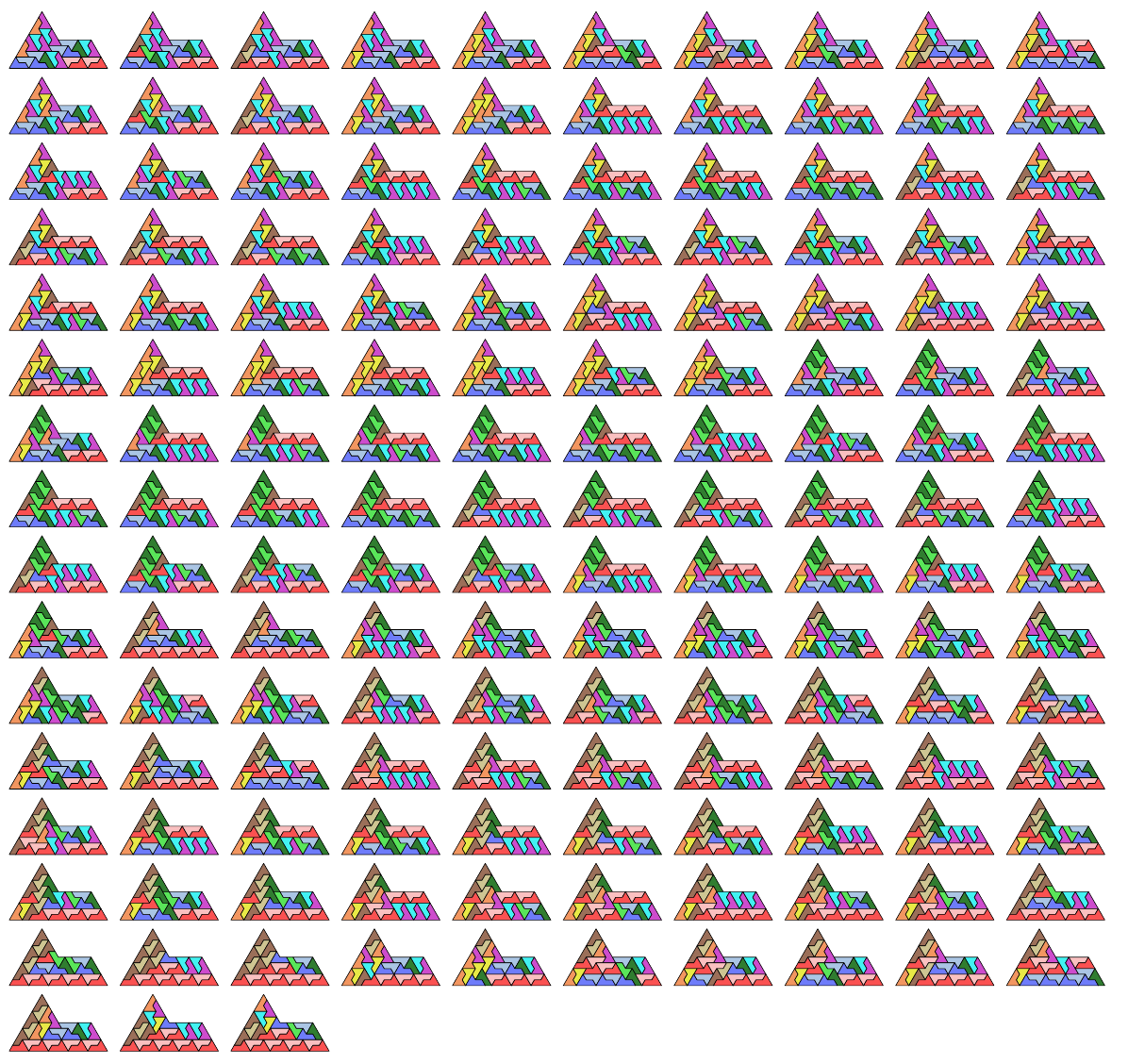}
\caption{The 153 sphinx tilings of an order-5 Sphinx frame.}
\label{fig:153tilingsorder5}
\end{figure*}

\begin{figure}[htbp]
\centering
\includegraphics[width=0.9\linewidth]{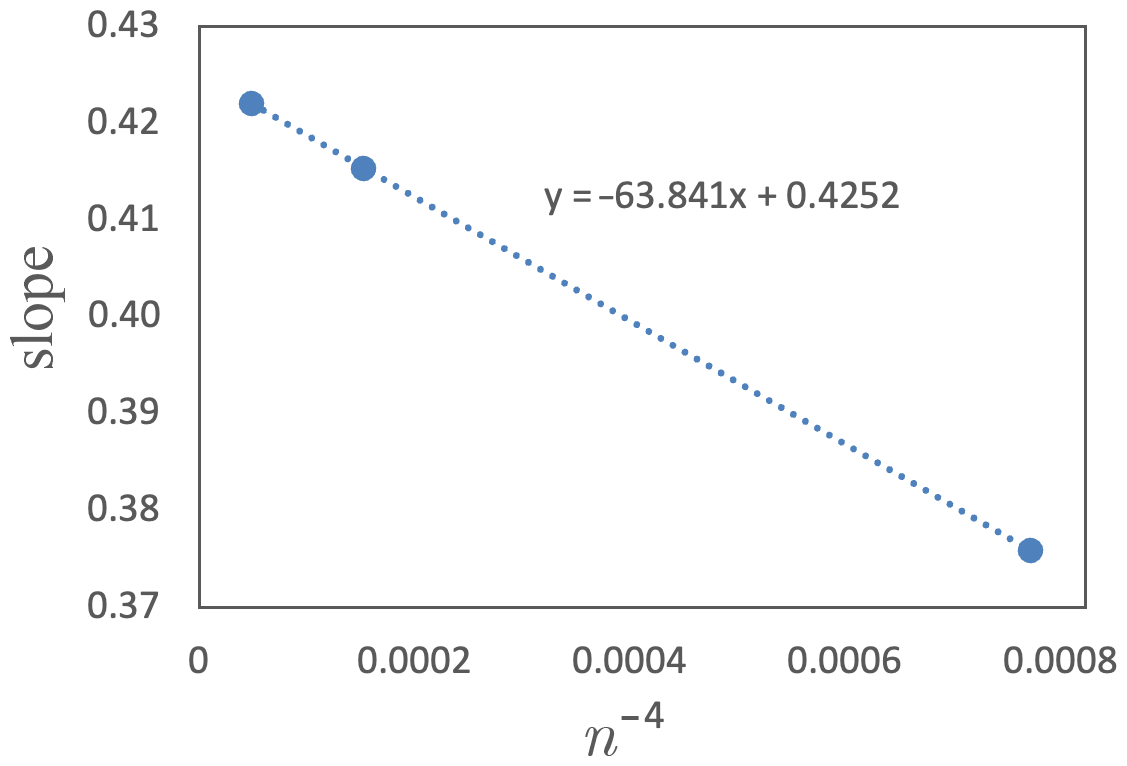}
\caption{A plot of the slopes in Fig.\ \ref{fig:ln-tilings} vs.\  $1/n^4$, implying for large $n$,  $S/k_B = \ln N_n \sim 0.425 n^2 = n^2 \ln 1.53$.}
\label{fig:SphinxSlope}
\end{figure}

We consider that the sphinx tilings obey the partition function
\begin{equation}
    Z(\beta) = \sum_{\{\tau\}} e^{-\beta E[\tau]}
\end{equation}
where $E[\tau]$ is an energy functional depending upon the tiling $\tau$ and $\beta \equiv 1/(k_B T)$.  Below, we consider $E[\tau]$ as an Ising-like or ice-type energy that depends only upon the properties of neighboring tiles, but first we consider the special case $\beta E =  0$, which implies $Z(0)= \sum_i 1 = N_n$ and the entropy is $S = k_B \ln N_n$.

The above special case reduces to a problem in combinatorial geometry: counting sphinx tilings. We mainly consider boundary frames being in the shape of a sphinx tile scaled by an integer factor (the order $n$) --- an {\it n-Sphinx}. 
In general, we define the order of a polyiamond as the greatest common divisor of its side lengths, measured in lattice units. 
All $n$-Sphinxes can be tiled by unit sphinxes, a result that can be shown via an inductive argument. Certain symmetric polyiamonds of unit order exhibit the analogous feature; for example, the first four polyiamonds in the top row of Fig.\ \ref{fig:Rep-tiles}. 
Other than the sphinx, however, there is no known asymmetric polyiamond that can tile every order of itself. This is an outstanding feature of the sphinx tile; it makes the sphinx unique among all asymmetric polyiamonds. It follows that a polyiamond consisting of two or more unit sphinxes (like the 3-by-2 parallelogram) can be tiled by sphinxes at all orders.
Such is not the case for many other frame shapes, where only certain orders can be tiled by sphinxes, such as triangular frames (12, 24, 36, $\ldots$) and regular hexagonal frames (6, 8, 10, $\ldots$). Considering all tilings of a Sphinx frame, it is clear that the number of tilings grows superexponentially with the order (see table\ \ref{tab:tilings} and Fig.\ \ref{fig:ln-tilings}). Previously, sphinx tilings of Sphinx frames have been enumerated up to order 8 \cite{HoriyamaOkamotoUehara15}. Here we extend the exact results to order 13, which allows important insights into this system.

\begin{figure}[htbp]
\centering
\includegraphics[width=0.9\linewidth]{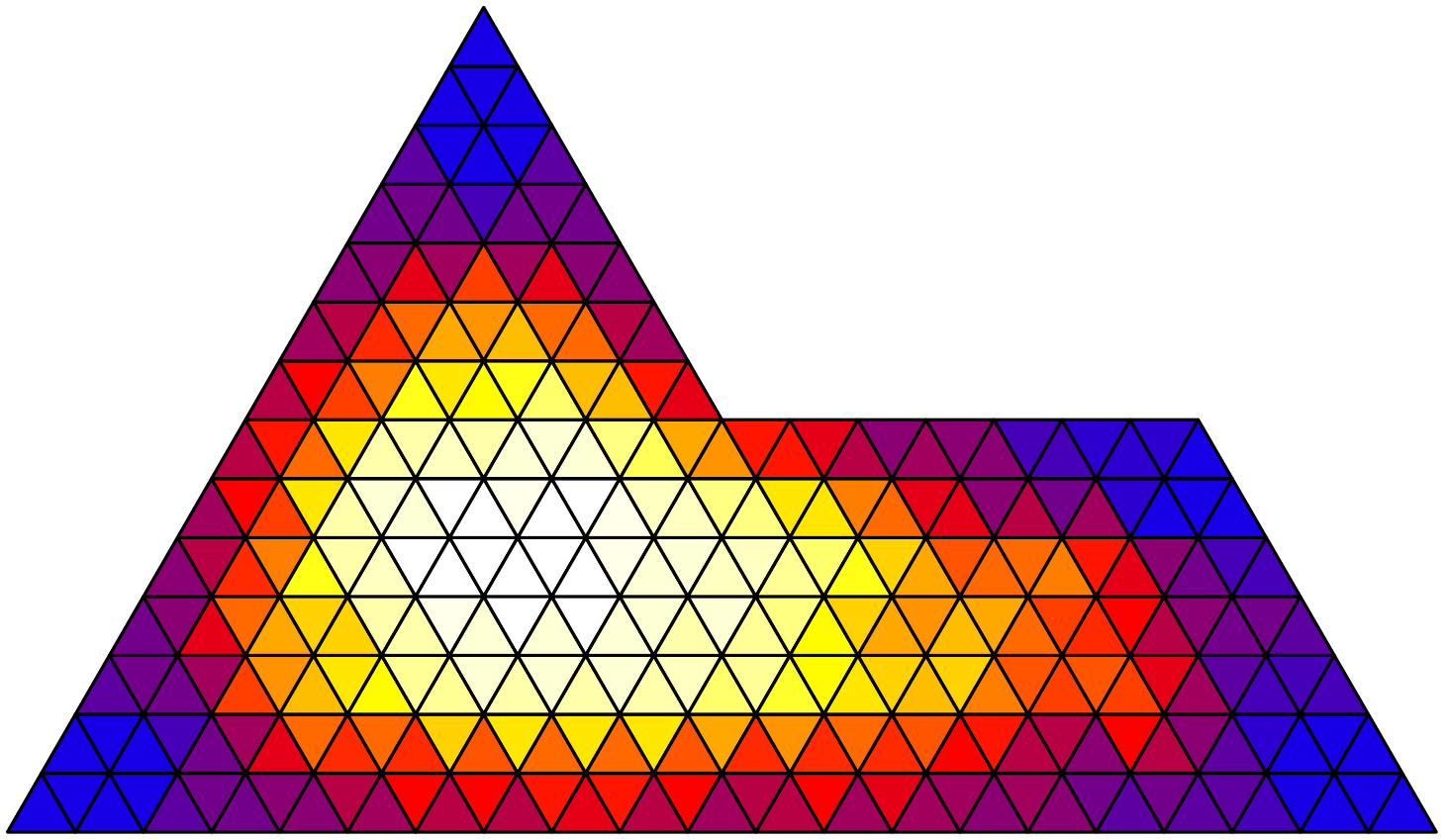}
\caption{The order-7 sphinx.  Colors represent for each triangle the  number of the 36 different ways in which it is covered by sphinx tiles, as shown in Fig.\ \ref{fig:AVtriangle}.} 
\label{fig:cell-covering-7}
\end{figure}

\section{Algorithms}

For the enumerations of tilings of Sphinx frames up to order 7, we used a ``backtracking" method, where we proceed across the triangles of the frame in a row-by-row manner and recursively go back to find other tilings.  This algorithm generates every possible complete tiling, and is limited in terms of the order it can handle. 

For higher-order frames, we have developed the {\it seam method}, a variant of meet-in-the-middle search.  While the Sphinx frame itself has no symmetry, it is decomposable into two symmetrical hemisphinx frames (trapezoids), identical save for their orientation and common edge (the line AB in Fig.\ \ref{fig:dangler-small}).  A seam is a self-avoiding lattice walk from vertex A to vertex B that does not leave the frame. Additionally, the walk must cross neither the red line, nor the green line two rows posterior. The number of seams can be reduced by certain conditions that must be satisfied so as to allow tilings of both areas --- see appendix\ \ref{sec:seam-method}. To make sure a tiling cannot belong to two different seams, we demand that all tiles of the left part of the frame cover at least one grid triangle on the left of the red line and that no tile of the right part crosses the red line. Whereas the number of self-avoiding walks increases very rapidly with the order of the sphinx, the number that are valid seams is vastly smaller.  For order 8, for instance, the number of seams is 1468, while the number of walks is 3\,523\,417. For each possible partition, we determine the number of tilings of the two areas and multiply the two numbers together, and then sum over all possible seams.  This method allows one to generate all tilings by convolution of the tilings from each side of the seams.

For $n > 11$, the severity of the memory constraints in the above methods led us to devise the {\it dangler} method. In principle, the new method produces seams for each row and onlyd the tilings between two consecutive seams have to be enumerated.
Consider initiating a tiling at the top row of Fig.\ \ref{fig:dangler-small} (middle panel) and continuing successively to the next rows.
For each row $R$, we require tilings to cover rows 1 to $R$ entirely, while allowing partial coverage in the next two rows. The shapes encountered in rows $R+1$ and $R+2$ (``danglers," seen in red in the right panel of Fig.\ \ref{fig:dangler-small}) can be represented by two bit strings, in which the $0$s and $1$s stand for empty and covered triangles, respectively.
Different tilings may lead to the same dangler, hence the number of danglers is in most cases smaller than the number of tilings. For each row $R$, we save the two binary integers of each dangler and the number of associated tilings.
Then the next row ($R+1$) is handled for all danglers of row $R$. Say $m$ tilings are associated with a certain dangler of row $R$ and $n$ tilings were found that lead to a certain new dangler of row $R+1$. Then the new dangler represents 
$m\cdot n$ tilings that cover the rows 1 to $R+1$ completely.
For any row, the number of danglers is less than or equal to the sum of all associated tilings.
For the first few rows, both numbers are increasing, but, at some stage, the number of danglers begins to decrease (at a row number dependent upon the frame and its orientation).
For example, the order-12 Sphinx frame in orientation 7 consists of 36 rows as shown in Fig.\ \ref{fig:dangler-state-7}.  For row 18, we found 509\,235 danglers and 347\,201\,208\,446\,538\,108\,883 tilings. 
It is a huge difference whether one must consider 509\,235 cases or 347\,201\,208\,446\,538\,108\,883 cases in the next step.
The number of different coverings of the empty triangles of a row is obtained by backtracking from left to right, ensuring no duplicates can occur.

With the dangler method, it takes less than a second on a desktop computer to calculate the number of sphinx tilings of an 8-Sphinx, and all results from our previous methods up to order 11 could be easily confirmed. 
Order 12 takes about 3 hours, and order 13 about 3 days using up to 6 cores. The bottleneck of the calculation is the memory required to save the danglers, which in practice limits this method to order 13.
The dangler method is much faster than the seam method, but the output of the algorithm is just the number of tilings and no tiling can be produced explicitly.
In contrast, the seam method can be split into several independent parts (corresponding to the different seams), 
and it has the advantage that the tilings of these parts can be saved and used for further study.

\begin{table}
\caption{Number of tilings $N_n$ of a Sphinx frame of order $n$.}
    \centering
    \begin{tabular}{|r|r|}
    \hline
        $n$ &  $N_n$ \\ \hline
        1 & 1 \\ 
        2 & 1 \\ 
        3 & 4 \\ 
        4 & 16 \\ 
        5 & 153 \\ 
        6 & 71\,838 \\ 
        7 & 5\,965\,398 \\ 
        8 & 2\,614\,508\,085 \\ 
        9 & 9\,822\,629\,511\,079 \\ 
        10 & 28\,751\,930\,151\,895\,611 \\ 
        11 & 162\,231\,215\,752\,303\,027\,270 \\ 
        12 & 32\,813\,942\,272\,624\,544\,838\,651\,213 \\ 
        13 & 1\,257\,159\,787\,425\,487\,037\,702\,548\,758\,466 \\ 
        \hline
    \end{tabular}
    \label{tab:tilings}
\end{table}

\begin{figure*}[htbp]
\centering
\includegraphics[width=0.465\linewidth]{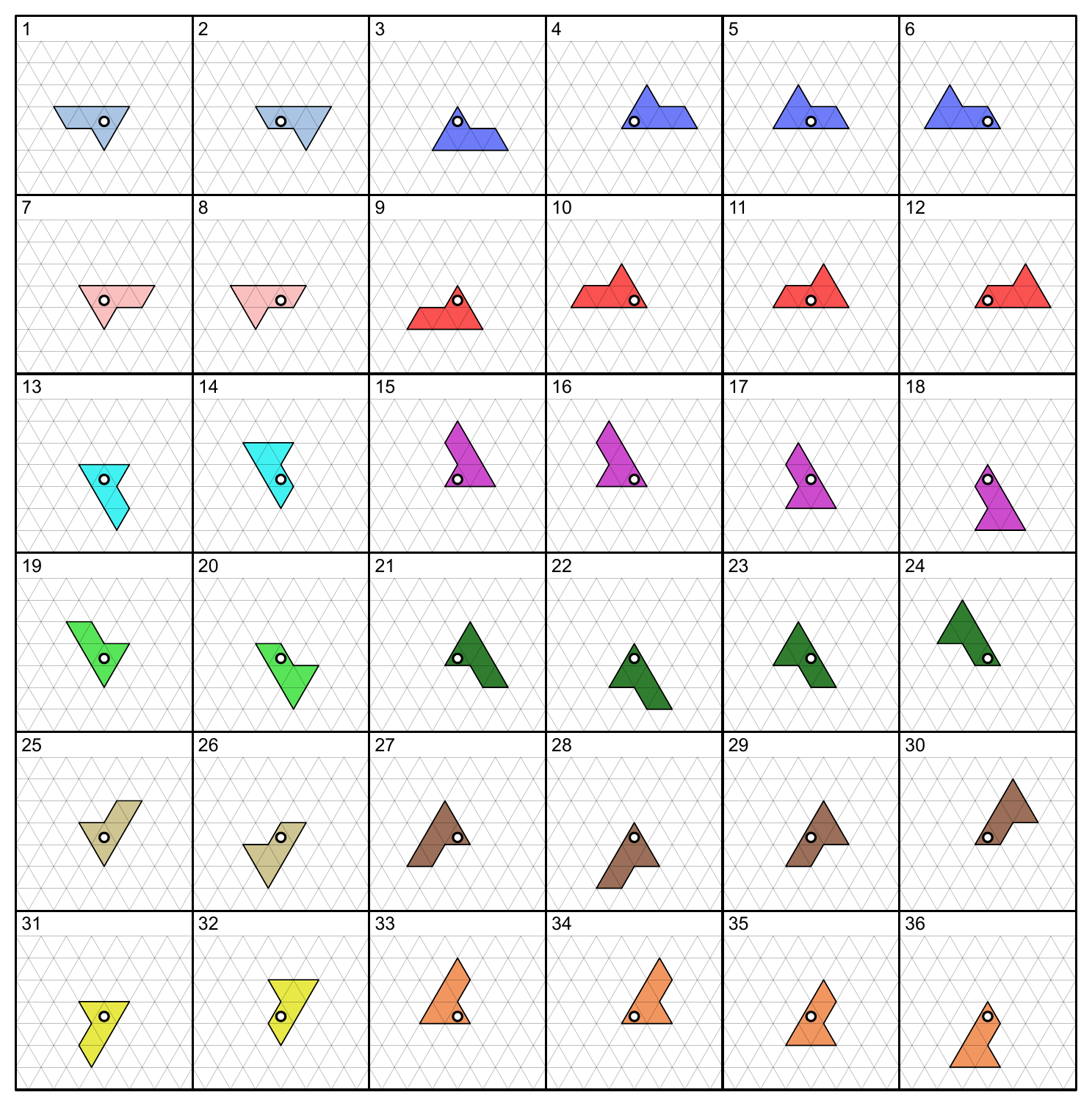}
\includegraphics[width=0.4\linewidth]{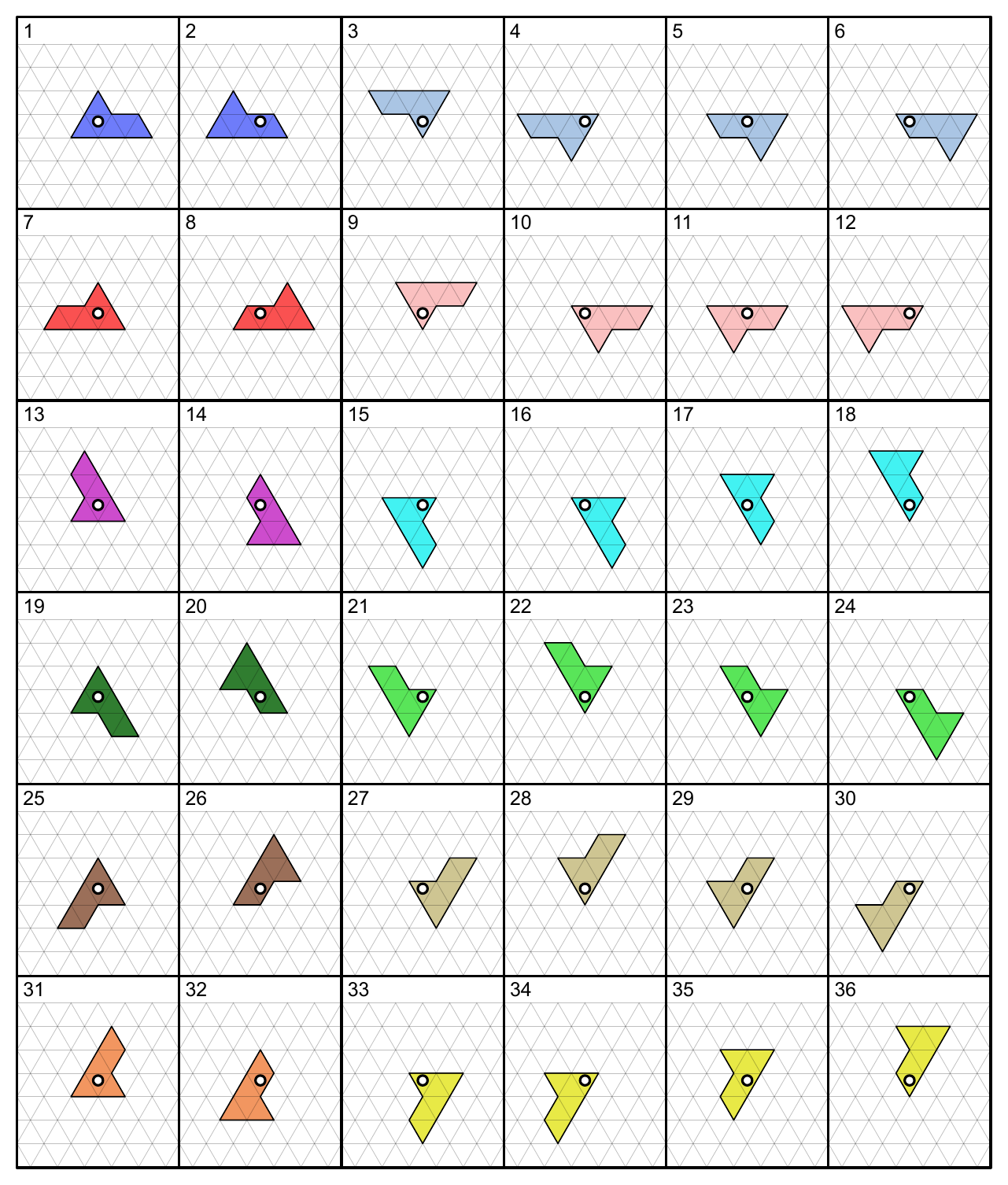}
\caption{The 72 different ways to cover a grid triangle by a certain face of a sphinx tile in a certain state, 36 for each orientation of the circled triangle (A on the left and V on the right).}
\label{fig:AVtriangle}
\end{figure*}

\section{Enumeration Results and Entropy} 

The exact numbers of tilings $N_n$ from the three methods combined are listed in table \ref{tab:tilings}. %and have been posted on \cite{OEISA279887}.
%The 153 tilings of 5-Sphinx are shown in Fig.\ \ref{fig:153tilingsorder5}.
For small $n$ all tilings of $n$-Sphinxes are shown in Fig.\ \ref{fig:sphinx-1-2-3} ($n = 1, 2, 3$), Fig.\ \ref{fig:tilings-order-4} ($n = 4$), Fig.\ \ref{fig:153tilingsorder5} ($n = 5$).
For $n=13$, the number of tilings is greater than $10^{30}$, highlighting the rapid growth in the number of tilings and the power of the dangler method.
The asymptotic behavior of $N_n$ can be surmised by plotting its logarithm as a function of $n^2$, the number of sphinx tiles in an $n$-Sphinx, as shown in  Fig.\ \ref{fig:ln-tilings}.
The data points (red circles) indicate a periodicity (mod 3), such that Sphinx frames of order $3k$ have more tilings than expected when looking at lower orders; we have no definitive explanation for this intriguing behavior.  

\begin{figure}[htbp]
\centering
\includegraphics[width=0.9\linewidth]{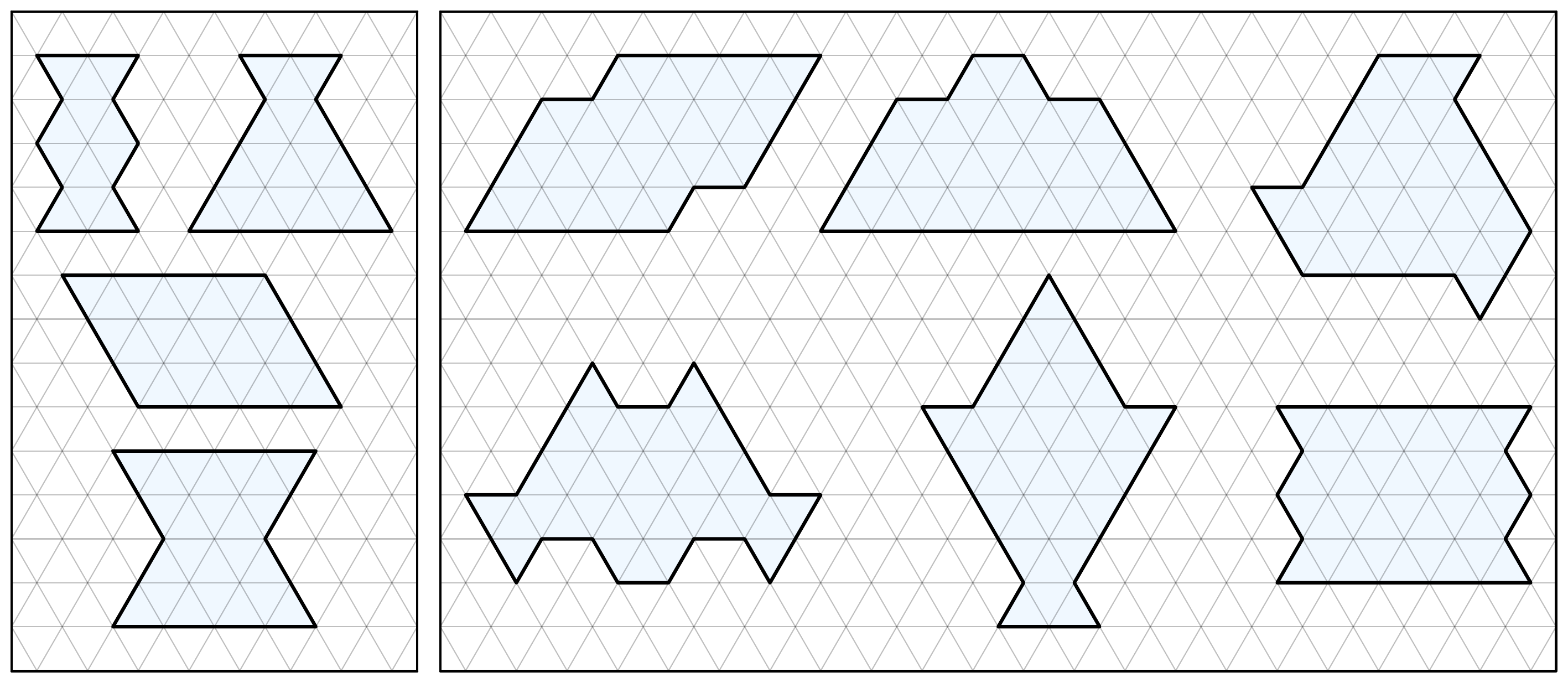}
\includegraphics[width=0.9\linewidth]{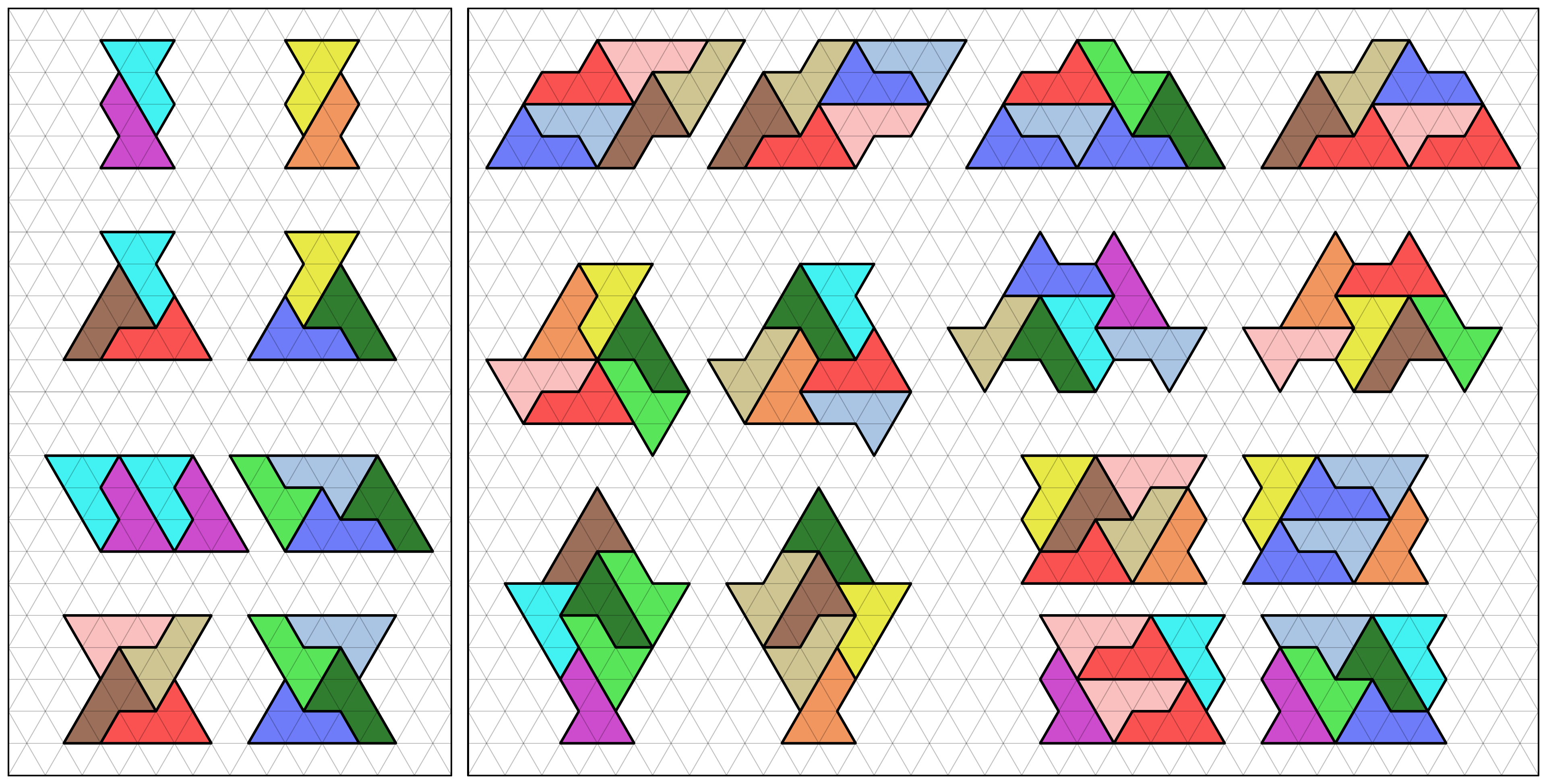}
\caption{Fundamental polyads of size 2, 3, 4, and 4 (upper left panel) and size 6 (upper right panel), and their multiple tilings (lower panels).}
\label{fig:fund-polyads-2-6}
\end{figure}

The equations of the linear fits in Fig.\ \ref{fig:ln-tilings} are:   $\ln N_n = 0.2288 n^2 - 0.7508$ ($n=$3--5),
$\ln N_n = 0.3759 n^2 - 2.5114$ ($n =$ 6--8), $\ln N_n = 0.4154 n^2 - 3.703$ ($n =$ 9--11), and  $\ln N_n = 0.4221 n^2 - 2.0353$ ($n=$ 12,13). 
Fig.\ \ref{fig:SphinxSlope} shows a plot of the three latter slopes as a function of $1/n^4$
implying $N_n \sim \exp(0.425 n^2-63.8/n^2)$ or an entropy that grows as
\begin{equation}
   S = k_B \ln N_n \sim k_B n^2 \ln \sigma
\end{equation}
where $\ln \sigma = 0.425$ is the entropy per tile, and $\sigma = \exp(0.425) = 1.53(1)$ is the ``sphinx constant."
The existence of this constant is not obvious, though not unexpected either.  It implies that the entropy is extensive.
 For comparison, the number of domino tilings in the Aztec diamond is given exactly by $N_n = (\sqrt{2})^A$, where $A = n(n+1)$ is the area of an order-$n$ diamond in domino units, implying $\sigma = \sqrt{2} \simeq 1.41421$ \cite{JockuschProppShor95,ElkiesKuperbergLarsenPropp92}. For a domino tiling of a rectangular boundary of dimensions $2n \times 2m$ containing $2 n m$ dominos \cite{TemperleyFisher61,Kasteleyn61,IzmailianPopoyanZiff19},  
 \begin{equation}
 \ln N_n \sim (2N+1)(2M+1) G/\pi + \ldots
 \end{equation}
 for large $N$ and $M$, where $G$ is the Catalan constant $\simeq 0.915966$, implying $\sigma = \exp(2 G/\pi) \simeq 1.79162$.
 For a square-rhombus tiling of an octagonal region, one has $\sigma = \exp(0.36021(3)) = 1.4466$ \cite{HutchinsonWidom15}.
 For a tiling of rhombuses within a hexagonal boundary on the triangular lattice, $\sigma = 3 \sqrt{3}/4 \simeq 1.29903$.  
 A random tiling of Penrose tiles was found to yield $\sigma = \exp(0.495) = 1.640$ \cite{ChenSpaepen88}.
For the ice-model, the entropy constant is $\sigma = 3/2$ in the mean-field limit \cite{Pauling35}, and  exactly $(4/3)^{3/2} \simeq 1.5396$ in two dimensions \cite{Lieb67b}, similar to the value we found here for sphinxes. Thus, the sphinx constant $\sigma$ falls generally in the range of other tiling entropy constants.

In Fig.\ \ref{fig:cell-covering-7}, we use a color map to show the number density of states out of the 36 possible states allowed a given triangular cell within the order-7 frame (Fig.\ \ref{fig:AVtriangle}). The numbers range from the maximum (36) shown in white to the minimum (4) shown in indigo. Note the possible emergence of an arctic-type interface for larger frames.

We note that once lower-order tilings have been identified, it is possible to carry out a general inflation or substitution process to generate higher-order sphinx tilings, as a generalization of the rep-tile process.  For example, each of the 4 tilings of the 3-Sphinx can be tiled with any of the 16 tilings of order 4, yielding $4\cdot16^4=262144$ tilings of the 12-Sphinx, among other possible inflations. However, all inflationary tilings taking together add up to an insignificant fraction of the $3.28\cdot10^{25}$ total tilings we found for the 12-Sphinx.

\section{Polyads}

\begin{figure}[htbp]
\centering
\includegraphics[width=0.5\linewidth]{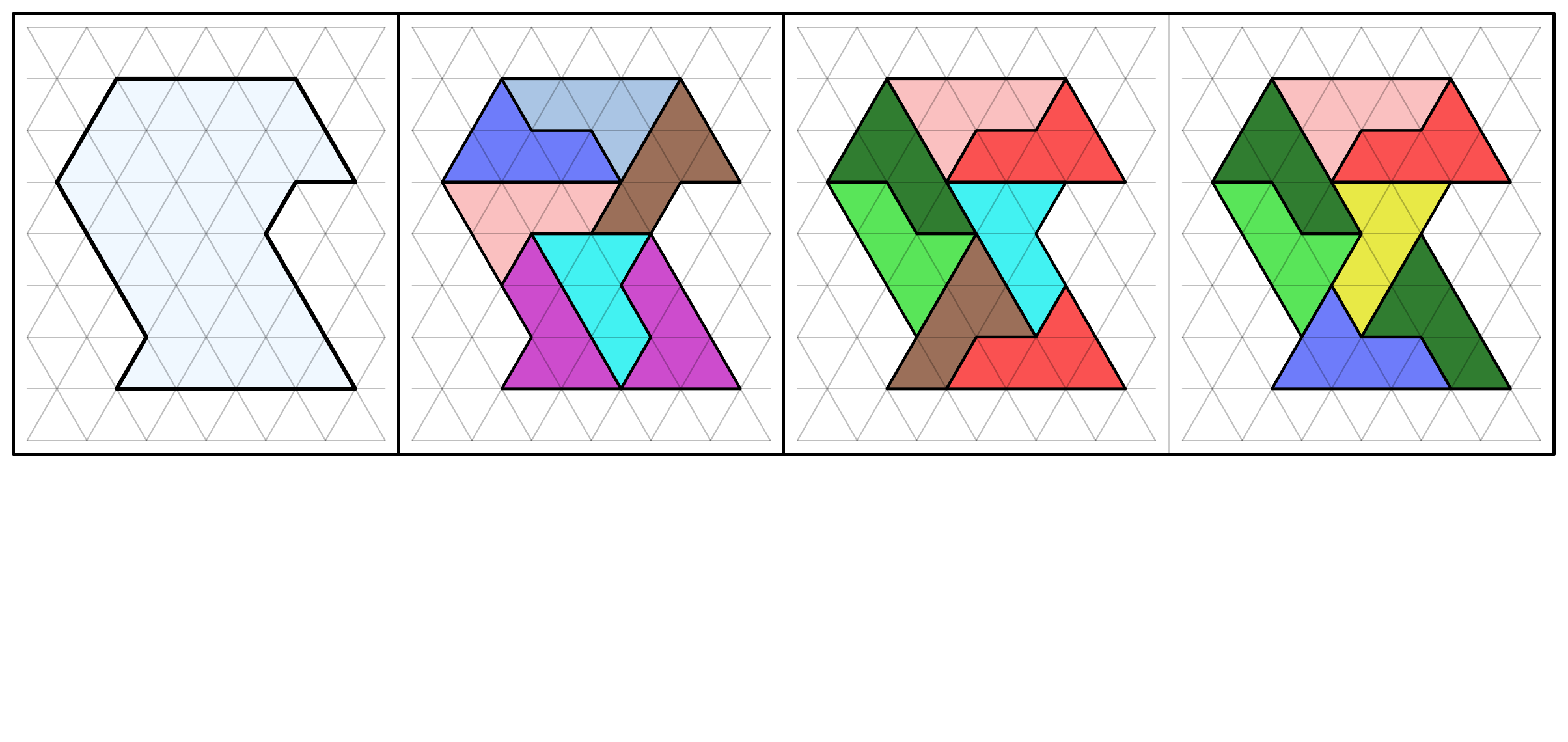}
\includegraphics[width=0.7\linewidth]{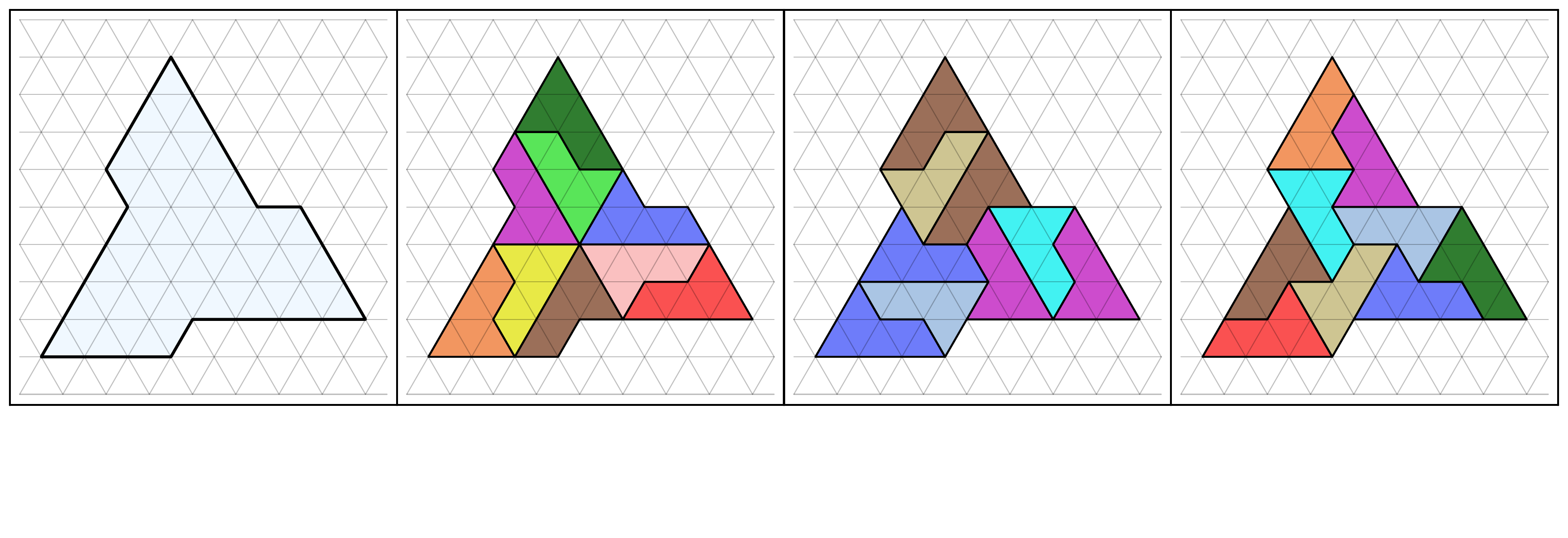}
\caption{Upper: An asymmetric fundamental heptad (polyad of size 7) with 3 tilings. The first one is disjoint to the others, but tilings 2 and 3 are not disjoint to each other. Lower: The ``mystical triode," a 3-fold symmetric fundamental ennead (polyad of size 9) with 3 mutually disjoint tilings.}
\label{fig:heptad-non-sym-triode}
\end{figure}

For the purposes of our discussion, we define a {\it polyad} of size $n$ as a simply connected polyiamond consisting of $6n$ triangular cells and tileable by $n$ sphinx tiles.  In other words, the size of a polyad is the area of that polyad measured in sphinx units. The polyad is a frame, and it does not refer to any specific tiling.
Two polyads are considered {\it equal} if they can be transformed into each other by rotation, reflection or translation. A polyad can have up to 12 different states, due to rotation and reflection.
Two tilings of a given polyad are {\it disjoint} if there are no sphinx tiles in the same position and orientation in the two tilings.
Two {\it sets of tilings} of the same polyad are disjoint if tilings of the first and second set are pairwise disjoint.
A {\it fundamental polyad} (or {\it f-polyad}) has more than one tiling that can be split into two disjoint sets of tilings.
Many small f-polyads have only two tilings and thus the two disjoint sets have only one element each (Fig.\ \ref{fig:fund-polyads-2-6}). Up to size 6, there exists only one f-polyad with more than two tilings: its 4 tilings appear in the lower-right of Fig.\ \ref{fig:fund-polyads-2-6}. The upper two tilings of that f-polyad belong to one set and the lower two tilings to the other set. As seen here, each set contains tilings that are not disjoint from each other. Both sets may be further split into disjoint sets. We have found only one f-polyad with more than two mutually disjoint sets of tilings, the ``mystical triode" of Fig.\ \ref{fig:heptad-non-sym-triode} (lower panel).

Rotation or mirror reflection of a given tiling of a symmetric frame may produce a distinct tiling, and such polyads can be fundamental (see Fig.\ \ref{fig:fund-polyads-2-6}). The smallest size where asymmetric f-polyads occur is size 7 (see an example in Fig.\ \ref{fig:heptad-non-sym-triode}).  As we shall see below, f-polyads will play a central role in our method to switch from one tiling to another without re-tiling the whole frame of the system.  

As an example of f-polyads in a different context, consider the domino tiling where a $2 \times 2$ square can be tiled with either two horizontal or two vertical dominos.  Thus, the square is an f-polyad for this system, and in fact it has been proven that by making successive changes in the tilings of these f-polyads, one can generate all possible domino tilings in a rectangular system of even side length \cite{KamioKoizumiKakazawa23}.  In Ref.\ \cite{LubyRandallSinclair01}, the authors have also considered larger regions (f-polyads) for this system.  Another example of an f-polyad is given by the tiling of a triangular lattice by lozenges (rhombuses), in which a hexagon of six triangles can be tiled by three lozenges in two different ways \cite{Kenyon93}.  By repeating this retiling, it is possible to generate all rhombic tilings of this system.

\begin{figure}[htbp]
\centering
\includegraphics[width= 0.8 \linewidth]{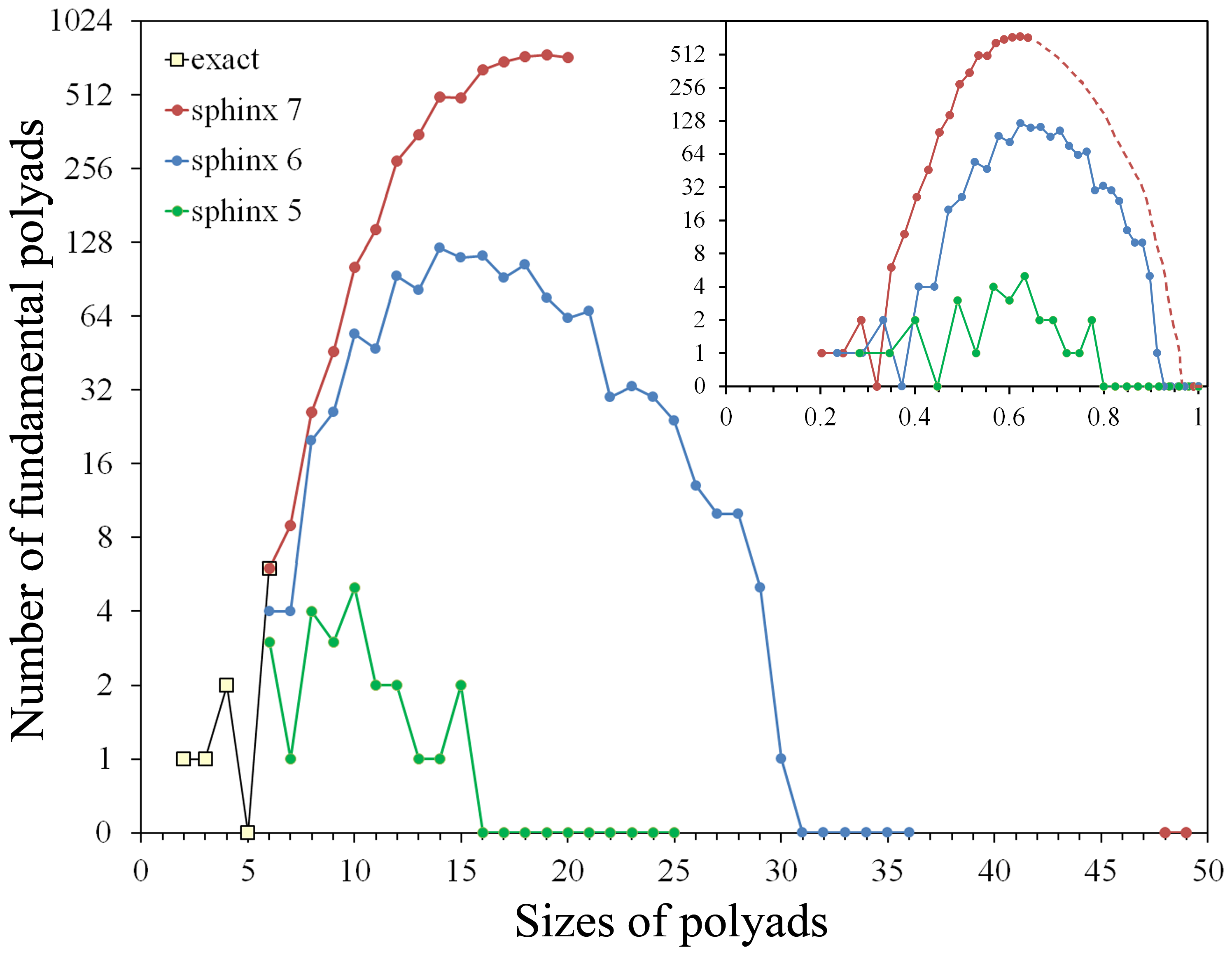}
\caption{Spectrum of fundamental polyads (number of distinct f-polyads for each size) found in the tilings of Sphinx frames of orders 5, 6 and 7. Inset shows a scaling plot of the number vs.\ $\sqrt{\hbox{size}}/$order, with a dashed curve connecting the measured data for size $\le 21$ to the two points at size 48 and 49 for order 7.  The vertical scale is logarithmic except for the point 0.}
\label{fig:fund-polyads-spectrum}
\end{figure}

\begin{figure*}[htbp]
\centering
\includegraphics[width=0.9\linewidth]{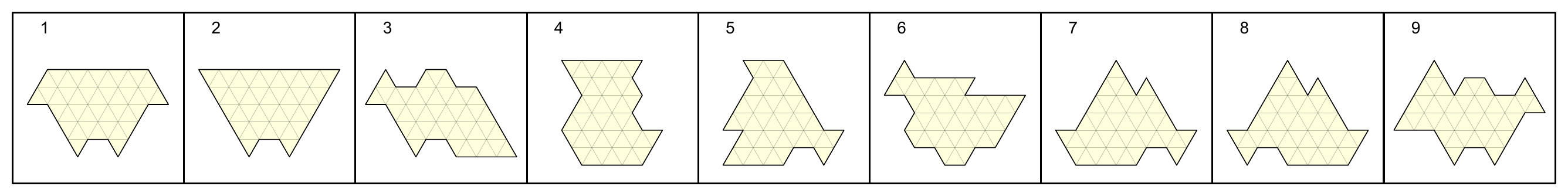}
\includegraphics[width=0.15\linewidth]{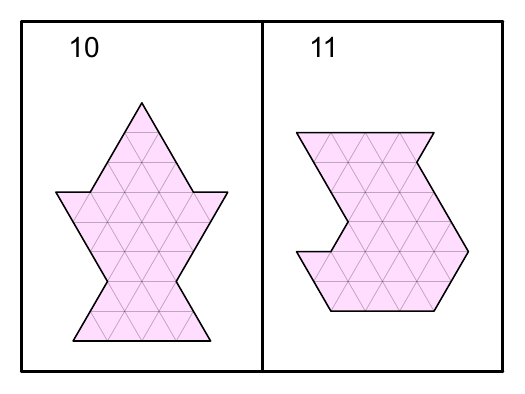}
\caption{Fundamental heptads (polyads of size 7). The nine heptads found in tilings of order-7 Sphinx frames (upper row), and two additional heptads not found in the order-7 tilings but are found in tilings of a sphinx of order 8 (lower row).}
\label{fig:fund-heptads}
\end{figure*}

\begin{figure*}[htbp]
\centering
\includegraphics[width=0.8\linewidth]{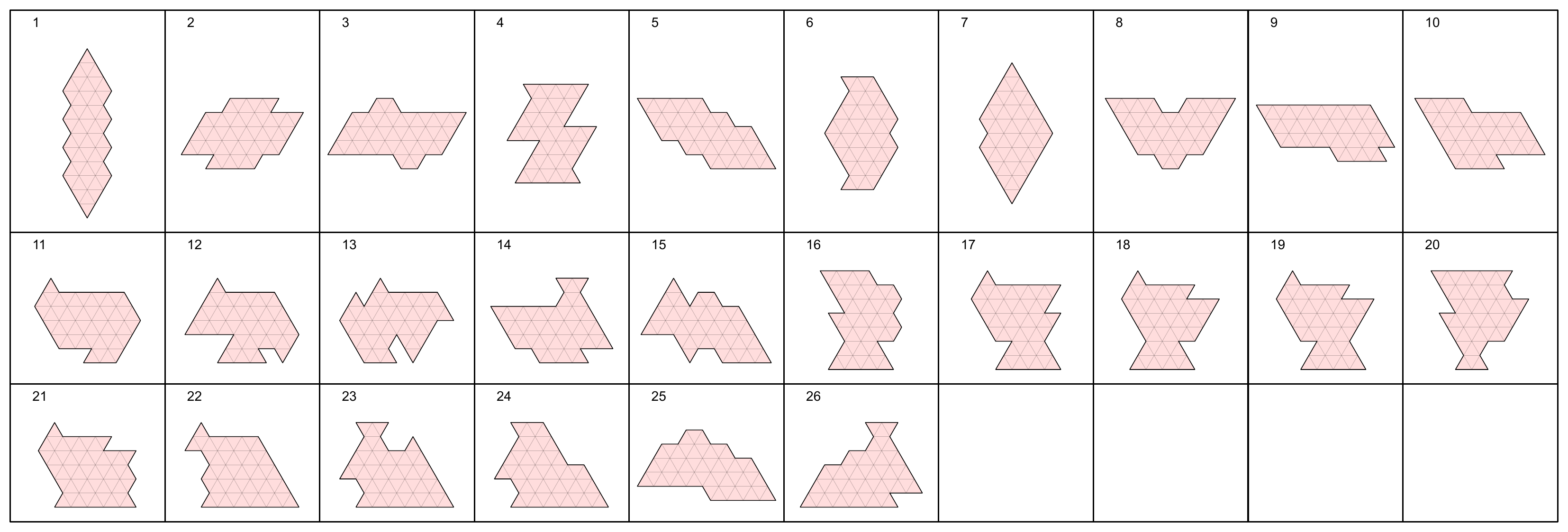}
\caption{Fundamental octads (polyads of size 8) found in tilings of the order-7 Sphinx.} 
\label{fig:fund-octads}
\end{figure*}

\begin{figure*}[htbp]
\centering
\includegraphics[width=1.0\linewidth]{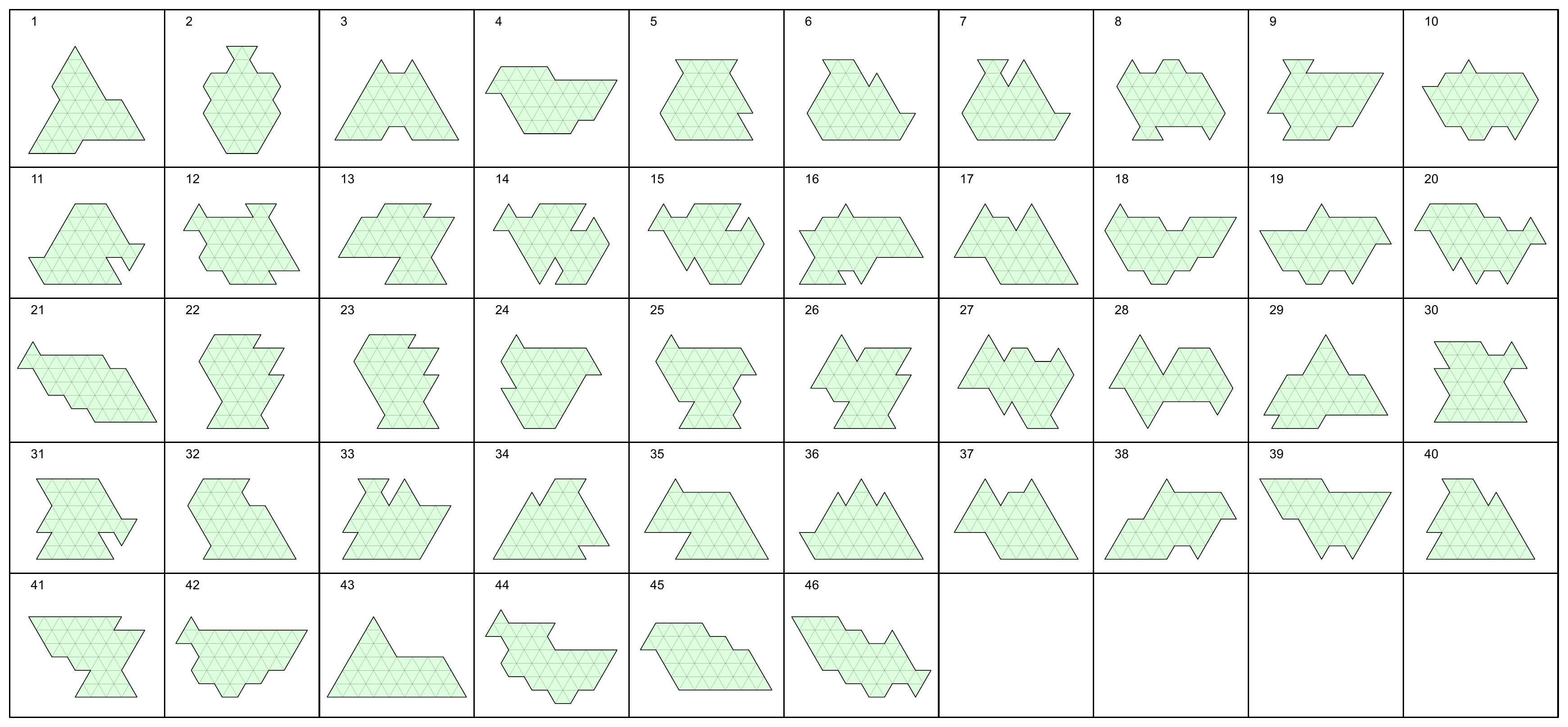}
\caption{Fundamental enneads (polyads of size 9) found in tilings of the order-7 Sphinx.
Number 1 is the ``mystical triode" that has three distinct tilings (Fig.\ \ref{fig:heptad-non-sym-triode}). Number 43 is the Sphinx frame of order 3 shown in Fig.\ \ref{fig:sphinx-1-2-3}, where the bottom right tiling has no common tiles with the other three tilings of the S3 frame, making it fundamental.}
\label{fig:fund-enneads}
\end{figure*}

\begin{figure*}[htbp]
\centering
\includegraphics[width=0.8\linewidth]{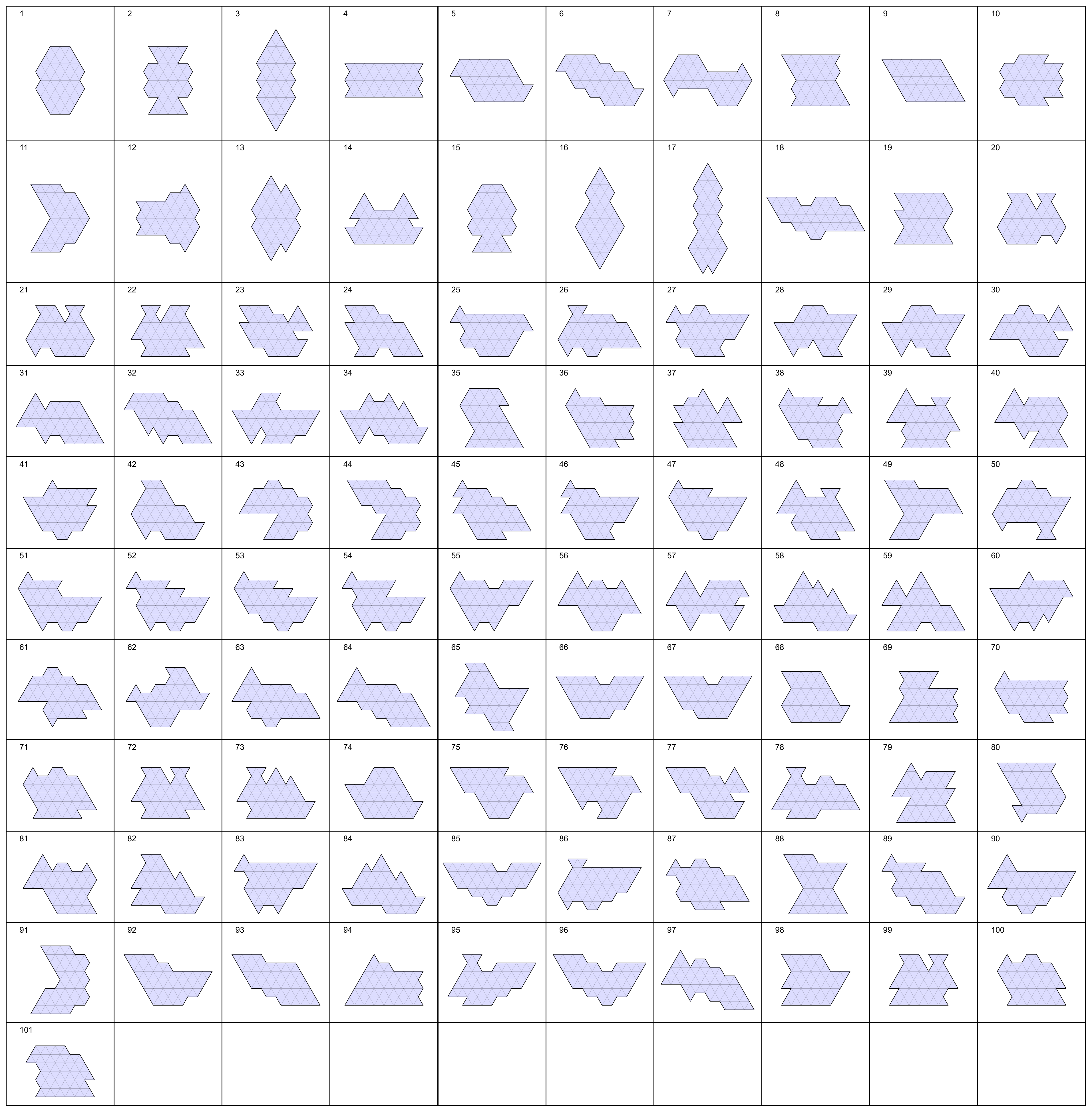}
\caption{Fundamental decads (polyads of size 10) found in tilings of the order-7 Sphinx.}
\label{fig:fund-decads}
\end{figure*}

All f-polyads up to size 6 were found exhaustively and their statistics are listed in table \ref{tab:polyad}. To efficiently find polyads it useful to describe them by closed self-avoiding walks and identify sections that do not allow disjoint tilings, thus eliminating those polyads. The f-polyad spectrum represents how many distinct f-polyads occur in a given frame as a function of their size.  The f-polyad spectra in tilings of Sphinx frames of order 5, 6 and partial results for order 7, are graphed in Fig.\ \ref{fig:fund-polyads-spectrum}. There are ten
fundamental polyads of size 1 to 6, as displayed in Fig.\ \ref{fig:fund-polyads-2-6}. 
Fundamental monads and pentads do not exist;
f-polyads for size 7, 8, 9, and 10 are given in Figs.\ \ref{fig:fund-heptads}, \ref{fig:fund-octads}, \ref{fig:fund-enneads}, and \ref{fig:fund-decads} respectively.

\begin{table}
\caption{Polyads of sizes $n=1$ to 6 and their properties.  The 46 dyads and their tilings are displayed in Figs.\ \ref{fig:dyadframes} and \ref{fig:dyadtilings}d.
}
    \centering
    \begin{tabular}{|c|c|r|r|r|}
    \hline
 $n$ & name 
   &   polyads \qquad  %put in for centering
    &\begin{tabular}{@{}c@{}}polyads with \\multiple tilings\end{tabular} 
   &\begin{tabular}{@{}c@{}}f-polyads \end{tabular}   \\
    \hline 
    1 & Monad  & 1 & 0 & 0 \\ 
    2 & Dyad  & 46  & 1 & 1  \\ 
    3 & Triad &  1\,868  & 25 & 1  \\ 
    4 & Tetrad & 98\,733  & 1\,940 & 2  \\ 
    5 & Pentad &  5\,449\,410  & 138\,865 & 0  \\ 
    6 & Hexad &  311\,784\,564 & 9\,816\,368 & 6 \\ \hline
    \end{tabular}
    \label{tab:polyad}
\end{table}

\section{Monte Carlo methods}

\begin{figure}[htbp]
\centering
\includegraphics[width=0.9\linewidth]{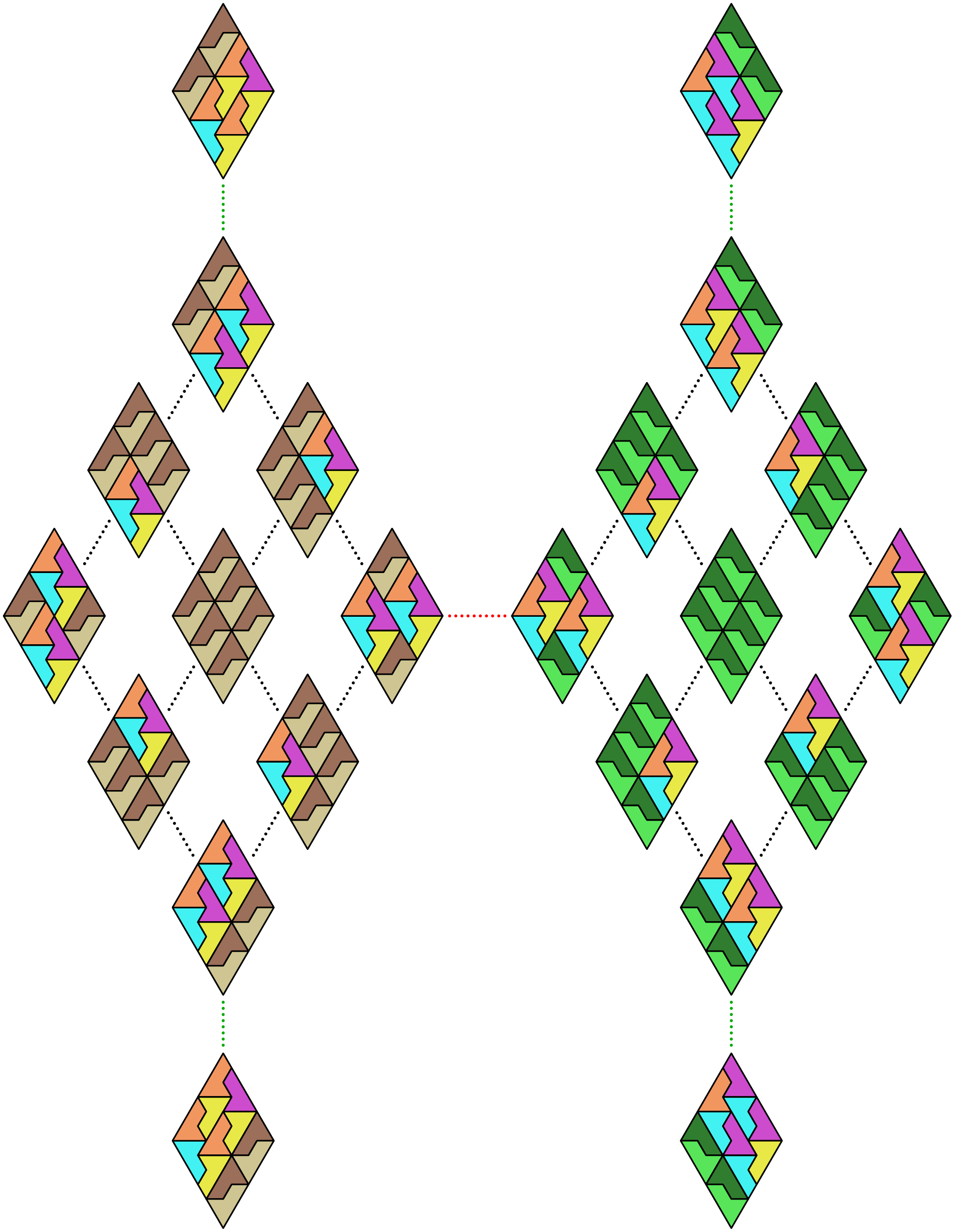}
\vspace*{10mm}
\includegraphics[width=0.4\linewidth]{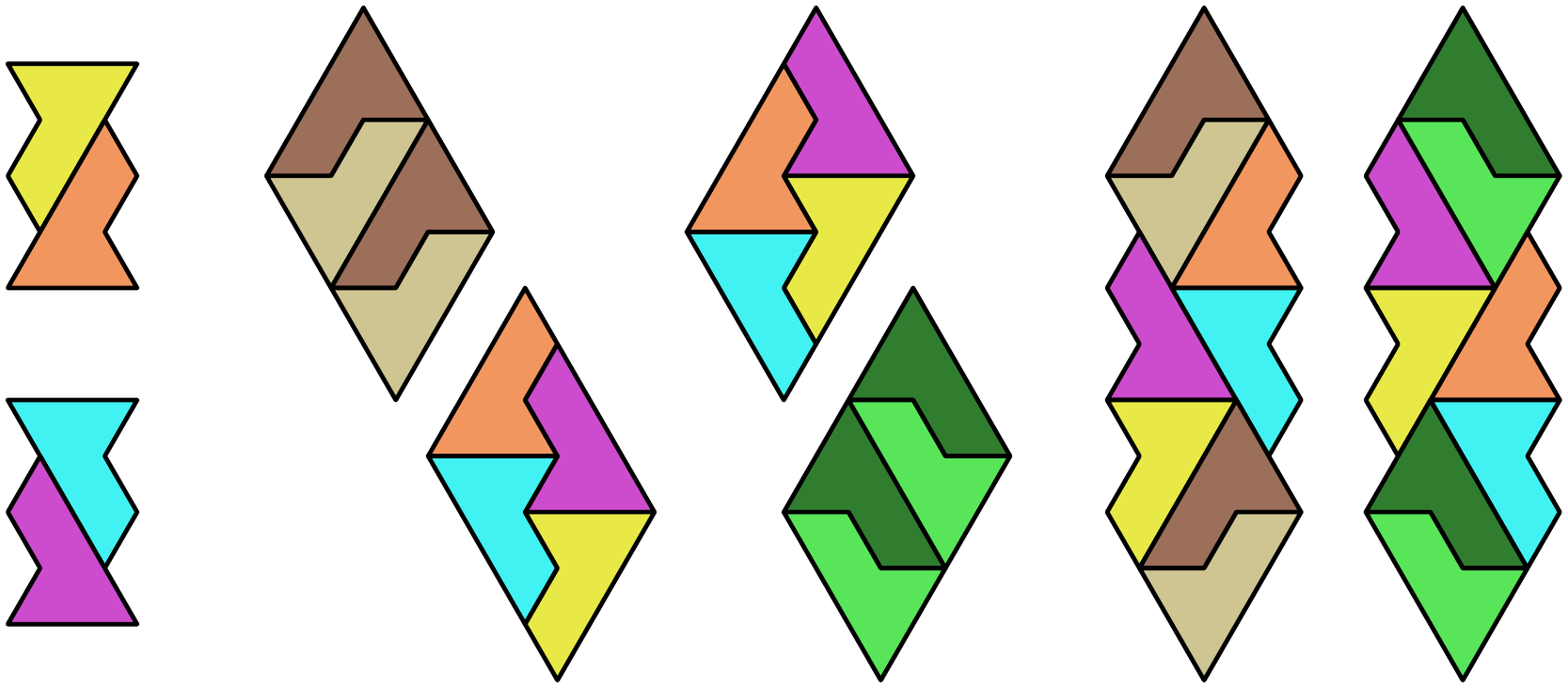}
\caption{Fundamental transitions between the 22 sphinx tilings of the order-6 Diamond (D6). 
These transitions allow the net chirality to change from $+12$ (center left) to $-12$ (center right) in 5 steps. The tilings of each of the 3 necessary f-polyads are shown 
on the bottom.
The black dotted lines show the 16 transitions using the $3 \times 4$ f-tetrad, the four green dotted lines represent the bowtie transition, while the central dotted red line indicates the transition of octad 1 from Fig.\ \ref{fig:fund-octads}.  That octad has mirror symmetry and the associated transition is just the mirror reflection (also a $180^\circ$ rotation) and has the net effect of switching all brown and light-brown tiles with green and light-green tiles, respectively, as seen in the lower-right bottom.}
\label{fig:diamond-6-transitions}
\end{figure}

\begin{figure}[htbp]
\centering
\includegraphics[width=0.9\linewidth]{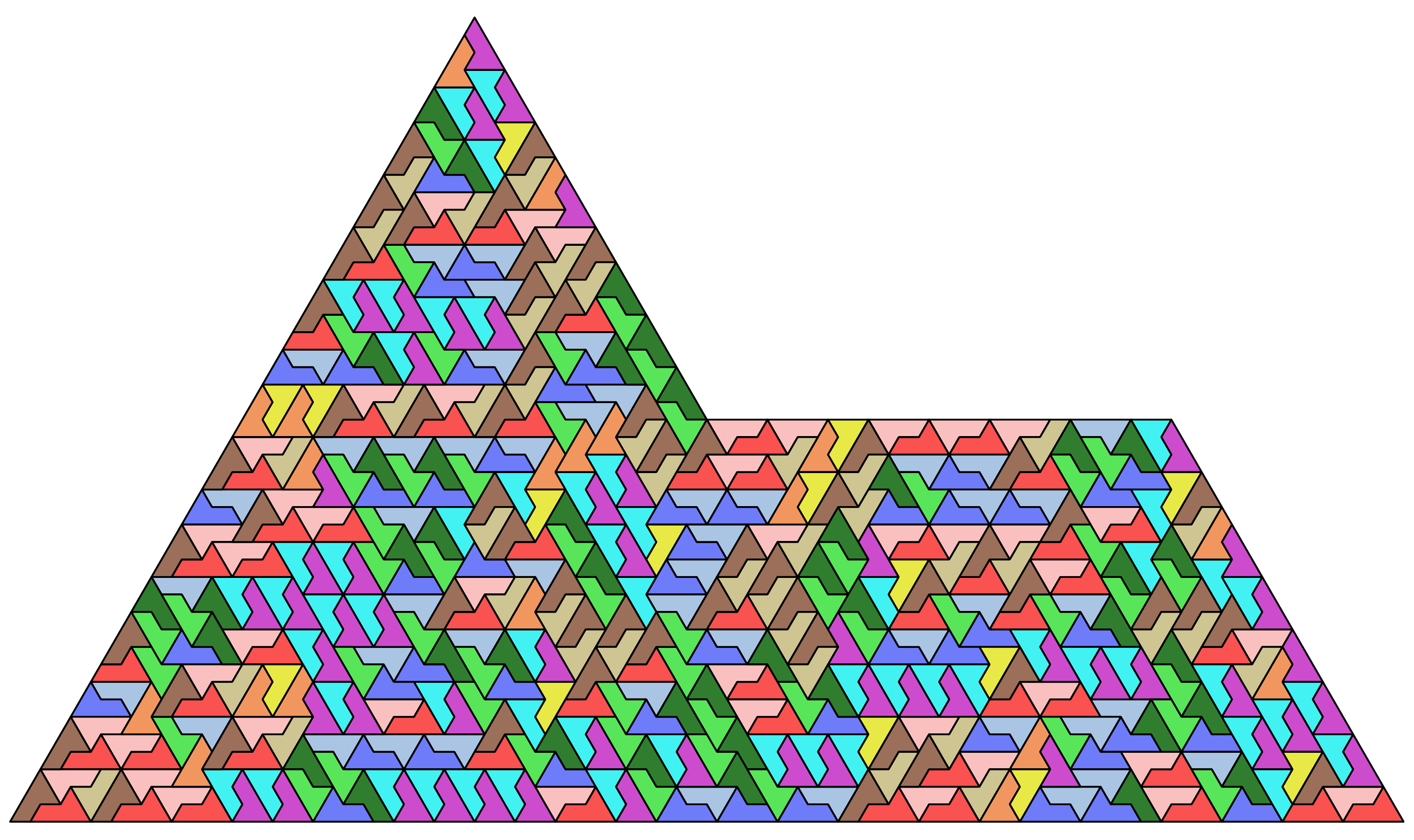}
\caption{A typical MC random tiling of a 23-Sphinx. }
\label{fig:random-tiling-23}
\end{figure}

\begin{figure*}[htbp]
\centering
\includegraphics[width=0.4\linewidth]{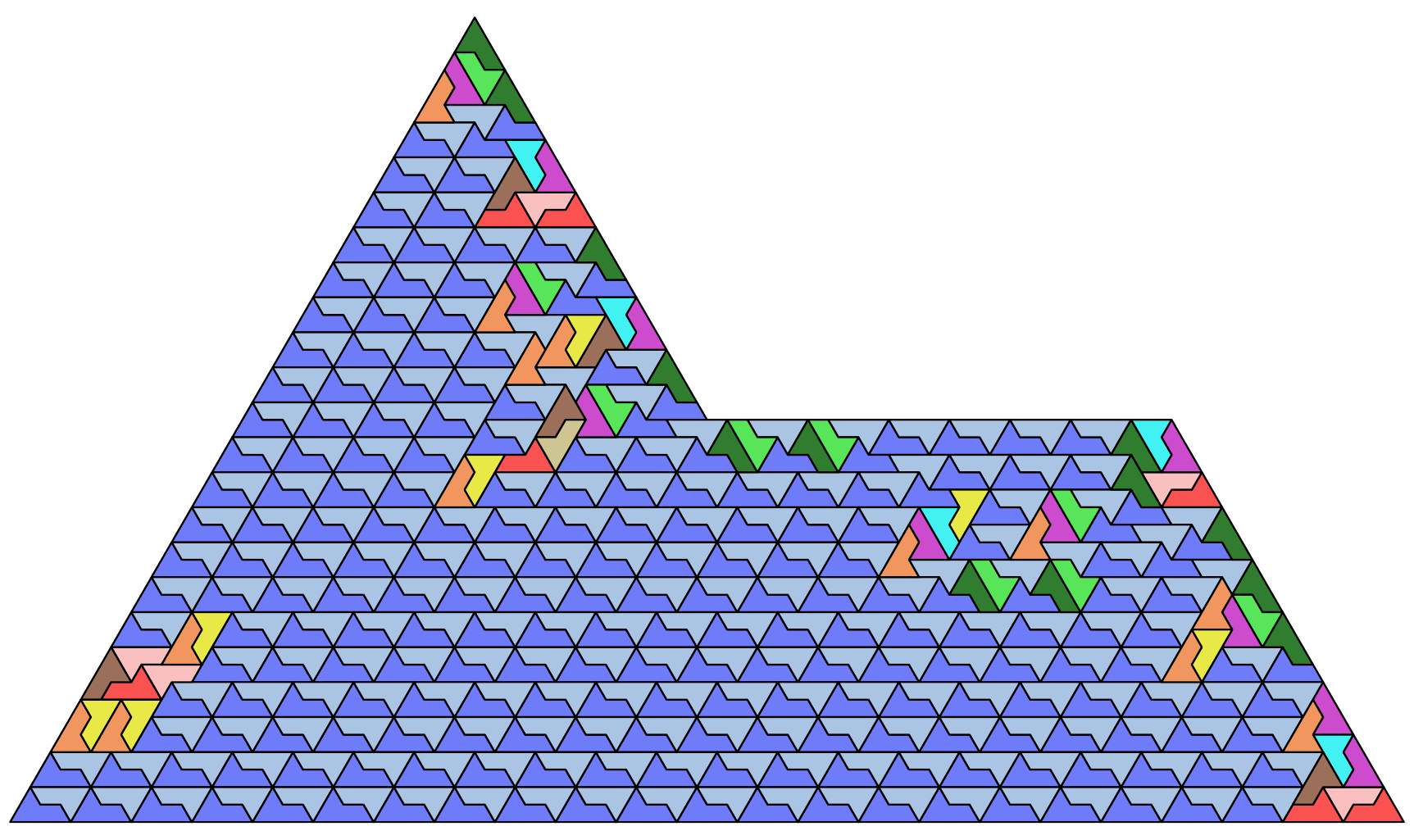}
\hspace{0.1\linewidth}
\includegraphics[width=0.4\linewidth]{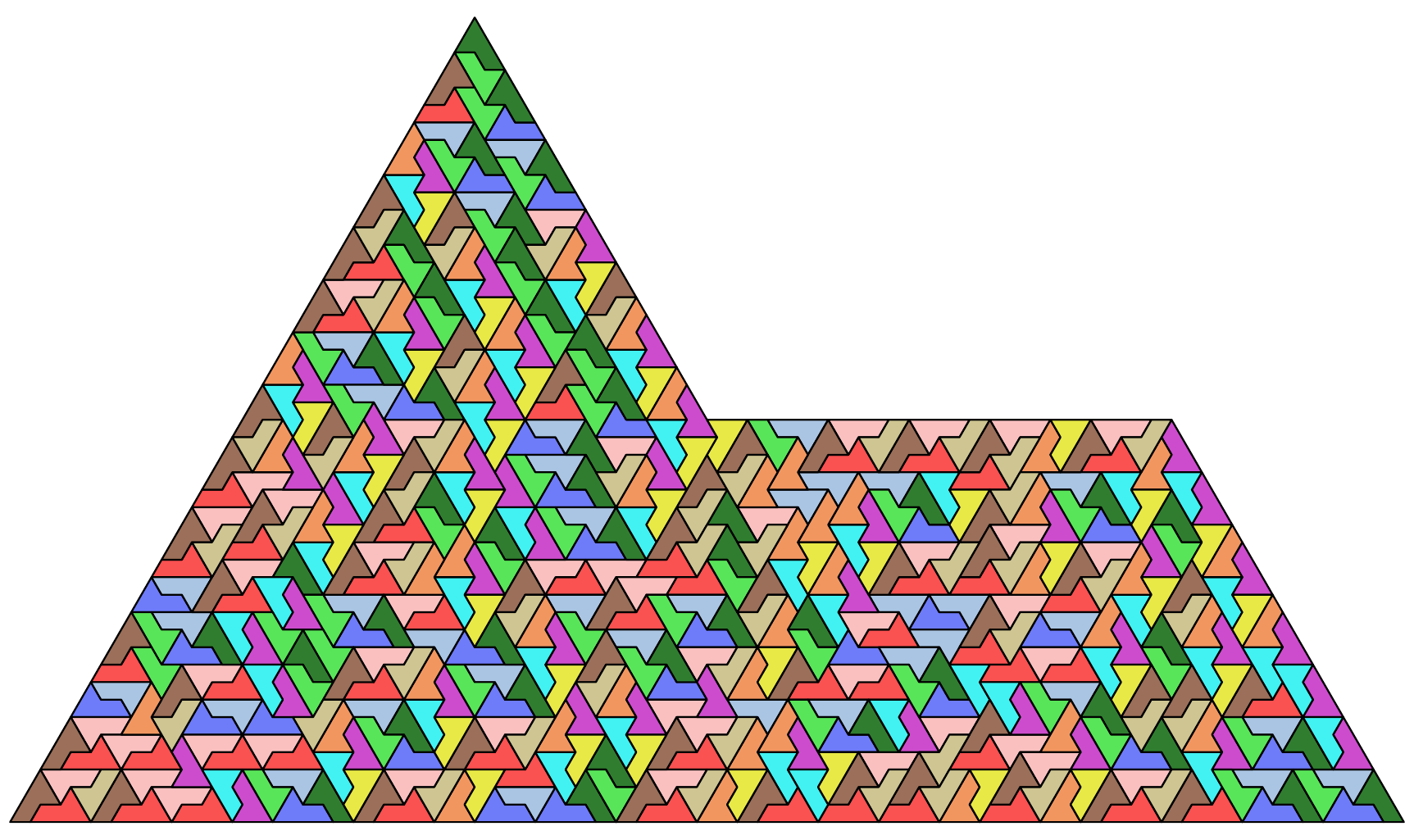}
\includegraphics[width=0.4\linewidth]{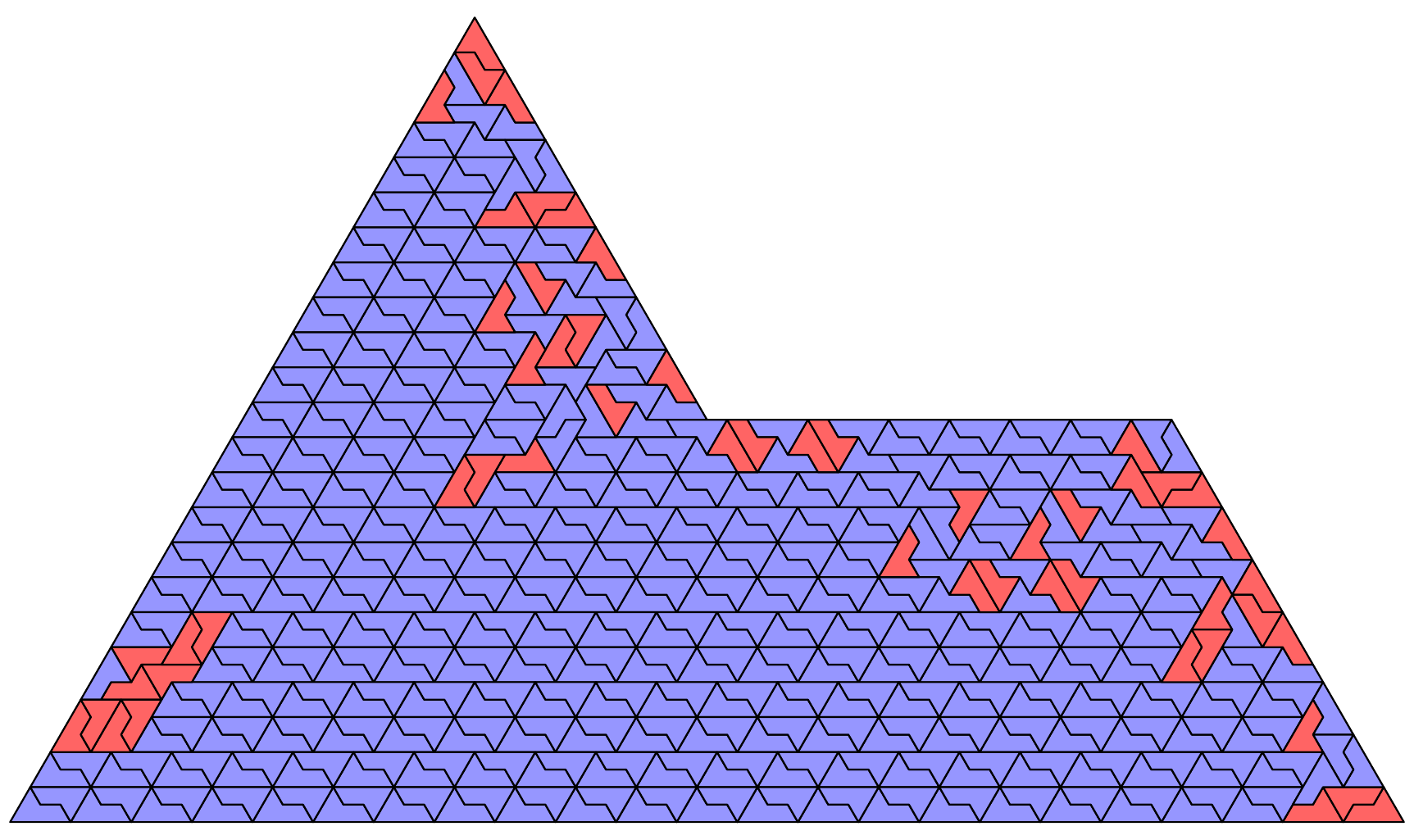}
\hspace{0.1\linewidth}
\includegraphics[width=0.4\linewidth]{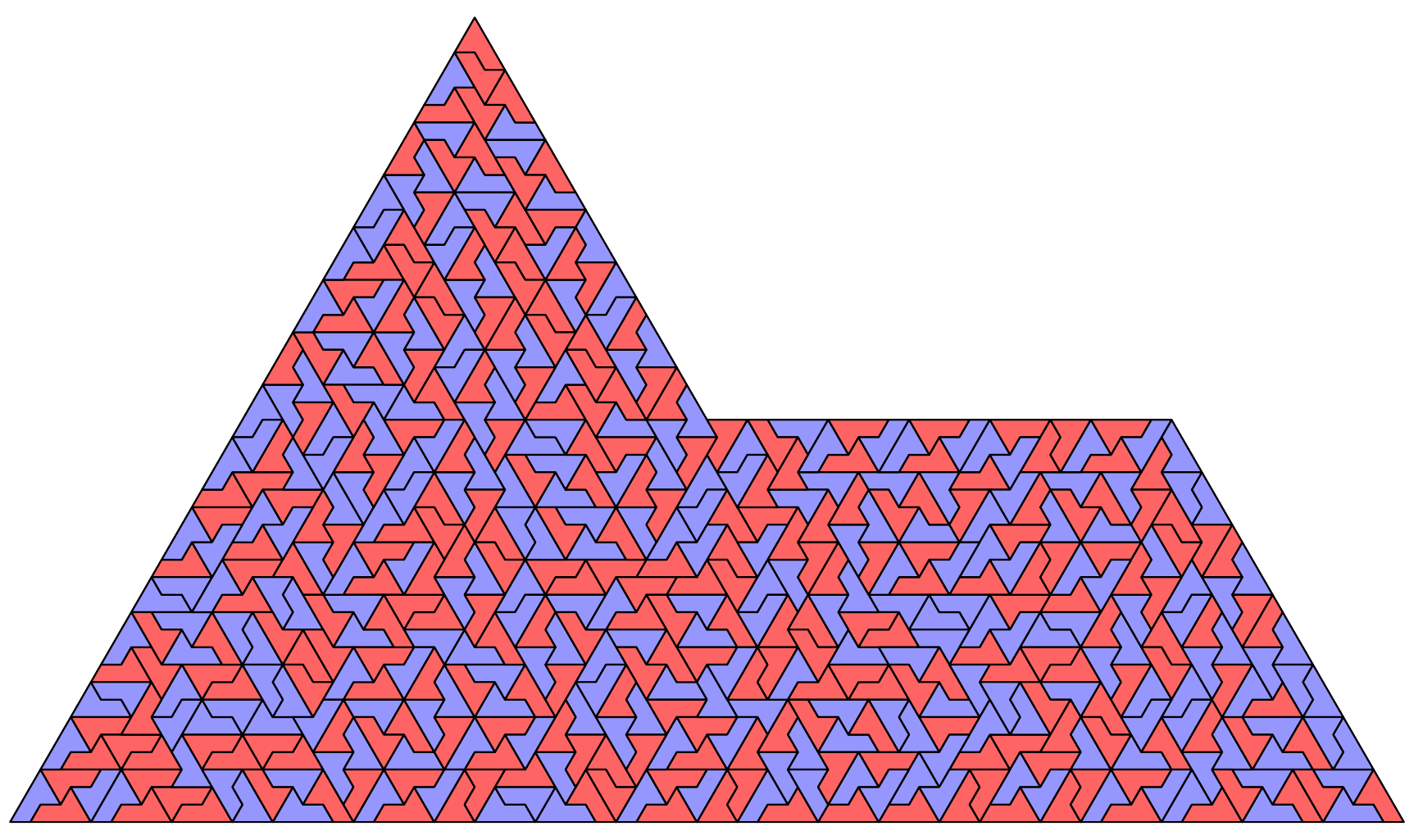}
\caption{Tiling realizations of an order-23 Sphinx frame for low and high chiral energy.
Top left: low energy, top right: high energy, with standard coloring of Fig.\ \ref{fig:standard-colors}. 
Bottom row: the same tilings, but with L-sphinxes shown in blue and R-sphinxes shown in red, the coloring of Fig\ \ref{fig:specialcolors}(middle).}
\label{fig:figure8}
\end{figure*}

\begin{figure}[htbp]
\centering
\includegraphics[width=1.0\linewidth]{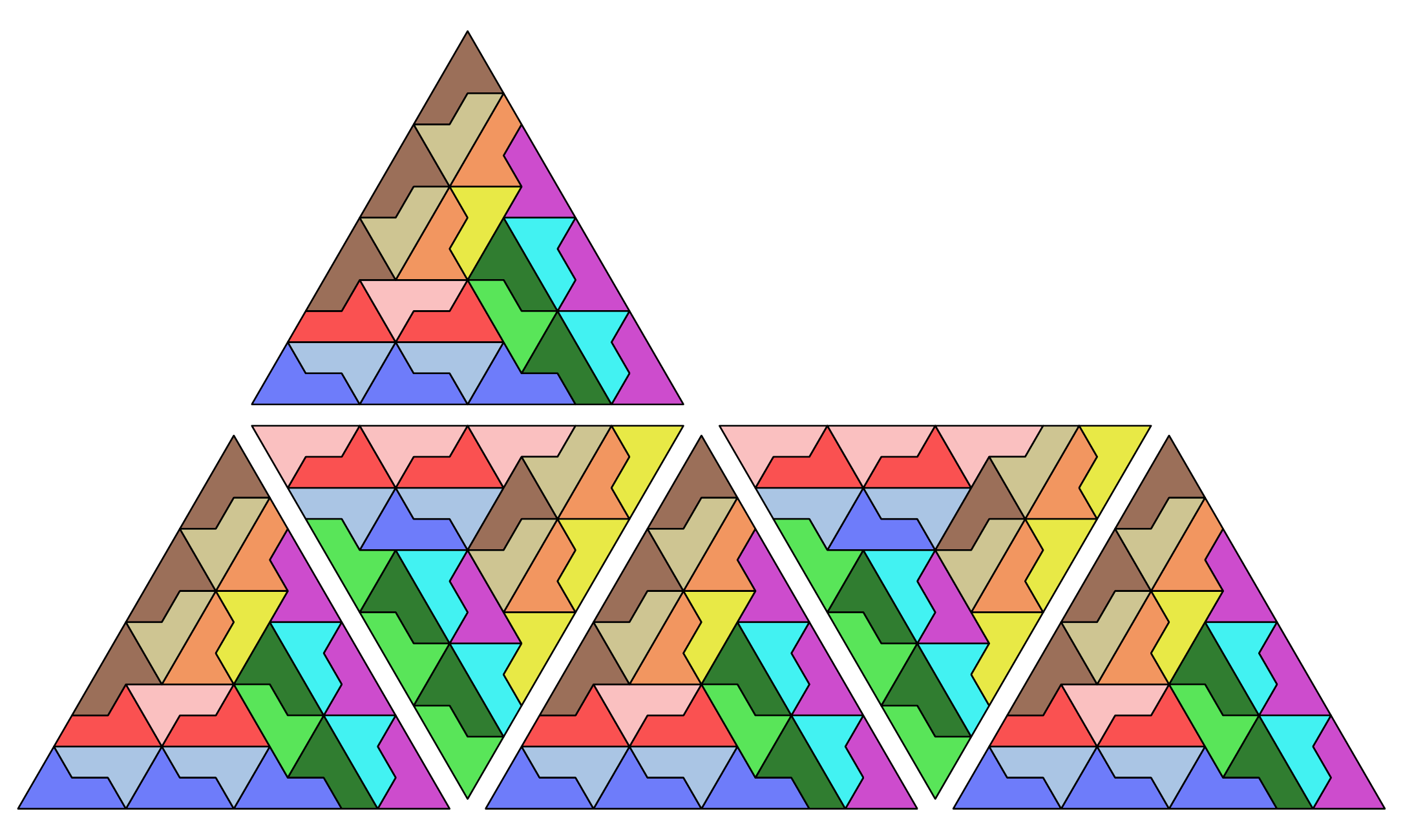}
\caption{A tiling of an order-12 Sphinx with 6 order-12 triangles as subframes.}
\label{fig:order12sphinx}
\end{figure}

\begin{figure*}[htbp]
\centering
\includegraphics[width=0.95\linewidth]{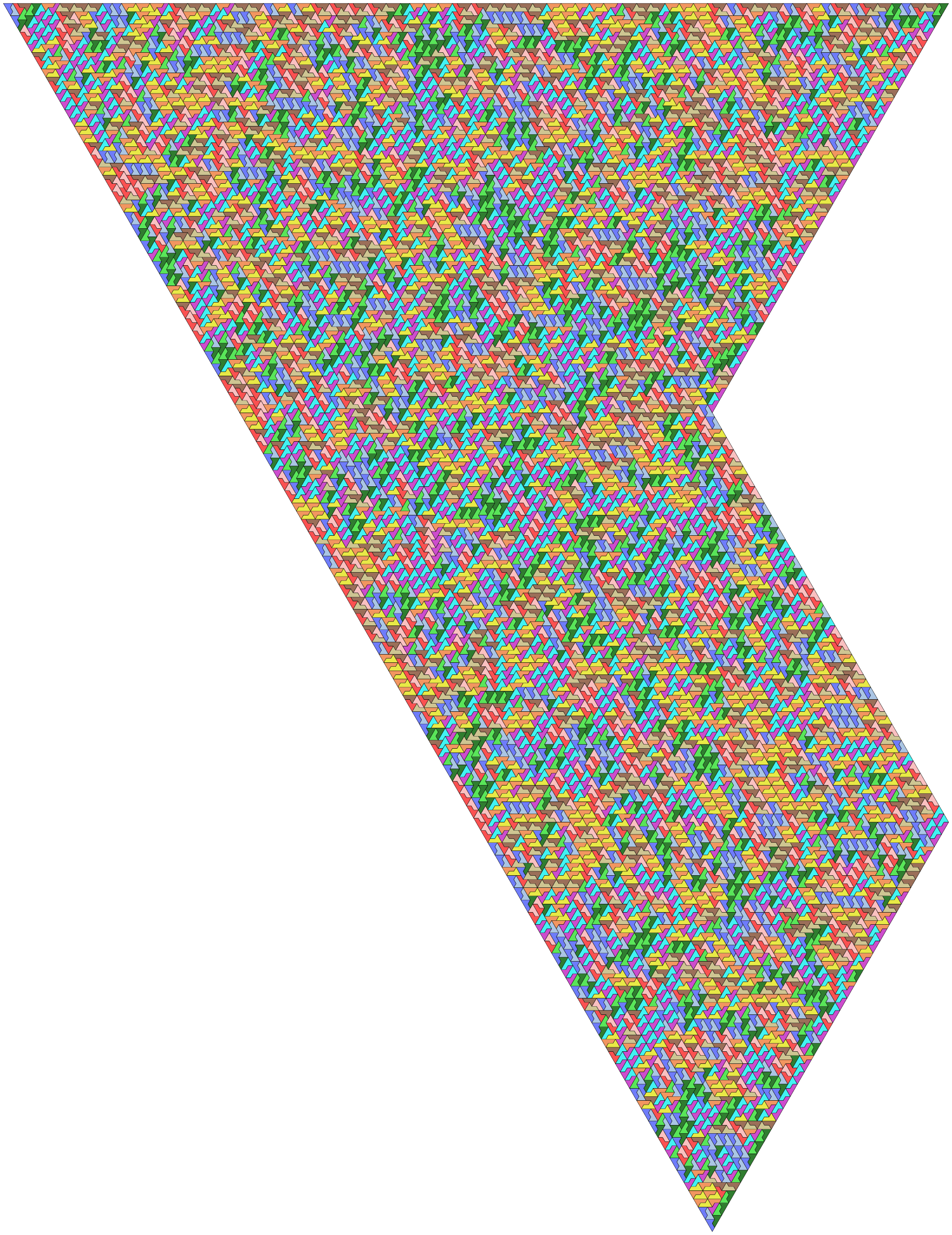}
\caption{A typical tiling of an order-100 Sphinx frame generated by the f-polyad MC algorithm. Notice patches of sphinxes aligned in the different orientations.}
\label{fig:tiling100rotated}
\end{figure*}

Using the concept of fundamental polyads, we developed a Monte Carlo method to generate new tilings from a given tiling. In the simulation, we pick one of the polyads at random and search for its occurrence.  If none is found, we consider a different polyad; otherwise we replace the polyad with another one of its tilings. A larger frame will typically include many fundamental polyads.  Polyads overlapping in space allow us to generate a sequence of different tilings. The power of our MC method stems from this fact: a sphinx tiling can be transformed to any other sphinx tiling of the same frame by a finite sequence of f-polyad transitions (see Fig.\ \ref{fig:diamond-6-transitions} for an example). We just need a small subset---polyads up to size 11---to find all tilings for frames of order up to six. Discussion of minimal f-polyads for a 7-Sphinx is given in Figs.\ S17 and S18. For larger frames, it is an open question whether a limited set of f-polyads is sufficient to find all the tilings.  In appendix \ref{sec:proof}, we prove that if all f-polyads can be used, the MC method will generate all possible tilings of a frame. A typical tiling of a 23-Sphinx is shown in Fig.\ \ref{fig:random-tiling-23},  which contains 103 f-polyads of sizes 2--6 and at least 34 larger ones.  Tilings with low and high chiral energy are shown in Fig.\ \ref{fig:figure8}.  A special tiling of a 12-Sphinx based upon triangle inflation is shown in Fig.\ \ref{fig:order12sphinx}. Finally, a typical tiling of a 100-Sphinx generated by the MC method is shown in Fig.\ \ref{fig:tiling100rotated}.  For the initial tiling for the MC method, we use a combination of known tilings or subsets of the sphinx and inflations.

%\vspace*{10mm}

\section{Chirality condensation}

\begin{figure}[htbp]
\centering
\includegraphics[width=0.9\linewidth]{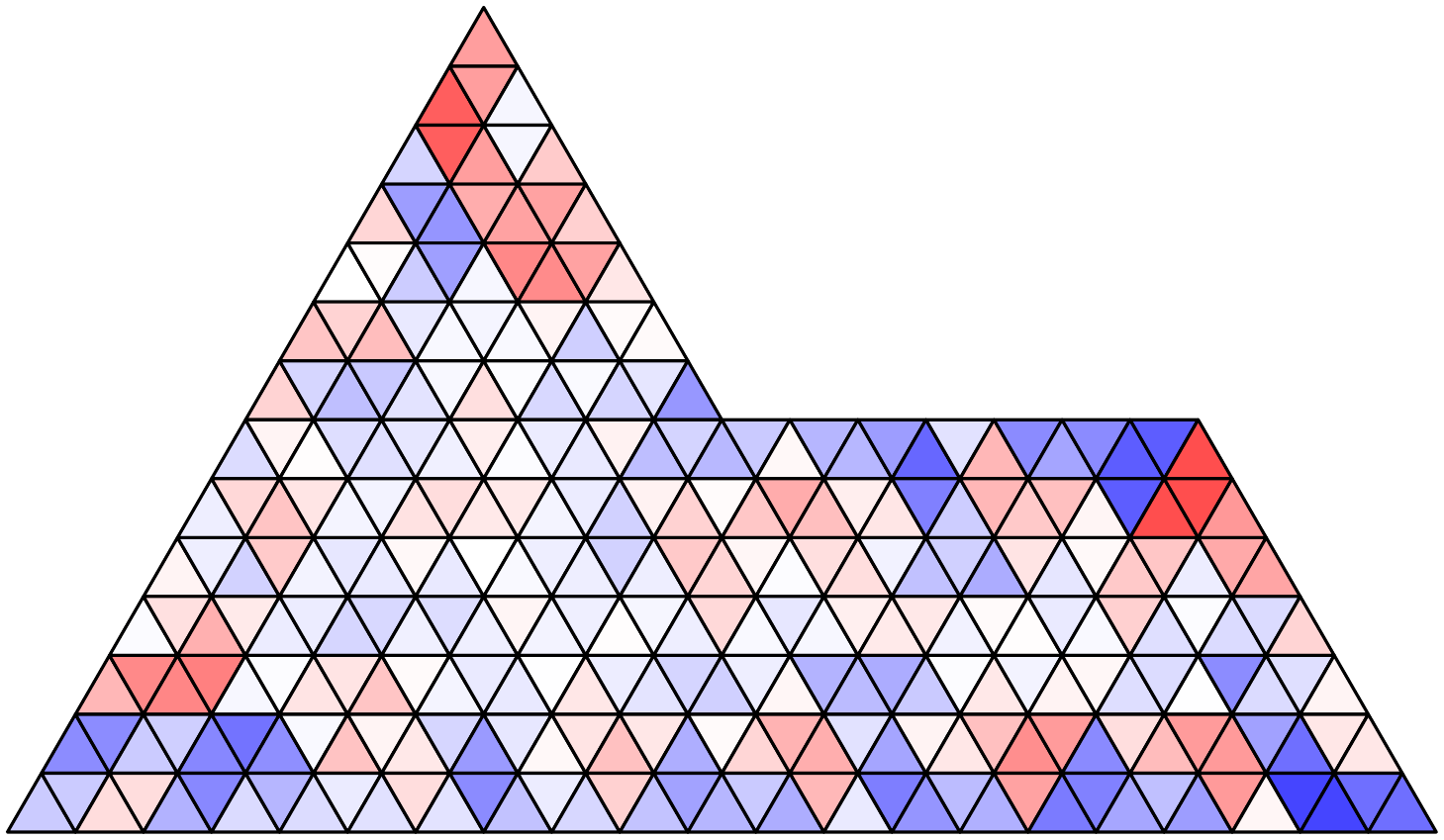}
\caption{Plot of average chirality of sphinxes over each triangle in a 7-Sphinx.  Blue (red) represents triangles that are covered more frequently with sphinxes of left (right) chirality.  For more examples, see Fig.\ \ref{fig:more-chirality}.}
\label{fig:Chirality7}
\end{figure} 

Surprisingly, the spatial distribution of average chirality in sphinx tilings
features regions of excess left- and right- handedness associated with specific corners of the frame. While this localization might perhaps be expected at low temperatures for asymmetric frames (like the Sphinx), it holds for symmetric frames also, even in the high-$T$ limit. In Fig.\ \ref{fig:Chirality7}, we show the spatial distribution of chirality on each triangle in a 7-Sphinx, averaged over all tilings. Localized preferences for one chirality over the other are precipitated by the shape of the boundary. For more details see appendix \ref{sec:chirality}.

\begin{figure}[htbp]
\centering
\includegraphics[width=1\linewidth]{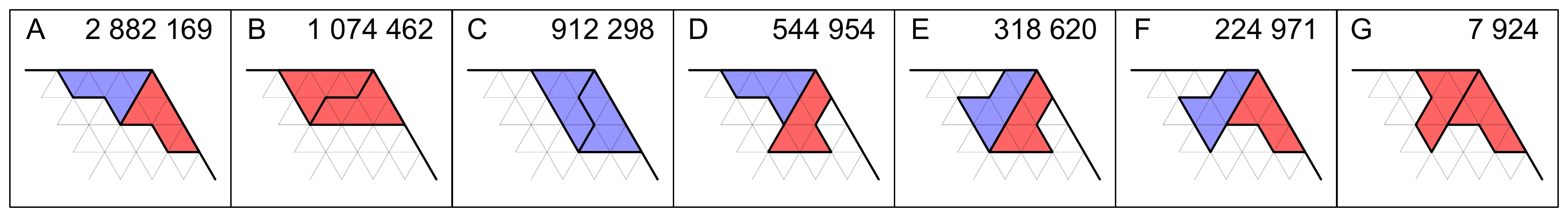}
\caption{There exist only 7 different sphinx constellations for the 120° angle of the 7-Sphinx.
The shown numbers tell us how often a constellation occurs in all  5 965 398  tilings of the 7-Sphinx.}
\label{fig:sphinx-7-angle-120}
\end{figure} 

\begin{figure*}[htbp]
\centering
\includegraphics[width=0.80\linewidth]{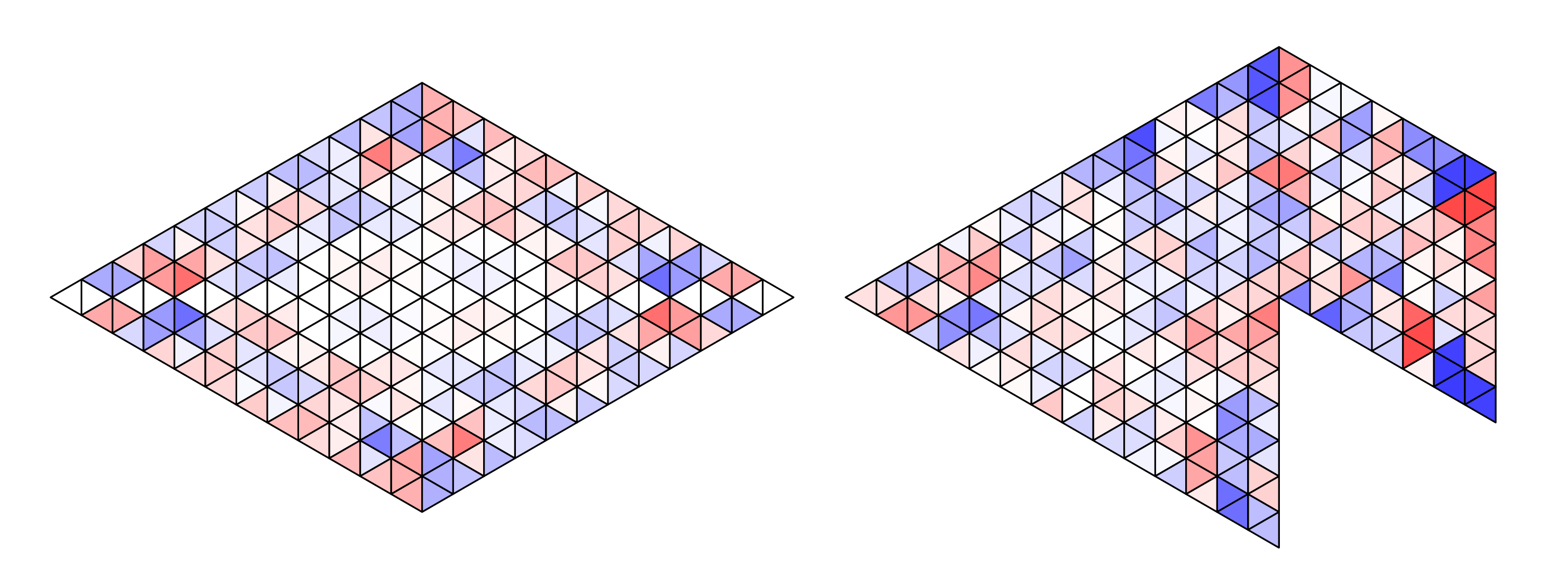}
\includegraphics[width=0.15\linewidth]{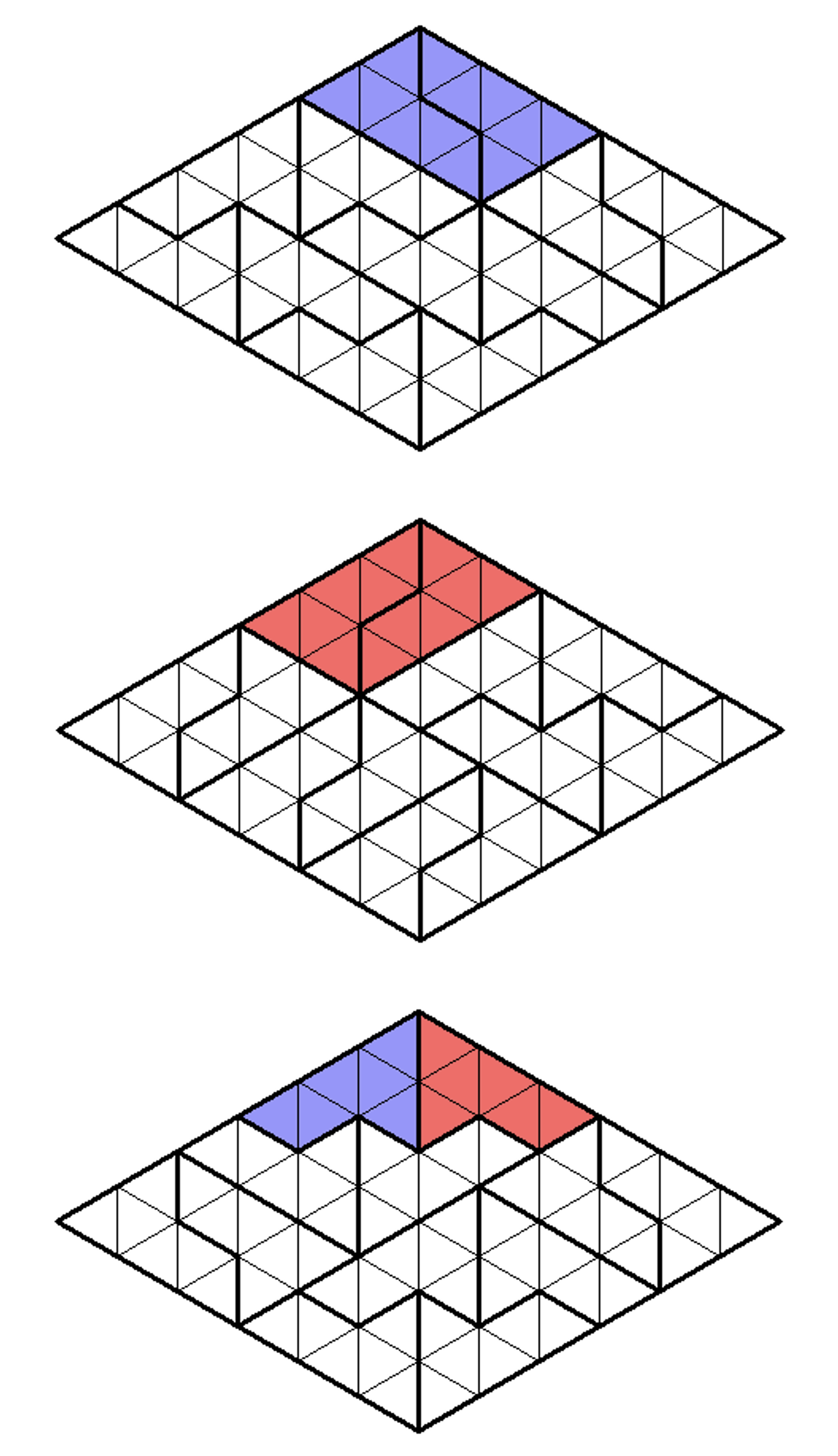}
\caption{Chirality distribution of a 12-Diamond of area 288 (left), tiling of an order-7 ``yacht" hexiamond of area 294 (middle), and a 
``microscopic'' explanation of the localization of chirality in a symmetric frame (right).}  These can be compared with the order-7 Sphinx chirality distribution as shown in Fig.\ \ref{fig:Chirality7}.
\label{fig:more-chirality}
\end{figure*}

There are 7 possible L-R-constellations in the 120° angle of the 7-Sphinx. They are shown in Fig.\ \ref{fig:sphinx-7-angle-120},
sorted according to the frequency in which they occur. A reflection with respect to the symmetry axis of the 120° angle not only interchanges L- and R- sphinx states, it simultaneously interchanges sphinx spatial locations as well.  Both effects combined allow condensation of chirality at {\it fixed} spatial locations, such as the appearance of chirality monopoles and dipoles, even for symmetric frames (see Fig.\ \ref{fig:more-chirality}).

\section{Chiral energetics}

\begin{figure}[htbp]
\centering
\includegraphics[width= 0.8 \linewidth]{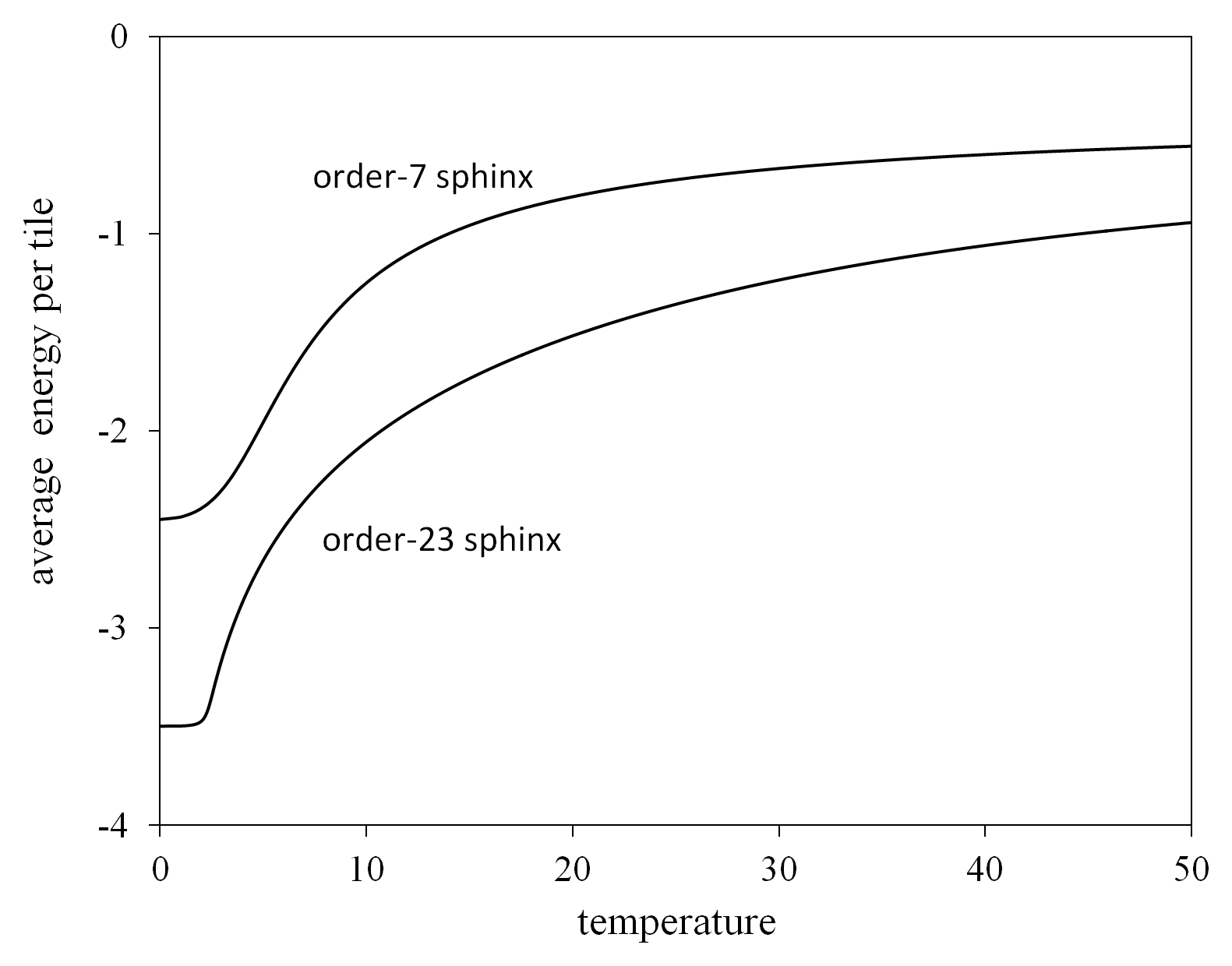}
 \caption{Energy per tile vs.\ $k_B T / J$ of tilings of a 23-Sphinx based on $10^9$ MC samples, and energy per tile of the 7-Sphinx based upon exact enumeration of all states.}
\label{fig:SphinxEnergy7-23-LR}
\end{figure}

\begin{figure}[htbp]
\centering
\includegraphics[width= 0.9 \linewidth]{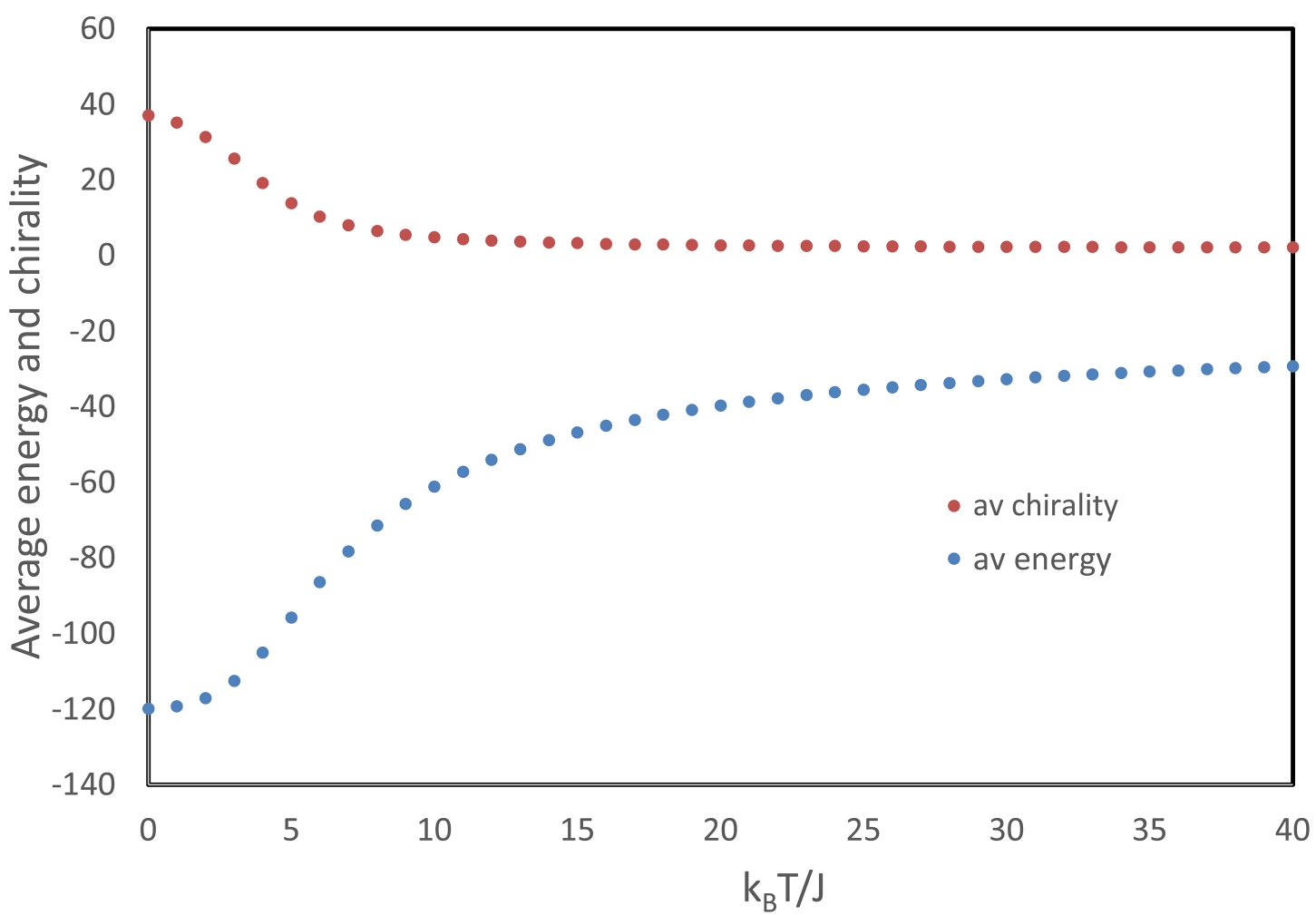}
\caption{The average energy and average chirality as a function of temperature, calculated from the exact enumeration of all 5965398 tilings of the 7-Sphinx.}
\label{fig:energy-chirality-7}
\end{figure}

We introduce an interaction energy in our system to study the effects of temperature on the equilibrium chirality of the tilings.  To define this energy, the L- and R- sphinx states are paramount, with $\chi_i = +1$ for L and $\chi_i = -1$ for R. We assign an energy $-J$ to each triangular lattice edge (length 1) between two touching sphinxes if the chiralities are the same, and $+J$ if the chiralities are different, and sum over all unit edges:
\begin{equation}
    E[\tau] = -J\!\!\sum_{\mathrm{edges}[\tau]}\! \chi_i \chi_j
\end{equation}
where $\tau$ is a particular tiling, and $i,j$ refer to the nearest-neighbor tiles adjacent an edge. 
For $J > 0$, we expect that at lower temperatures the system will condense into a phase of a single chirality, if the boundary frame allows it to occur; otherwise, there should be a phase with a majority of one chirality. Now we carry out the same polyad-based MC simulation as above, but add the Metropolis acceptance criterion based upon the changes in energy associated with a given trial move (accept a move with probability min$(e^{-\beta \Delta E},1)$ \cite{MetropolisEtAl53}).

The resulting average energy is plotted in Fig.\ \ref{fig:SphinxEnergy7-23-LR}. To find this curve for the order-23 sphinx, we carried out our MC algorithm at high temperatures, so that all states were equally likely, and sampled the distribution $n(E)$ of the number of states at energy $E$, which we used to calculate $E(T)$.  For low $E$ of this system, we extrapolated the behavior to our estimated minimum energy of %$ -4(23-1)^2 =J
$-1850 J$, while the lowest energy we found via MC was $-894 J$.  Exact results for the net chirality as a function of temperature of the sphinx of order 7 are graphed in Fig.\ \ref{fig:energy-chirality-7}.  We expect that, as the system size gets larger, a sharp chirality transition occurs, but this is an area for future study.

\section{Conclusions}

As far as we are aware, this is the first statistical-mechanics study of a tiling problem for tiles that are chiral and have no symmetry.   We have included an interaction energy and studied the chirality and average energy as a function of temperature.

Our paper builds a foundation for future tiling research.
It is possible to apply f-polyad and Monte Carlo methods to develop a statistical mechanics for tiles other than the sphinx tile. Such a tile could be any polyiamond, polyomino, hat polykite, member of the spectre family, or any other polygon. Some of our methods do not even require working on a lattice.

In the high-temperature limit, where there are no interaction effects in the system, the system becomes simply a tiling problem, and by carrying out extensive enumeriations we are able to find the behavior of the entropy of the tilings, which is compared to the behavior of the dimers and other tiling models.

As the temperature is lowered, the energetics and chirality are affected by the interaction and the behavior becomes more step-like at a finite temperature as the order increases.  Perhaps for even larger systems a sharp transition will become evident.

We developed exact and Monte Carlo methods to explore the statistical mechanics of, and to expose the chiral nature inherent in, ensembles of densely packed chiral tiles subject to finite spatial boundaries. Spatial localization of chirality associated with particular boundary regions, even in racemic systems with zero net chirality, appears to be a generic feature of some interest.

\section*{Acknowledgments}  We thank Artem Ripatti checking $Z(0) = N_n$ up to order $n=12$, and also thank Douglas Hofstadter, Federico Ricci-Tersenghi, Toshiki Saitoh and Matt Skoss for helpful correspondence.

\appendix
\section{Sphinx Nomenclature}
\label{sec:nomenclature}
\begin{figure}[htbpbp]
\centering
\includegraphics[width=0.7\linewidth]{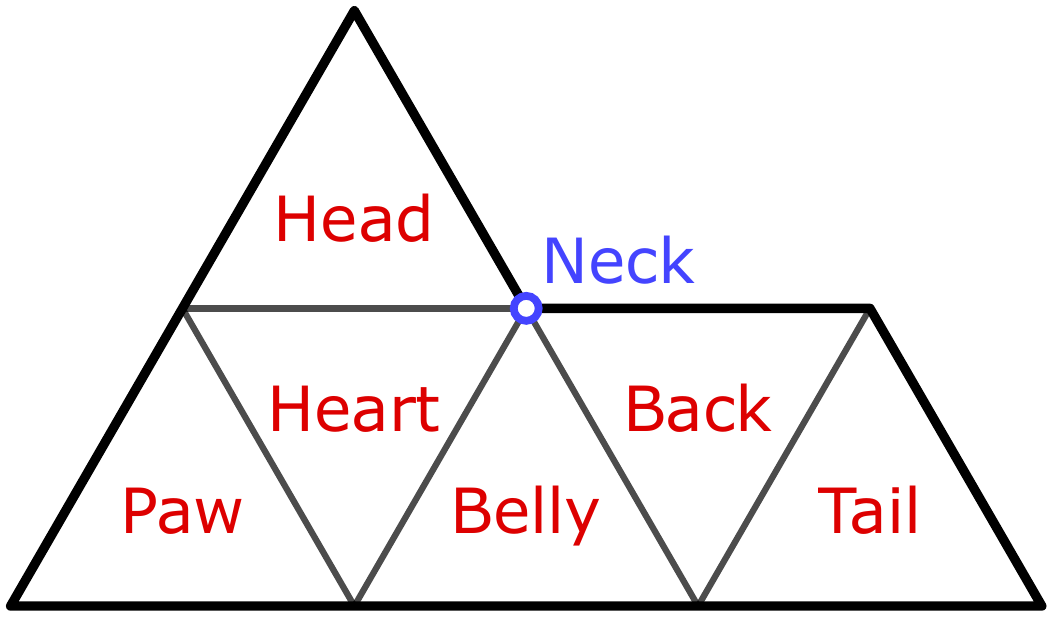}
\caption{The six faces of the sphinx hexiamond in its canonical state, and anatomical names that we have assigned to them.}
\label{fig:six-faces}
\end{figure}

\textbf{Sphinx Anatomy} (Fig.\ \ref{fig:six-faces})\\
The sphinx lives on the triangular lattice. It has 6 triangle faces, connected along their edges.

\begin{figure}[htbpbp]
\centering
\includegraphics[width=0.7\linewidth]{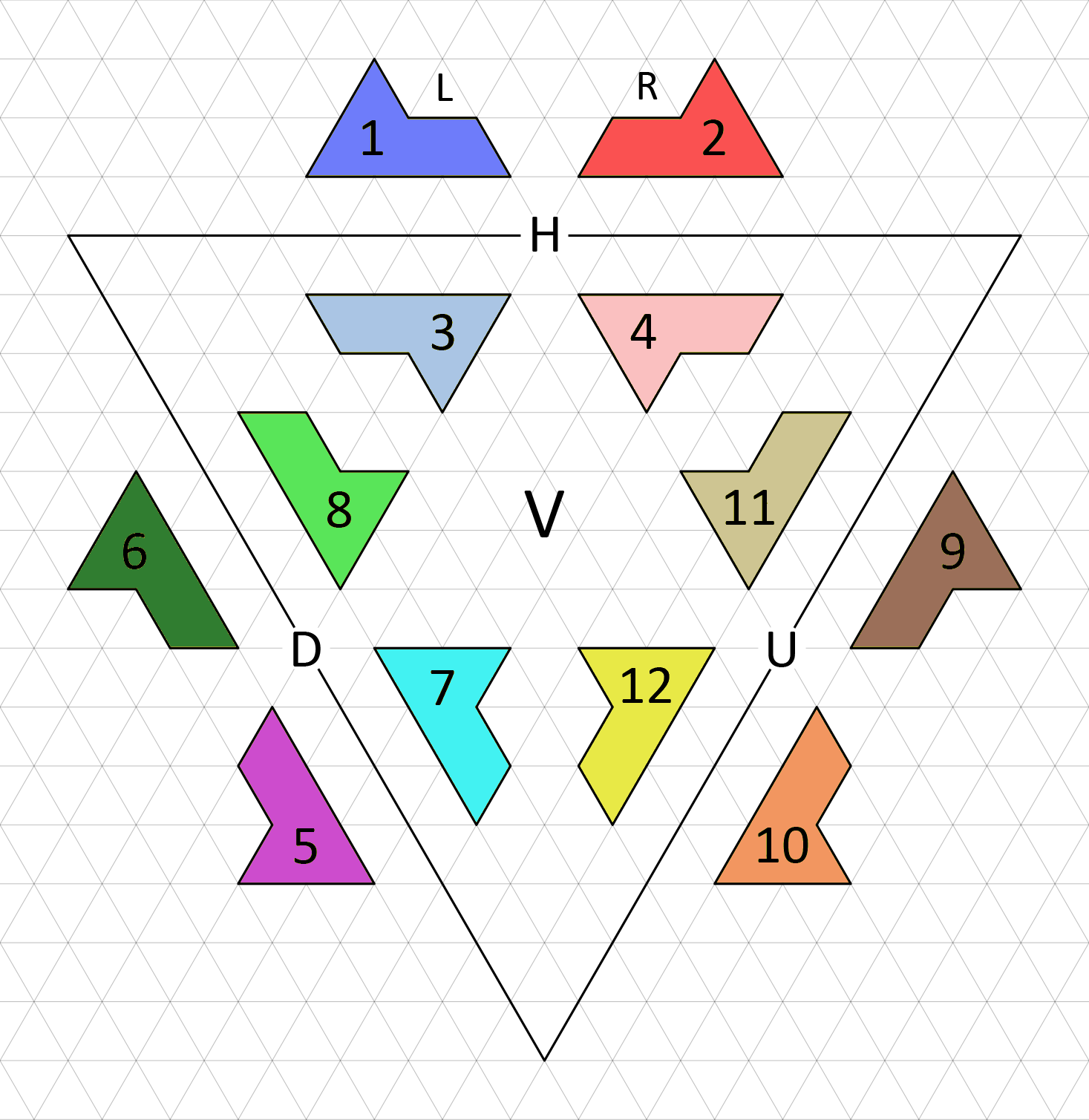}
\caption{Standard colors and numerical designations for the 12 different states of a sphinx tile. The letter labels refer to the sphinx properties and states shown in Fig.\ \ref{fig:specialcolors}.}
\label{fig:standard-colors}
\end{figure}

\textbf{Lattice orientation and colors} (Fig.\ \ref{fig:standard-colors})\\
We always orient the triangular lattice so that there are horizontal grid lines.
Besides the \underline{h}orizontal grid lines (H) there are rising lines (U) which go \underline{u}p from left to right and \underline{d}escending lines (D).
In this lattice the sphinx can have 12 different orientations (states).
To distinguish the states we use 12 different colors (standard colors).

The state of a sphinx tile can be described by 3 parameters, as follows.\\

\begin{figure}[htbpbp]
\centering
\includegraphics[width=0.62\linewidth]{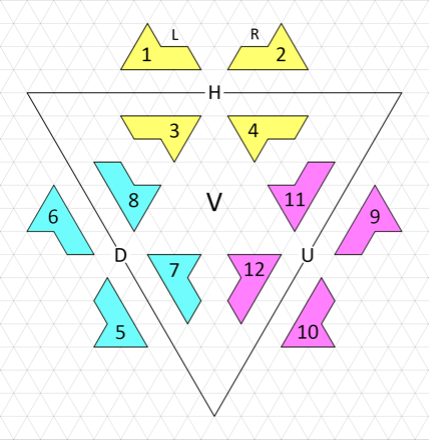}
\includegraphics[width=0.62\linewidth]{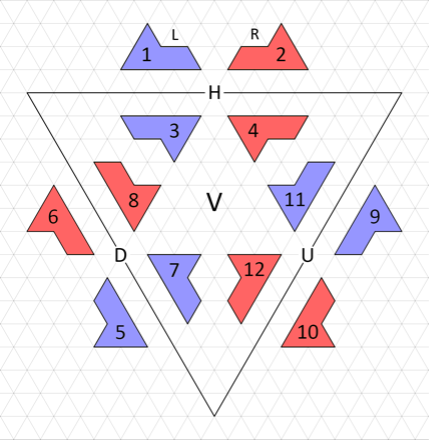}
\includegraphics[width=0.62\linewidth]{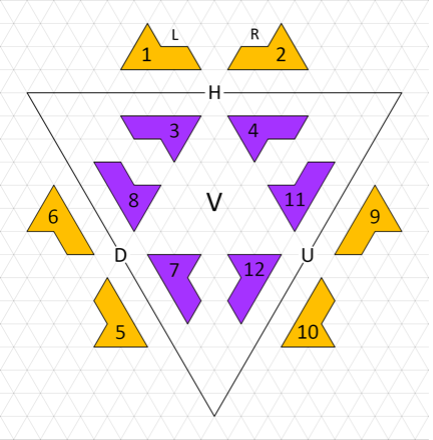}
\caption{Special colors used to distinguish certain tile properties. Upper: Base slope, 0° (H, horizontal) (yellow), $+60$° (U, up) (magenta), $-60$° (D, down) (turquoise); Middle: Chirality,  L = left (blue), R = right (red); Lower: Charge, A ($+$) (amber), V ($-$) (violet).}
\label{fig:specialcolors}
\end{figure}

\textbf{Base slope} (Fig.\ \ref{fig:specialcolors} top) \\
There are 3 different slopes for the base (longest edge) of the sphinx, depending on which type of grid line the base lies,
Thus we can describe the base orientation by H, U or D. (H-sphinx, U-sphinx, D-sphinx)

\textbf{Chirality} (fig.\ \ref{fig:specialcolors} middle)\\
The sphinx can have two different handednesses (chiralities), which we call left (L) and right (R) based on the position of the ``head'' of the sphinx.
Consider a walk along the boundary of a sphinx. Start at the neck and surround the head first. We speak of a L-sphinx if the head lays on the left when it is passed, in this case the walk is counterclockwise. Otherwise (head on the right, clockwise walk) we speak of a R-sphinx.
If we want to distinguish sphinxes by their chirality we use the color b\underline{l}ue for L-sphinxes and \underline{r}ed for R-sphinxes. 

\textbf{Charge} (Fig.\ \ref{fig:specialcolors} bottom)\\
From the 6 triangular faces of the sphinx there can be 4 upright or only 2 upright, depending on the orientation of the sphinx.
Because the letters A and V look like different oriented triangles,
we speak of an A-sphinx if it consists of 4 upright triangles and otherwise of a V-sphinx.
The colors \underline{a}mber and \underline{v}iolet may be used to distinguish these types of sphinxes.

\begin{figure}[htbp]
\centering
\includegraphics[width=1\linewidth]{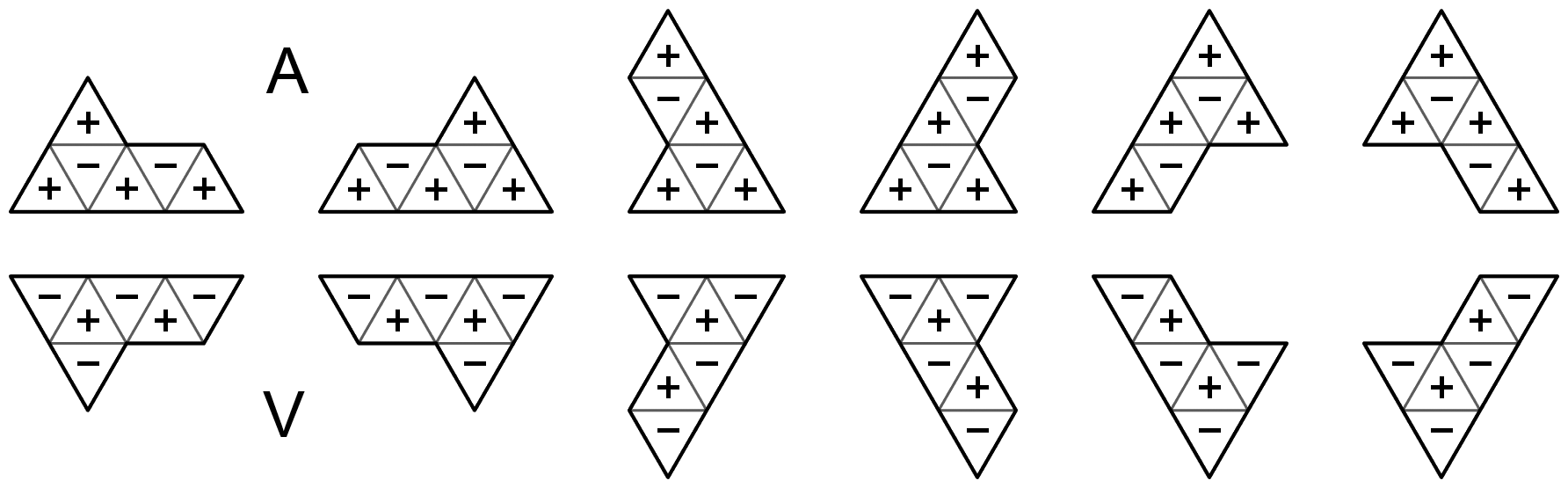}
\caption{Illustrations of the A- and V-tile orientations, where $+$ and $-$ represent A and V triangles respectively. Sphinx tiles with states shown in the upper row are called A-sphinxes and those in the lower row are called V-sphinxes.}
\label{fig:a-v-charge}
\end{figure}

We give A-triangles the charge $q=+1$ and V-triangles the charge $q=-1$; then a sphinx has the charge $|q|=2$, with
$q$(A-sphinx)$ =+2$, $q$(V-sphinx)$ =-2$. See Fig.\ \ref{fig:a-v-charge}.

\textbf{Transformations of the sphinx states}\\
The sphinx can be transformed from a certain state to another by rotations and reflections.
Here we restrict the discussion on 3 fundamental transformations.
\begin{itemize}
\item[(a)]
$120^{\circ}$-Rotation (counter clockwise)\\
changes the base slope (H$\rightarrow$D$\rightarrow$U$\rightarrow$H),\\ 
preserves chirality and charge.
\item[(b)]
$180^{\circ}$-Rotation (point reflection)\\
changes the sign of the charge (A$\leftrightarrow$V),\\
preserves chirality and base slope.
\item[(c)]
Reflection at an axis vertical to the base\\
changes the chirality (L$\leftrightarrow$R),\\
preserves base slope and charge.
\end{itemize}

\begin{figure}[htbp]
\centering
\includegraphics[width=0.8\linewidth]{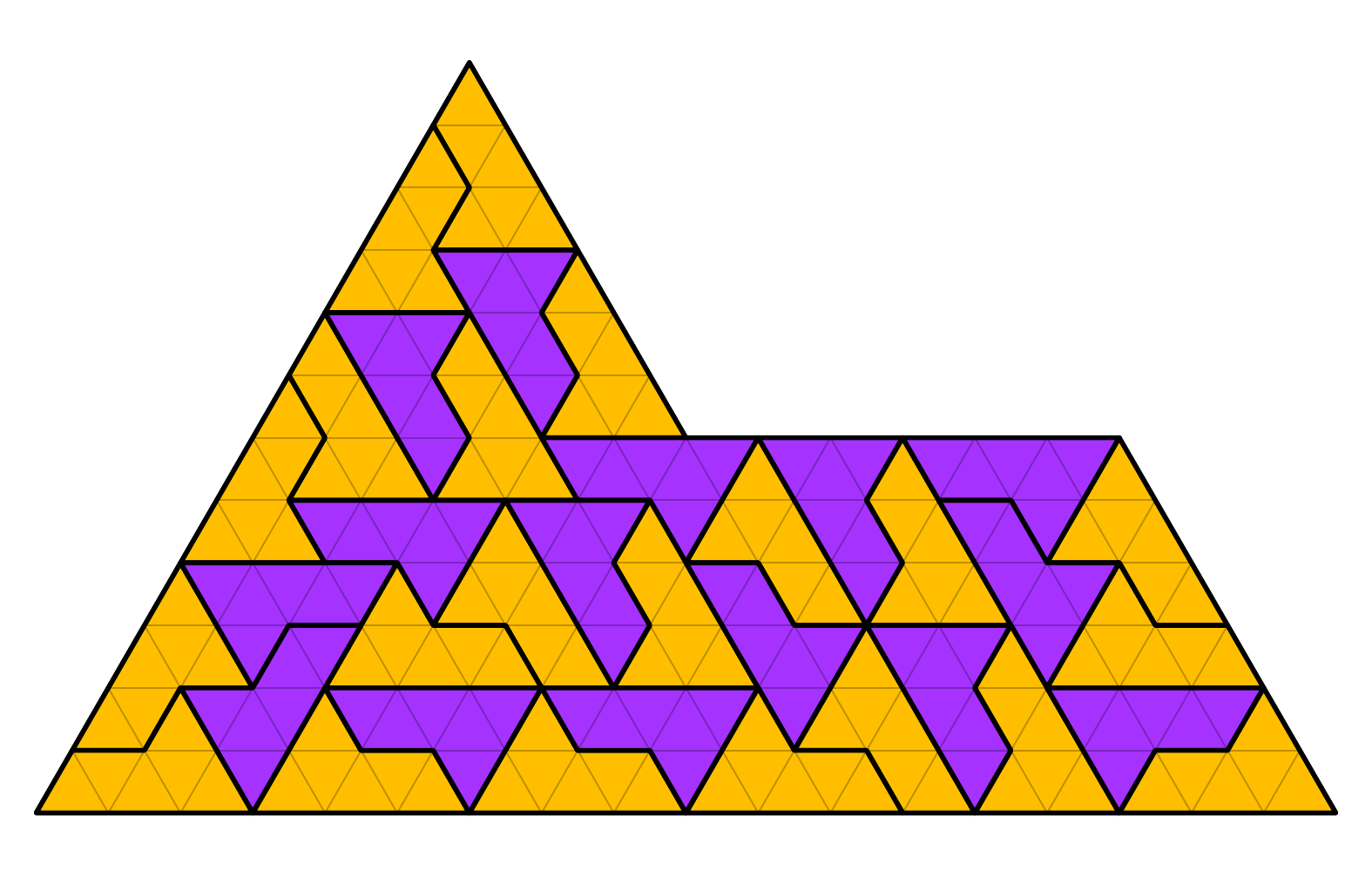}
\caption{Each tiling of an order-6 sphinx in the canonical state (state 1) consists of 21 A-sphinxes (colored in amber) and 15 V-sphinxes (colored in violet).}
\label{fig:AVsphinx6}
\end{figure}

\textbf{Tiling Criteria}\\
Consider a given frame with a certain orientation (state) in the triangular grid. Call $t$ the number of triangles (area) of the frame, $t_A$ the number of A-triangles and $t_V$ the number of V-triangles.
Then the charge of the frame is $q = t_A - t_V$. The frame only can be tiled by sphinxes if the following 3 conditions are satisfied. These conditions are necessary but not sufficient.
\begin{itemize}
\item[(1)]
$t$ has to be a multiple of 6.\\
\textit{Because the area of a sphinx tile is 6.}
\item[(2)]
$q$ has to be even.\\
\textit{Because the charge of a sphinx tile is $2$ or $-2$.}
\item[(3)]
$t / 6$ and $q / 2$ have to have the same parity.\\
\textit{Say $n_A$ and $n_V$ are the numbers of A- and V-sphinxes in the tiling.
Then $n_A + n_V = t / 6$ and $n_A - n_V = q / 2$.
Thus $2n_A = t / 6 + q / 2$.
Therefore $t / 6$ and $q / 2$ have to be even both or odd both.}
\end{itemize}
Corollary:\\
All sphinx tilings of a certain frame with certain orientation consist of the same number of A-sphinxes and the same number of V-sphinxes.\\
An example with $t = 36$ and $q = 12$ is shown in Fig.\ \ref{fig:AVsphinx6}.\\
The charge of a sphinx frame of order $n$ is $2n$.

\textbf{Different ways to cover a grid triangle by a sphinx tile.} (fig.\ \ref{fig:AVtriangle})\\ 
Each `A' ($\triangle$) and `V' ($\nabla$) grid triangle and can be covered in 36 ways by a sphinx tile.
Depending on its orientation, a sphinx tile has 2 (or 4) A-faces and 4 (or 2) V-faces. A given
A- or V-triangle may be covered by one of the 6 faces of a sphinx tile in one of 12 states.
The prevalence of each triangular face and each state of all tiles in the tilings of S7 is shown in Fig.\ \ref{fig:cell-covering-7}.

\begin{figure*}[htbp]
\centering
\includegraphics[width=1.0\linewidth]{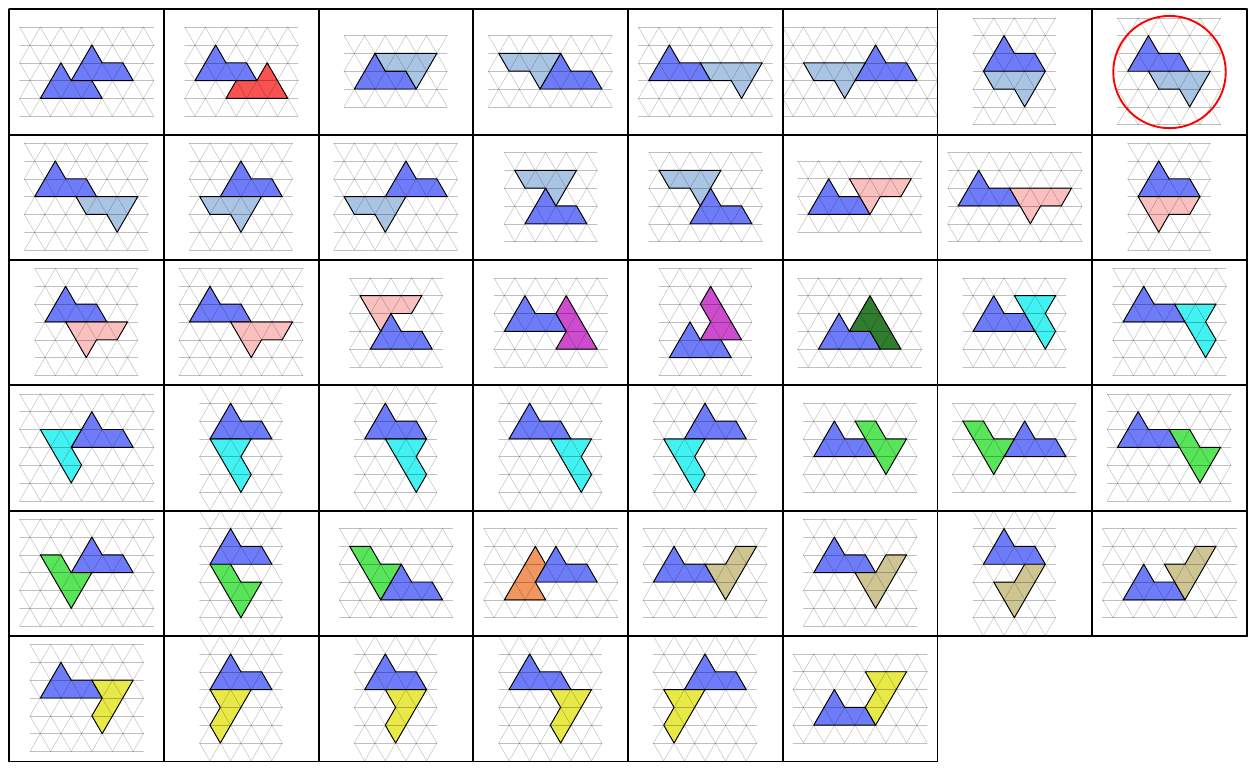}
\caption{The free tilings of the 46 sphinx dyads are shown in the color spectrum of Fig\ \ref{fig:standard-colors}. The dark blue sphinx tile is in the canonical state. Here the dyads are sorted by the state of the second tile. The bowtie (circled in red) is the only dyad with two different fixed tilings; the second tiling is a mirror reflection of the one shown here as shown in Fig.\ \ref{fig:fund-polyads-2-6}.}
\label{fig:dyadtilings}
\end{figure*}

\section{Proof of ergodicity of MC method using fundamental transitions}
\label{sec:proof}

Definition of a fundamental transition:

\begin{itemize} 
\item A transition of a sphinx tiling to a different sphinx tiling of the same frame is called fundamental if all sphinx tiles that do not have the same position and orientation in both tilings build a general fundamental
polyad. (A general fundamental polyad can have a hole.)
\end{itemize}

Main Theorem:
\begin{itemize}
\item A sphinx tiling can be transformed to any other sphinx tiling of the same frame  by a finite sequence of fundamental transitions.
\end{itemize}

This theorem guarantees that, in principle, the MC method using fundamental transitions will access all states and will thus be ergodic, assuming all general f-polyads are known for the given system. For a small frame this is shown in Fig.\ \ref{fig:diamond-6-transitions}. We know that for sphinx frames up to order 6, small sets of f-polyads (without hole) are sufficient. The question remains how we can determine effective, suitable sets of f-polyads that allow all states of a given (large) system to be found.

Proof by induction:\\
A frame that can be tiled by sphinx tiles is a polyad.
Say $u$ and $v$ are two different tilings of the same polyad $\Phi$.  Define $\sigma$ as the size of $\Phi$ and define $M$ as the set of all tilings of $\Phi$. $u \ne v$  and  $u, v \in M \implies |M| \geq 2$  and $\sigma \geq 2$.  ($\sigma > 1$ because a polyad of size 1 has only one tiling.)

Base case $\sigma  = 2$,\\
There are 46 polyads of size 2 (dyads) (Fig.\ \ref{fig:dyadtilings}) but only one (the bowtie) has two tilings. The bowtie is a fundamental polyad because it has exactly two disjoint tilings. $\Phi$ is the bowtie $ \implies \Phi$ is fundamental $\implies$ the transition from $u$ to $v$ is fundamental. The theorem is true for $\sigma = 2$

Induction step. To show:\\
The theorem is true for $\sigma = n+1$ if it is true for $\sigma = 2, 3, 4, ..., n.$

Case 1:  $u$ and $v$ are not disjoint.\\
We delete all tiles which have same position and orientation in both tilings. At least one tile is deleted because $u$ and $v$ are not disjoint. There remain one or more polyads, each with disjoint tilings in $u$ and $v$ and each with size $\leq n$.   Successively one after the other polyad shall be handled and the tiling of the polyad in $u$ shall be transformed to the tiling of this polyad in $v$. If such a polyad is fundamental it contributes one fundamental transition to the searched sequence. Otherwise it contributes a finite sub-sequence of fundamental transitions, because the polyad satisfies the conditions of the theorem with size $\leq n$.

Case 2:  $u$ and $v$ are disjoint.

Case 2.1:  $|M| = 2$	($\Phi$ has only 2 tilings).\\
In this case $\Phi$ is fundamental $\implies$ the transition from $u$ to $v$ is fundamental.

Case 2.2:  $|M| > 2$	($\Phi$ has more than 2 tilings).\\
(We try to find a chain of tilings from $u$ to $v$ where no neighbors are disjoint.)\\
Recurrence definition of two sequences of sets $R_i \subseteq M$ and $S_i \subset M$\\
$R_0 := M$  and  $S_0 := \{v\}$.\\
$R_i := R_{i-1} \backslash S_{i-1}$ and  $S_i := \{r \in R_i \mid  \exists \, t \in S_{i-1}$ with $r$ and $t$ not disjoint\}\\ 
($R_i$ consists of all (remaining) tilings of M that are not used before in any $S_j$ with $j < i$.)\\
($S_i$ consists of all remaining tilings that have a tile in same position and orientation as a tiling of $S_{i-1}$.)\\
Determine $R_i$  and $S_i$ consecutively until $i = k$ with  $S_k = \{ \}$  or  $u \in S_k$. This always happens because $M$ is finite.

Case 2.2.1:  $S_k = \{ \}$	(a chain does not exist).\\
Then  $u \in R_k$  and  $v \in M\backslash R_k = S_0 \cup S_1 \cup S_2 \cup \ldots \cup S_{k-1}$\\
$\implies$ $M$ is split into two disjoint sets of tilings  $R_k$  and  $M\backslash R_k$
$\implies \Phi$ is fundamental $\implies$ the transition from $u$ to $v$ is fundamental.
  
Case 2.2.2:  $u \in S_k$  	(a chain exists).\\
$t_k := u$\\
For $i = k-1$ down to 0 successively choose a tiling $t_i \in S_i$ with $t_i$ not disjoint to $t_{i+1}$\\ 
($t_i$ exists according to the definitions of $S_i$ and $S_{i+1}$.)\\
As $t_i$ and $t_{i+1}$ are not disjoint they satisfy the condition of case 1 and can be transformed in each other by a finite sequence of fundamental transitions.\\ 
As $t_k$ = $u$ and $t_0 = v$ we achieve a sequence of fundamental transitions from $u$ to $v$. $\blacksquare$

%End of proof.

% ****************  End of the proof ****************************

%----------------------------------------------------------------------------------------------------
%\setcounter{figure}{0}
\section{Tilings of sphinx frames of order 4 and 5}
The 16 tilings of an order-4 sphinx frame are presented in Fig.\ \ref{fig:tilings-order-4}.
Here the canonical state (state 1) of the sphinx frame is shown. The standard colors of the tiles are used. The color of a tile depends on its state. See details for states and colors in Fig.\ \ref{fig:standard-colors}.
All 153 tilings of an order-5 sphinx frame (in state 1) are shown with standard colors in Fig.\ \ref{fig:153tilingsorder5}.

%-------------------------------------------------------------------------------------------------------

%-------------------------------------------------------------------------------------------------------
%\setcounter{figure}{0}
\section{Seam method}
\label{sec:seam-method}
\begin{figure*}[htbp]
\centering
\includegraphics[width=0.127\linewidth]{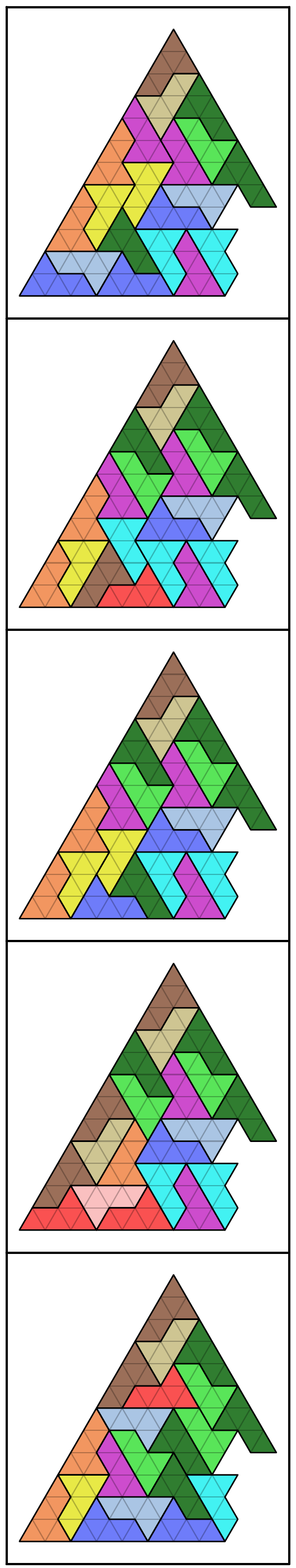}
\includegraphics[width=0.83\linewidth]{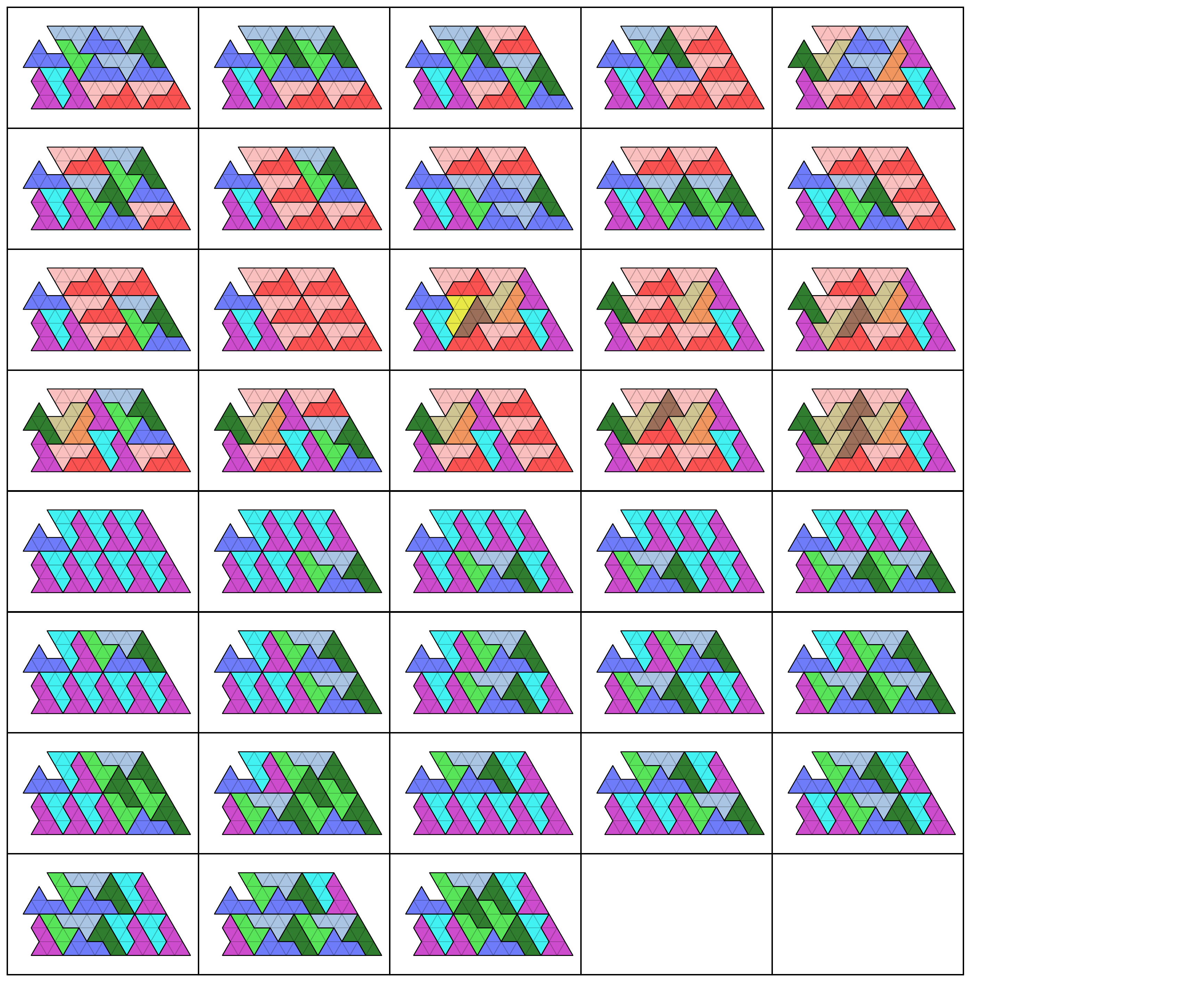}
\caption{Illustration of the seam method for a sphinx of order 6. The considered seam allows 5 tilings for the left part and 38 tilings for the right part, yielding a total of $5\cdot38 = 190$ possible tilings for this one seam.}
\label{fig:seammethod}
\end{figure*}

For an order-6 sphinx, there are a total of 145 valid seams and a total of 71\,838 tilings.
One of this seams was shown in Fig.\ \ref{fig:dangler-small} on the left side.
For this seam all tilings of both parts are presented in Fig.\ \ref{fig:seammethod}. Each combination of one of the 5 left parts with one of the 38 right parts is a tiling of the whole frame. Thus this seam contributes  $5\cdot38 = 190$ tilings.

\begin{figure*}[htbp]
\centering
\includegraphics[width=0.8\linewidth]{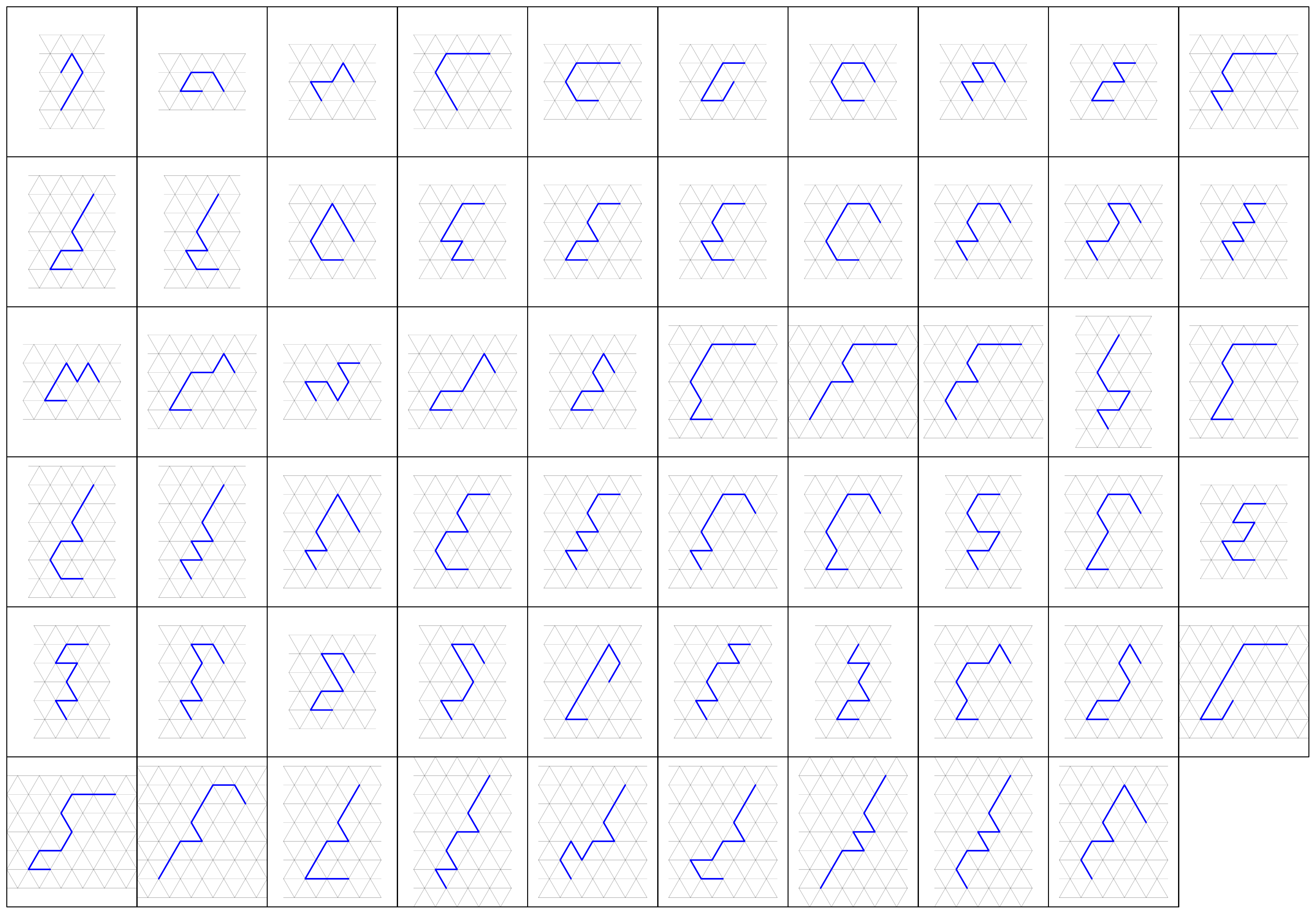}
\caption{Self-avoiding partial walks that do not allow a tiling for one or both parts of a seam on an order-6 sphinx frame, and thus can be ruled out immediately}. 
\label{fig:not-tilable-seam-parts}
\end{figure*}

Some necessary conditions for seams so as to allow tilings of both parts of the frame:
\begin{itemize}
\item The number of grid triangles on both sides of the seam have to be multiples of 6.

\item Say $t_A$ is the number of A-triangles and $t_V$ the number of V-triangles for the left or right part from the seam, then $t_A$ and $t_V$ have to be even. Moreover it can be shown that $2\cdot t_A - t_V$ has to be multiple of 6.

\item There are self-avoiding partial walks that do not allow a tiling with sphinx tiles. All seams which contain such a part are non-valid. Fig.\ \ref{fig:not-tilable-seam-parts}
\end{itemize}

It is not a problem if non-valid seams are considered in the program. If no tiling exists for one part, then the value 0 is added to the current number of tilings.

%---------------------------------------------------------------------
%\setcounter{figure}{0}
\section{Dangler method}
\label{sec:dangler method}
For the dangler method we work with sphinx frames in state 7, Fig.\ \ref{fig:dangler-state-7}.
All danglers hanging in the first row are distinct. Here the number of tilings is equal to the number of danglers, table \ref{tab:dangler}. The maximum number of danglers is reached in row 6. Note that the number of tilings represented by the danglers is already more than 1000 times greater than the number of danglers of row 6. Here the amount of memory required to store the danglers reaches its maximum. After row 12 the number of danglers is less than 1 million and the remaining calculation time is negligibly short.

\begin{table*}
\caption{Number of dangler shapes and tilings for an order-12 Sphinx frame}
    \centering
    \begin{tabular}{|r|r|r|r|}
    \hline
    & Number of \ \   & Number of tilings that cover all & \\
    & dangler shapes & rows from $1$ to $R$ completely & Maximum number \ \ \\
     & in rows $R+1$ & and allow tiles to reach from & of tilings per dangler \\
     Row $R$ & and $R+2$ & row $R$ to rows $R+1$ or $R+2$ & \\
    \hline
    1 & 1563660 & 1563660 & 1 \\
    2 & 2425788 & 6749488 & 136 \\
    3 & 44808275 & 185916194 & 377 \\
    4 & 134472174 & 3783718296 & 3024 \\
    5 & 183606640 & 38805692479 & 820432 \\
    6 &  261688546 & 278986845030 & 1903023 \\
    7 & 124226131 & 2299176633379 & 28857388 \\
    8 & 50764453 & 12906149275568 & 410155553 \\
    9 & 16451093 & 65764091664663 & 5011013238 \\
    10 & 5652931 & 342742865380799 & 21543804545 \\
    11 & 2151320 & 1705127038597723 & 879954404572 \\
    12 & 1306886 & 9277068222902118 & 1853428696911 \\
    13 & 957858 & 62760561968254764 & 14126586516272 \\
    14 & 770269 & 313908741372537769 & 115717404631817 \\
    15 & 625220 & 1767697678819240349 & 534458519097758 \\
    16 & 568380 & 11265688433944695738 & 2223905305929387 \\
    17 & 541625 & 63347947744620011957 & 26055598644871087 \\
    18 & 509235 & 347201208446538108883 & 112210412559783398 \\
    19 & 499270 & 2183684053476827440910 & 574138948225372195 \\
    20 & 496092 & 12301442561700289999794 & 4593417385095581151 \\
    21 & 489835 & 70101592808398613837681 & 23676181631639159289 \\
    22 & 425316 & 372819101729163840549119 & 116072981473250019901 \\
    23 & 161882 & 858648007102714918220990 & 888328915722460383112 \\
    24 & 98913 & 2972400093641788177161127 & 1821936225977560494508 \\
    25 & 34962 & 6954282621282954785418968 & 9817471633252747653333 \\
    26 & 10363 & 12243537479285116578416905 & 62178029366375487047045 \\
    27 & 3427 & 22157545493614557462734950 & 138476696952227099253046 \\
    28 & 768 &  33152355783977067609383308 & 399511925424970024549877 \\
    29 & 192 &  32058089044124212675313227 & 1703964867539070271230296 \\
    30 & 61 &  49926904977648125717224170 & 2814151538058596107313074 \\
    31 & 12 &  33708952511830048478624562 & 6143214589592323184596454 \\
    32 & 4 &  32813942272624544838651213 & 11091382736397093501411140 \\
    33 & 4 &  32813942272624544838651213 & 11091382736397093501411140 \\
    34 & 1 & \, 32813942272624544838651213 & \, 32813942272624544838651213 \\
    \hline
    \end{tabular}
    \label{tab:dangler}
\end{table*}

\section{Sphinx dyads}
\begin{figure*}[htbp]
\centering
\includegraphics[width=1.0\linewidth]{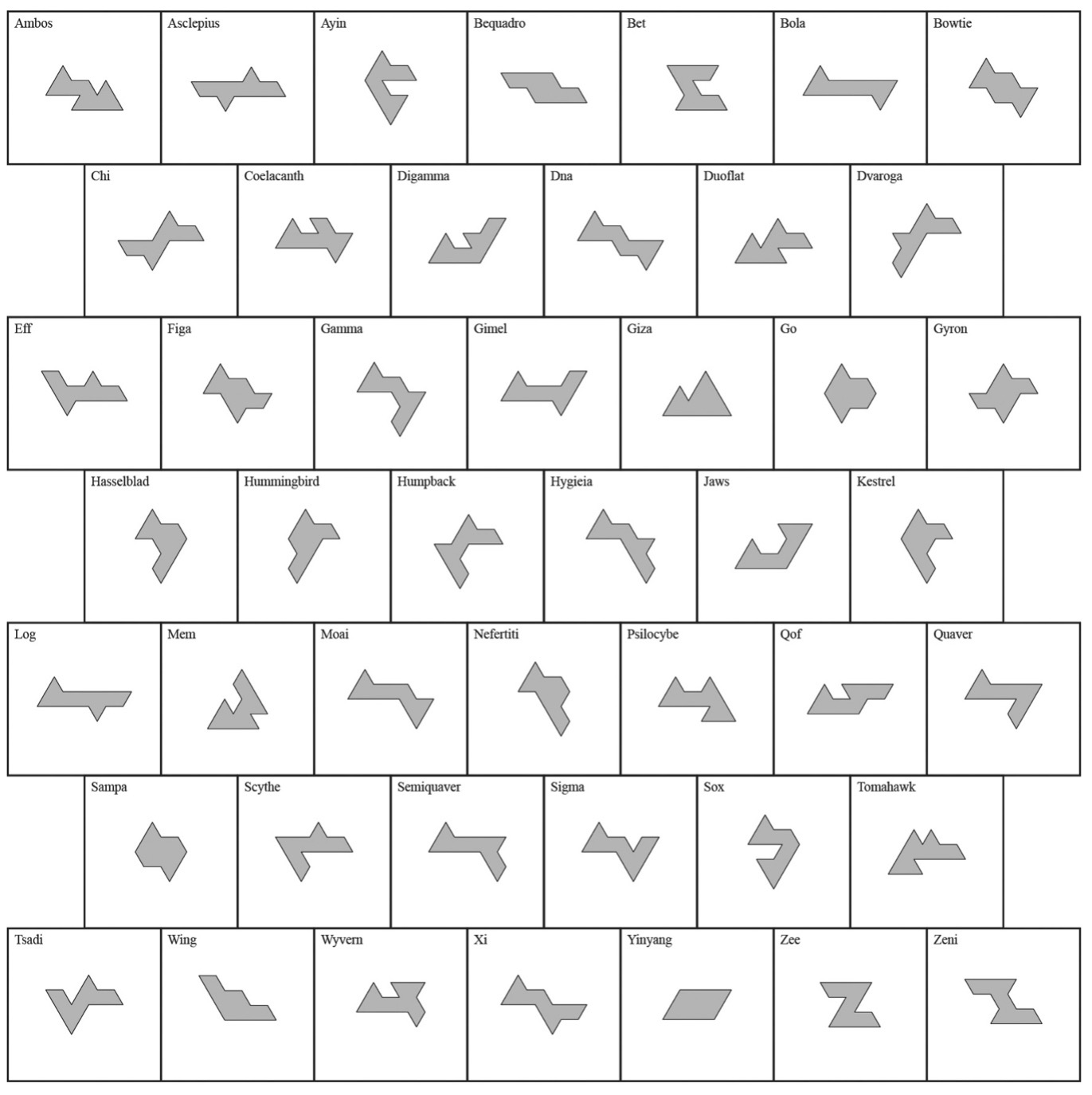}
\caption{The 46 different free sphinx dyads arranged alphabetically by the names we have given them.}
\label{fig:dyadframes}
\end{figure*}

A sphinx dyad is a polyad of size 2, meaning that it is a size-2 polyiamond with area $2\cdot6 = 12$ (in units of triangular area), which can be tiled by two sphinx tiles. 
There exist 46 free sphinx dyads.
``Free" means distinct up to rotation and reflection (Fig.\ \ref{fig:dyadframes}).
Although one dyad (the bowtie) has two different tilings, the number of free tilings is also 46, because the tilings of the bowtie are mirror images of each other, see first two tilings in Fig.\ \ref{fig:fund-polyads-2-6}.
All 46 free dyad tilings are shown in Fig.\ \ref{fig:dyadtilings}\\
Find the numbers of polyads with sizes from 3 to 6 (triads, tetrads, pentads, hexads) in table \ref{tab:polyad}.

%-------------------------------------------------------------------------------------------------------

%\setcounter{figure}{0}
\section{Fundamental polyads found in tilings of an order-7 sphinx frame}
In order to find f-polyads with size $\geq 7$ we consider all pairs of tilings of an order-7 sphinx frame. In such a pair we delete all tiles that occur in the same position and orientation in both tilings. In most cases the remaining part is not fundamental. But as we know all tilings of the order-7 sphinx, all f-polyad that occur in an order-7 tiling will remain at least in one pair of tilings. Of course there also are f-polyads that are not part of any order-7 sphinx tiling. Hence this search is not exhaustive. See Figs.\ \ref{fig:fund-heptads}, \ref{fig:fund-octads}, \ref{fig:fund-enneads}, and \ref{fig:fund-decads}.

\begin{figure*}[htbp]
\centering
\includegraphics[width=0.3\linewidth]{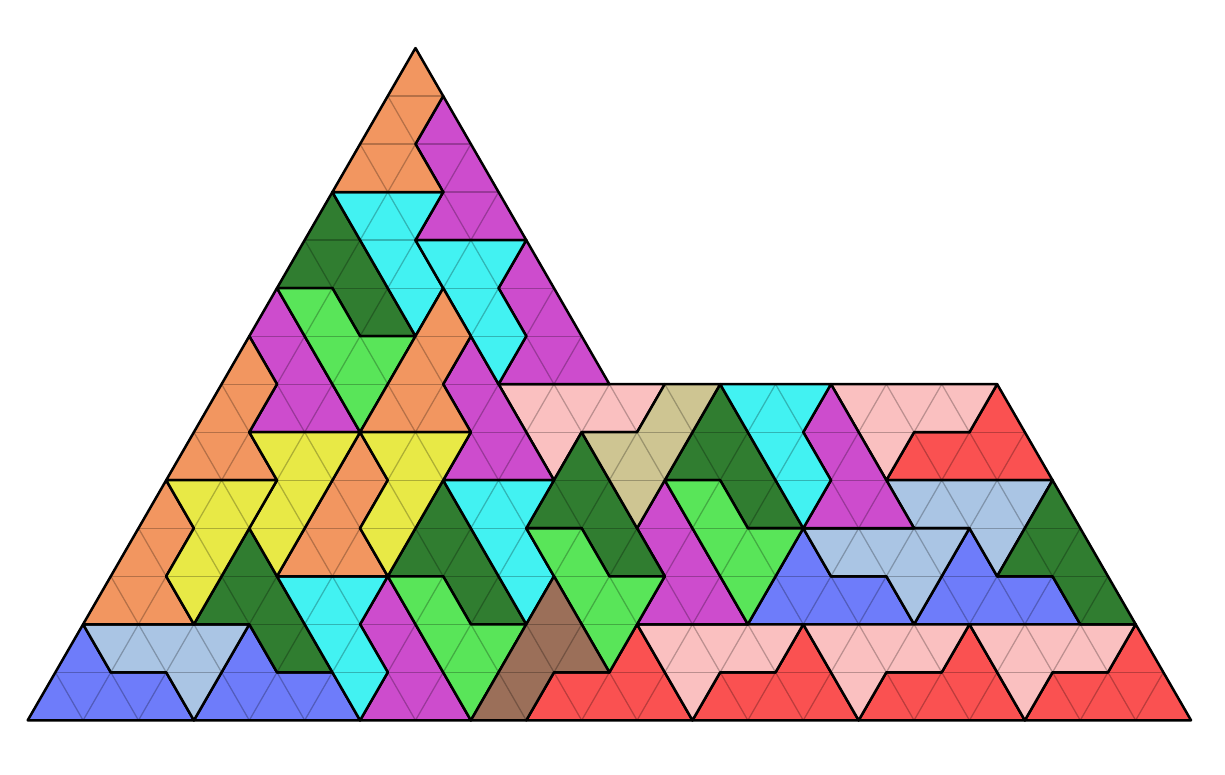}
\includegraphics[width=0.114\linewidth]{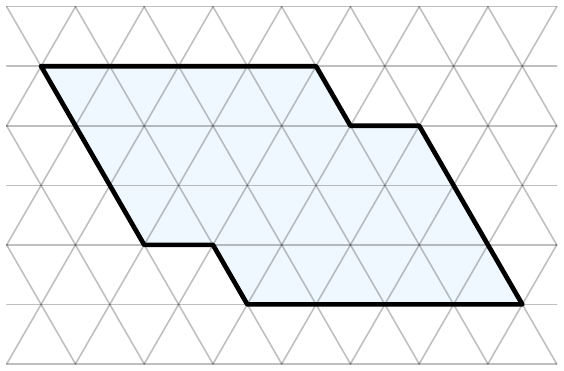}
\includegraphics[width=0.3\linewidth]{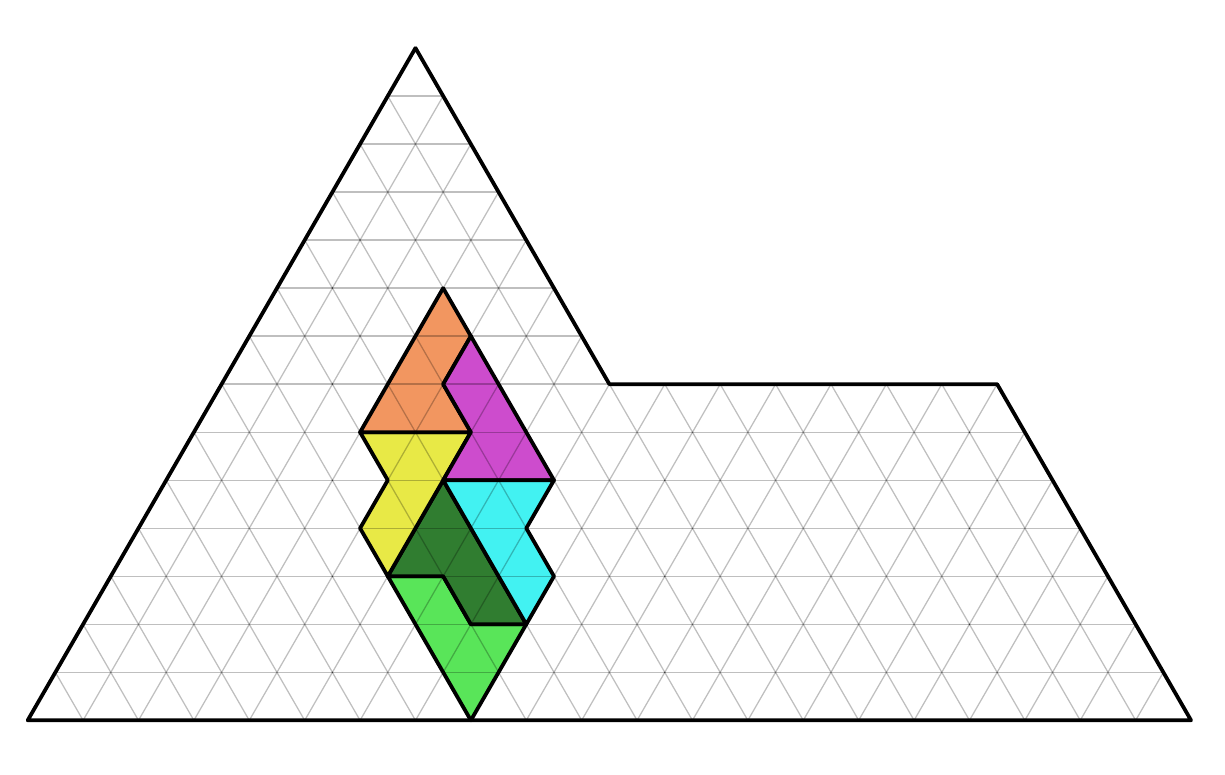}
\includegraphics[width=0.3\linewidth]{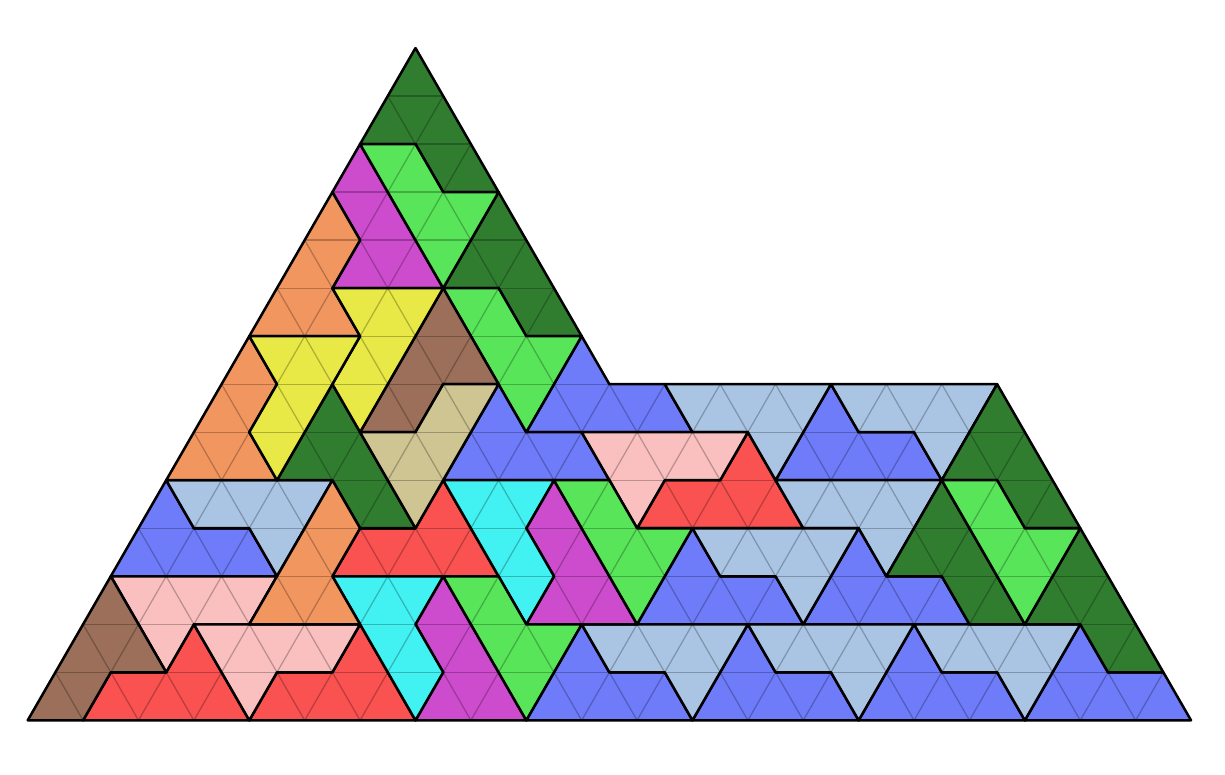}
\includegraphics[width=0.114\linewidth]{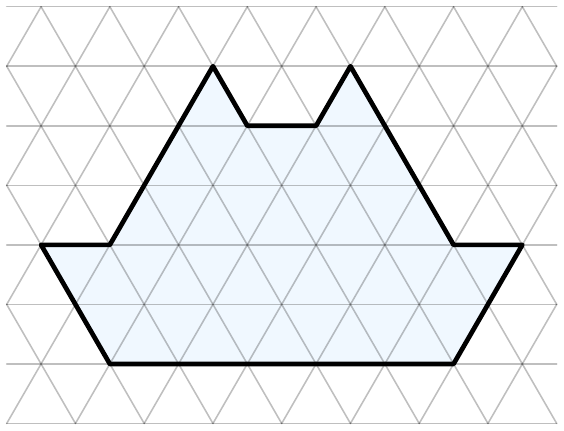}
\includegraphics[width=0.3\linewidth]{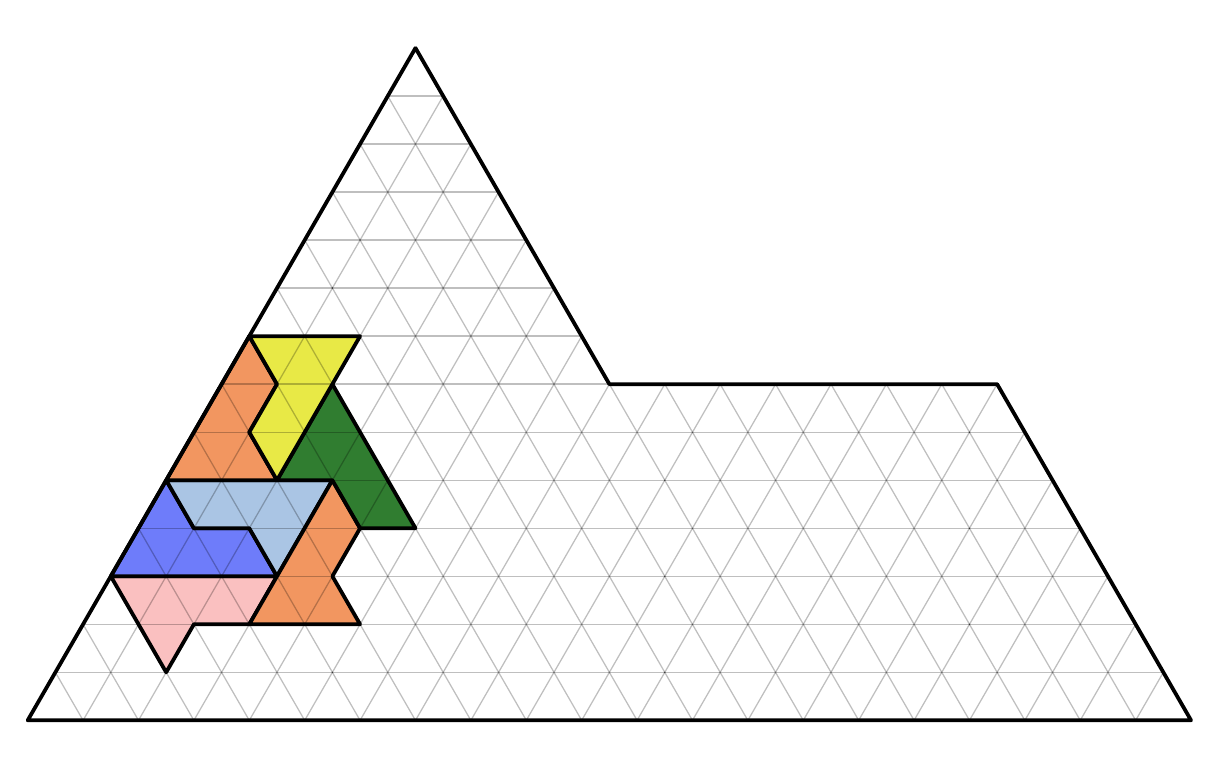}
\caption{The only two tilings of an S7 frame that do not contain any of the six fundamental polyads of Fig.\ \ref{fig:SixCommonFundPolyads}.  Here, the tilings contain a fundamental hexad (upper figure) and heptad (lower figure) as displayed to the right of each tiling.}
\label{fig:figure11}
\end{figure*}

\begin{figure*}[htbp]
\centering
\includegraphics[width=0.6\linewidth]{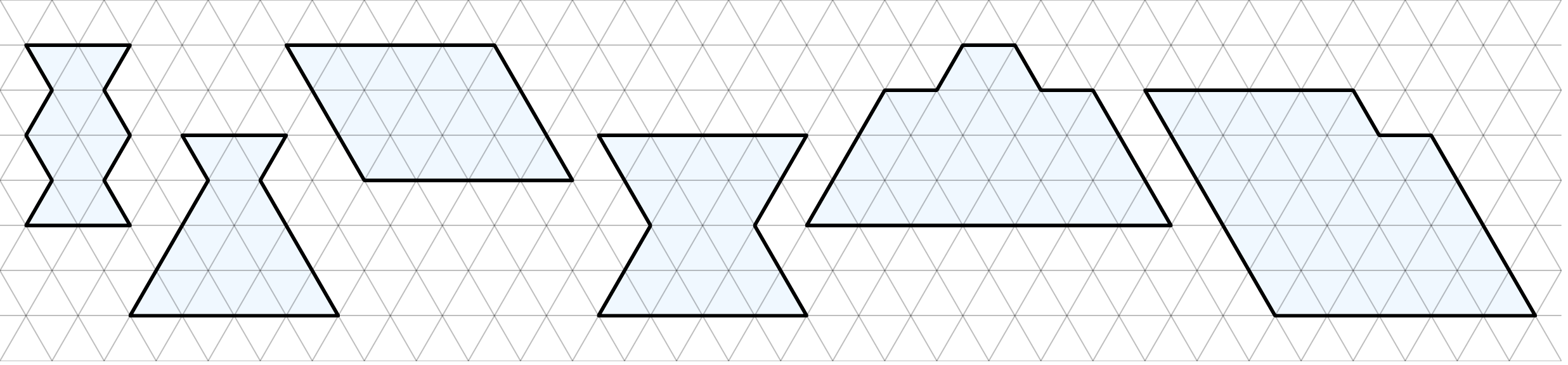}
\caption{Except for the two tilings in Fig.\ \ref{fig:figure11}, all 5 965 398 sphinx tilings of an order-7 Sphinx frame contain at least one of these 6 fundamental polyads (of orders 2, 3, 4, 4, 6 and 8 respectively). Furthermore, there are only seven tilings of the order-7 Sphinx lacking the first 5 polyads above. %all of 
}
\label{fig:SixCommonFundPolyads}
\end{figure*}

Except for the two tilings in Fig.\ \ref{fig:figure11}, all 5\,965\,398 sphinx tilings of an order-7 Sphinx frame contain at least one of the 6 fundamental polyads (of orders 2, 3, 4, 4, 6 and 8 respectively) that are shown in Fig.\ \ref{fig:SixCommonFundPolyads}. Furthermore, there are only seven tilings of the order-7 Sphinx lacking the first 5 polyads of Fig.\ \ref{fig:SixCommonFundPolyads}.

\section{Chirality distribution}
\label{sec:chirality}
In order to study the chirality distribution of a frame that is tilable by sphinx tiles we consider all tilings of this frame and count how often each grid triangle in the frame is covered by L- and R-sphinxes. The chirality of a grid triangle is defined by $\chi = \frac{L - R}{L + R}$, where $L$ ($R$) is the number of coverings by L-sphinxes (R-sphinxes). For each grid triangle $L + R$ has the same value, the total number of tilings of the frame. In order to demonstrate the chirality distribution triangles with $\chi > 0$ are colored blue and such with $\chi < 0$ are colored red. Dependent on the size of $|\chi|$ in the range from 0 to 1 the color is lighter or darker. White color means $\chi = 0$.
The chirality distribution of the order-7 Sphinx (5\,965\,398 tilings) was shown in Fig.\ \ref{fig:Chirality7}. This can be compared with a Diamond of order 12 (29\,014\,790 tilings) and the Hexiamond ``Yacht" of order 7. The areas of the frames are comparable: the 7-Sphinx and the 7-Yacht are tiled by 49 sphinx tiles while the 12-Diamond contains 48 sphinx tiles. See Fig.\ \ref{fig:more-chirality}
Beside the Sphinx the Yacht is the only Hexiamond of that can be tiled in order 7 by elementary sphinxes.

On the right of Fig.\ \ref{fig:more-chirality}, we give a ``microscopic'' explanation of the localization of chirality even within a symmetric frame, as illustrated by three small Diamond frames. Consider the 120$^\circ$ corners of the Sphinx and Diamond frames, where chirality ``dipoles" are apparent. The upper corner (and, by symmetry, the lower corner) of a 6-Diamond (D6) admits only the three possibilities shown, out of all 22 tilings of D6.  There are 7, 7, and 8 tilings, respectively, that contain these three possible corners. The wing dyad (Fig.\ \ref{fig:dyadframes}) is primarily responsible for this seemingly paradoxical situation, because it alone is invariant under the reflection isometry. The mirror reflection of such a symmetric frame has two effects: it changes not only the chirality of individual tiles, it also switches their spatial positions, allowing for this type of locally asymmetric structure in the average chirality field. In general, other types of local distributions of chiralty are allowed and associated with corners of the frame boundary; for instance, monopoles and quadrupoles.

\bigskip

\section{General substitution method}
Twelve is the minimal order of a Sphinx frame (S12) that admits tilings wherein each of its six triangular subframes (in this case, T12) is tileable by sphinxes, since Triangle frames smaller than T12 allow no sphinx tilings, Fig.\ \ref{fig:order12sphinx}. This substitution method, using all 830 tilings of T12, generates $830^6$ distinct sphinx tilings of S12. Although this accounts for over $3.2\times 10^{17}$ tilings, that is just one hundred millionth of all the S12 tilings.  Of course, any polyiamond (by definition, composed of unit equilateral triangles) can be inflated by a factor of 12, used as a frame, and then tiled by sphinxes analogously. The set of tilings generated in this manner will likewise be a minute subset of the set of all sphinx tilings forced by that frame.

\section{Sample tilings with low and high chiral energy}
Beside other applications of the Monte Carlo method we also search for tilings with low and high chiral energy in sphinx frames that are too large for exhaustive treatings. In Fig.\ \ref{fig:figure8} two samples are shown one for low and one for high chiral energy. Both tilings are shown in standar coloring as well as in chiral coloring with L-sphinxes in blue and R-sphinxes in red.

\section{Large sphinx tiling}
The Monte Carlo method using fundamental polyads makes it possible to generate random tilings of sphinx frames as large as order 100 and beyond, as illustrated in Fig.\ \ref{fig:tiling100rotated}.

\bibliography{letter}

\begin{thebibliography}{45}%
\makeatletter
\providecommand \@ifxundefined [1]{%
 \@ifx{#1\undefined}
}%
\providecommand \@ifnum [1]{%
 \ifnum #1\expandafter \@firstoftwo
 \else \expandafter \@secondoftwo
 \fi
}%
\providecommand \@ifx [1]{%
 \ifx #1\expandafter \@firstoftwo
 \else \expandafter \@secondoftwo
 \fi
}%
\providecommand \natexlab [1]{#1}%
\providecommand \enquote  [1]{``#1''}%
\providecommand \bibnamefont  [1]{#1}%
\providecommand \bibfnamefont [1]{#1}%
\providecommand \citenamefont [1]{#1}%
\providecommand \href@noop [0]{\@secondoftwo}%
\providecommand \href [0]{\begingroup \@sanitize@url \@href}%
\providecommand \@href[1]{\@@startlink{#1}\@@href}%
\providecommand \@@href[1]{\endgroup#1\@@endlink}%
\providecommand \@sanitize@url [0]{\catcode `\\12\catcode `\$12\catcode
  `\&12\catcode `\#12\catcode `\^12\catcode `\_12\catcode `\%12\relax}%
\providecommand \@@startlink[1]{}%
\providecommand \@@endlink[0]{}%
\providecommand \url  [0]{\begingroup\@sanitize@url \@url }%
\providecommand \@url [1]{\endgroup\@href {#1}{\urlprefix }}%
\providecommand \urlprefix  [0]{URL }%
\providecommand \Eprint [0]{\href }%
\providecommand \doibase [0]{https://doi.org/}%
\providecommand \selectlanguage [0]{\@gobble}%
\providecommand \bibinfo  [0]{\@secondoftwo}%
\providecommand \bibfield  [0]{\@secondoftwo}%
\providecommand \translation [1]{[#1]}%
\providecommand \BibitemOpen [0]{}%
\providecommand \bibitemStop [0]{}%
\providecommand \bibitemNoStop [0]{.\EOS\space}%
\providecommand \EOS [0]{\spacefactor3000\relax}%
\providecommand \BibitemShut  [1]{\csname bibitem#1\endcsname}%
\let\auto@bib@innerbib\@empty
%</preamble>
\bibitem [{\citenamefont {Visheratina}\ \emph {et~al.}(2022)\citenamefont
  {Visheratina}, \citenamefont {Kumar},\ and\ \citenamefont
  {Kotov}}]{VisheratinaKumarKotov22}%
  \BibitemOpen
  \bibfield  {author} {\bibinfo {author} {\bibfnamefont {A.}~\bibnamefont
  {Visheratina}}, \bibinfo {author} {\bibfnamefont {P.}~\bibnamefont {Kumar}},\
  and\ \bibinfo {author} {\bibfnamefont {N.}~\bibnamefont {Kotov}},\ }\bibfield
   {title} {\bibinfo {title} {Engineering of inorganic nanostructures with
  hierarchy of chiral geometries at multiple scales},\ }\href
  {https://doi.org/10.1002/aic.17438} {\bibfield  {journal} {\bibinfo
  {journal} {AIChE Journal}\ }\textbf {\bibinfo {volume} {68}},\ \bibinfo
  {pages} {e17438} (\bibinfo {year} {2022})}\BibitemShut {NoStop}%
\bibitem [{\citenamefont {Fowler}\ and\ \citenamefont
  {Rushbrooke}(1937)}]{FowlerRushbrooke37}%
  \BibitemOpen
  \bibfield  {author} {\bibinfo {author} {\bibfnamefont {R.~H.}\ \bibnamefont
  {Fowler}}\ and\ \bibinfo {author} {\bibfnamefont {G.~S.}\ \bibnamefont
  {Rushbrooke}},\ }\bibfield  {title} {\bibinfo {title} {An attempt to extend
  the statistical theory of perfect solutions},\ }\href
  {https://doi.org/10.1039/TF9373301272} {\bibfield  {journal} {\bibinfo
  {journal} {Trans. Faraday Soc.}\ }\textbf {\bibinfo {volume} {33}},\ \bibinfo
  {pages} {1272} (\bibinfo {year} {1937})}\BibitemShut {NoStop}%
\bibitem [{\citenamefont {Temperley}\ and\ \citenamefont
  {Fisher}(1961)}]{TemperleyFisher61}%
  \BibitemOpen
  \bibfield  {author} {\bibinfo {author} {\bibfnamefont {H.~N.~V.}\
  \bibnamefont {Temperley}}\ and\ \bibinfo {author} {\bibfnamefont {M.~E.}\
  \bibnamefont {Fisher}},\ }\bibfield  {title} {\bibinfo {title} {Dimer problem
  in statistical mechanics---an exact result},\ }\href
  {https://doi.org/10.1080/14786436108243366} {\bibfield  {journal} {\bibinfo
  {journal} {Phil. Mag.}\ }\textbf {\bibinfo {volume} {6}},\ \bibinfo {pages}
  {1061} (\bibinfo {year} {1961})}\BibitemShut {NoStop}%
\bibitem [{\citenamefont {Kasteleyn}(1961)}]{Kasteleyn61}%
  \BibitemOpen
  \bibfield  {author} {\bibinfo {author} {\bibfnamefont {P.}~\bibnamefont
  {Kasteleyn}},\ }\bibfield  {title} {\bibinfo {title} {The statistics of
  dimers on a lattice: I. {T}he number of dimer arrangements on a quadratic
  lattice},\ }\href
  {https://doi.org/https://doi.org/10.1016/0031-8914(61)90063-5} {\bibfield
  {journal} {\bibinfo  {journal} {Physica}\ }\textbf {\bibinfo {volume} {27}},\
  \bibinfo {pages} {1209} (\bibinfo {year} {1961})}\BibitemShut {NoStop}%
\bibitem [{\citenamefont {Kac}\ and\ \citenamefont {Ward}(1952)}]{KacWard52}%
  \BibitemOpen
  \bibfield  {author} {\bibinfo {author} {\bibfnamefont {M.}~\bibnamefont
  {Kac}}\ and\ \bibinfo {author} {\bibfnamefont {J.~C.}\ \bibnamefont {Ward}},\
  }\bibfield  {title} {\bibinfo {title} {A combinatorial solution of the
  two-dimensional {I}sing model},\ }\href
  {https://doi.org/10.1103/PhysRev.88.1332} {\bibfield  {journal} {\bibinfo
  {journal} {Phys. Rev.}\ }\textbf {\bibinfo {volume} {88}},\ \bibinfo {pages}
  {1332} (\bibinfo {year} {1952})}\BibitemShut {NoStop}%
\bibitem [{\citenamefont {Lieb}(1967{\natexlab{a}})}]{Lieb67}%
  \BibitemOpen
  \bibfield  {author} {\bibinfo {author} {\bibfnamefont {E.~H.}\ \bibnamefont
  {Lieb}},\ }\bibfield  {title} {\bibinfo {title} {Solution of the dimer
  problem by the transfer matrix method},\ }\href
  {https://doi.org/10.1063/1.1705163} {\bibfield  {journal} {\bibinfo
  {journal} {J. Math. Phys.}\ }\textbf {\bibinfo {volume} {8}},\ \bibinfo
  {pages} {2339} (\bibinfo {year} {1967}{\natexlab{a}})}\BibitemShut {NoStop}%
\bibitem [{\citenamefont {Priezzhev}\ and\ \citenamefont
  {Ruelle}(2008)}]{PriezzhevRuelle08}%
  \BibitemOpen
  \bibfield  {author} {\bibinfo {author} {\bibfnamefont {V.~B.}\ \bibnamefont
  {Priezzhev}}\ and\ \bibinfo {author} {\bibfnamefont {P.}~\bibnamefont
  {Ruelle}},\ }\bibfield  {title} {\bibinfo {title} {Boundary monomers in the
  dimer model},\ }\href {https://doi.org/10.1103/PhysRevE.77.061126} {\bibfield
   {journal} {\bibinfo  {journal} {Phys. Rev. E}\ }\textbf {\bibinfo {volume}
  {77}},\ \bibinfo {pages} {061126} (\bibinfo {year} {2008})}\BibitemShut
  {NoStop}%
\bibitem [{\citenamefont {Loebl}(2010)}]{Loebl10}%
  \BibitemOpen
  \bibfield  {author} {\bibinfo {author} {\bibfnamefont {M.}~\bibnamefont
  {Loebl}},\ }\bibinfo {title} {2d {I}sing and dimer models},\ in\ \href
  {https://doi.org/10.1007/978-3-8348-9329-1_9} {\emph {\bibinfo {booktitle}
  {Discrete Mathematics in Statistical Physics: Introductory Lectures}}}\
  (\bibinfo  {publisher} {Vieweg+Teubner},\ \bibinfo {address} {Wiesbaden},\
  \bibinfo {year} {2010})\ pp.\ \bibinfo {pages} {157--171}\BibitemShut
  {NoStop}%
\bibitem [{\citenamefont {Pham}(2020)}]{Pham20}%
  \BibitemOpen
  \bibfield  {author} {\bibinfo {author} {\bibfnamefont {A.~M.}\ \bibnamefont
  {Pham}},\ }\bibfield  {title} {\bibinfo {title} {A {P}faffian formula for the
  {I}sing partition function of surface graphs},\ }\href
  {https://doi.org/10.1088/1742-5468/aba1e4} {\bibfield  {journal} {\bibinfo
  {journal} {J. Stat. Mech.: Th. Exp.}\ }\textbf {\bibinfo {volume} {2020}},\
  \bibinfo {pages} {083103} (\bibinfo {year} {2020})}\BibitemShut {NoStop}%
\bibitem [{\citenamefont {Dhar}(1990)}]{Dhar90}%
  \BibitemOpen
  \bibfield  {author} {\bibinfo {author} {\bibfnamefont {D.}~\bibnamefont
  {Dhar}},\ }\bibfield  {title} {\bibinfo {title} {Self-organized critical
  state of sandpile automaton models},\ }\href
  {https://doi.org/10.1103/PhysRevLett.64.1613} {\bibfield  {journal} {\bibinfo
   {journal} {Phys. Rev. Lett.}\ }\textbf {\bibinfo {volume} {64}},\ \bibinfo
  {pages} {1613} (\bibinfo {year} {1990})}\BibitemShut {NoStop}%
\bibitem [{\citenamefont {Jockusch}\ \emph {et~al.}(1995)\citenamefont
  {Jockusch}, \citenamefont {Propp},\ and\ \citenamefont
  {Shor}}]{JockuschProppShor95}%
  \BibitemOpen
  \bibfield  {author} {\bibinfo {author} {\bibfnamefont {W.}~\bibnamefont
  {Jockusch}}, \bibinfo {author} {\bibfnamefont {J.}~\bibnamefont {Propp}},\
  and\ \bibinfo {author} {\bibfnamefont {P.}~\bibnamefont {Shor}},\ }\bibfield
  {title} {\bibinfo {title} {Random domino tilings and the arctic circle
  theorem},\ }\href {arxiv:math/9801068} {\bibfield  {journal} {\bibinfo
  {journal} {preprint arXiv:}\ }\textbf {\bibinfo {volume} {math/9801068}}
  (\bibinfo {year} {1995})}\BibitemShut {NoStop}%
\bibitem [{\citenamefont {Elkies}\ \emph {et~al.}(2023)\citenamefont {Elkies},
  \citenamefont {Kuperberg}, \citenamefont {Larsen},\ and\ \citenamefont
  {Propp}}]{ElkiesKuperbergLarsenPropp92}%
  \BibitemOpen
  \bibfield  {author} {\bibinfo {author} {\bibfnamefont {N.}~\bibnamefont
  {Elkies}}, \bibinfo {author} {\bibfnamefont {G.}~\bibnamefont {Kuperberg}},
  \bibinfo {author} {\bibfnamefont {M.}~\bibnamefont {Larsen}},\ and\ \bibinfo
  {author} {\bibfnamefont {J.}~\bibnamefont {Propp}},\ }\bibfield  {title}
  {\bibinfo {title} {Alternating-sign matrices and domino tilings. {I}},\
  }\href {https://doi.org/10.1023/A:1022420103267} {\bibfield  {journal}
  {\bibinfo  {journal} {J. Algebraic Combinatorics}\ }\textbf {\bibinfo
  {volume} {1}},\ \bibinfo {pages} {111–132} (\bibinfo {year}
  {2023})}\BibitemShut {NoStop}%
\bibitem [{\citenamefont {Ferrari}\ and\ \citenamefont
  {Spohn}(2006)}]{FerrariSpohn06}%
  \BibitemOpen
  \bibfield  {author} {\bibinfo {author} {\bibfnamefont {P.~L.}\ \bibnamefont
  {Ferrari}}\ and\ \bibinfo {author} {\bibfnamefont {H.}~\bibnamefont
  {Spohn}},\ }\bibfield  {title} {\bibinfo {title} {Domino tilings and the
  six-vertex model at its free-fermion point},\ }\href
  {https://doi.org/10.1088/0305-4470/39/33/003} {\bibfield  {journal} {\bibinfo
   {journal} {J. Phys. A: Math. Gen.}\ }\textbf {\bibinfo {volume} {39}},\
  \bibinfo {pages} {10297} (\bibinfo {year} {2006})}\BibitemShut {NoStop}%
\bibitem [{\citenamefont {Kenyon}\ and\ \citenamefont
  {Wilson}(2011)}]{KenyonWilson11}%
  \BibitemOpen
  \bibfield  {author} {\bibinfo {author} {\bibfnamefont {R.~W.}\ \bibnamefont
  {Kenyon}}\ and\ \bibinfo {author} {\bibfnamefont {D.~B.}\ \bibnamefont
  {Wilson}},\ }\bibfield  {title} {\bibinfo {title} {Boundary partitions in
  trees and dimers},\ }\href {https://doi.org/10.1090/S0002-9947-2010-04964-5}
  {\bibfield  {journal} {\bibinfo  {journal} {Trans. Amer. Math. Soc.}\
  }\textbf {\bibinfo {volume} {363}},\ \bibinfo {pages} {1325} (\bibinfo {year}
  {2011})}\BibitemShut {NoStop}%
\bibitem [{\citenamefont {Pauling}(1935)}]{Pauling35}%
  \BibitemOpen
  \bibfield  {author} {\bibinfo {author} {\bibfnamefont {L.}~\bibnamefont
  {Pauling}},\ }\bibfield  {title} {\bibinfo {title} {The structure and entropy
  of ice and of other crystals with some randomness of atomic arrangement},\
  }\href {https://doi.org/10.1021/ja01315a102} {\bibfield  {journal} {\bibinfo
  {journal} {J. Am. Chem. Soc.}\ }\textbf {\bibinfo {volume} {57}},\ \bibinfo
  {pages} {2680} (\bibinfo {year} {1935})}\BibitemShut {NoStop}%
\bibitem [{\citenamefont {Temperley}\ and\ \citenamefont
  {Lieb}(1971)}]{TemperleyLieb71}%
  \BibitemOpen
  \bibfield  {author} {\bibinfo {author} {\bibfnamefont {H.~N.~V.}\
  \bibnamefont {Temperley}}\ and\ \bibinfo {author} {\bibfnamefont {E.~H.}\
  \bibnamefont {Lieb}},\ }\bibfield  {title} {\bibinfo {title} {Relations
  between the `percolation’ and `colouring’ problem and other
  graph-theoretical problems associated with regular planar lattices: Some
  exact results for the `percolation’ problem},\ }\href
  {https://doi.org/10.1098/rspa.1971.0067} {\bibfield  {journal} {\bibinfo
  {journal} {Proc. Roy. Soc. London. Ser. A}\ }\textbf {\bibinfo {volume}
  {322}},\ \bibinfo {pages} {251} (\bibinfo {year} {1971})}\BibitemShut
  {NoStop}%
\bibitem [{\citenamefont {Wang}(1961)}]{HWang1961}%
  \BibitemOpen
  \bibfield  {author} {\bibinfo {author} {\bibfnamefont {H.}~\bibnamefont
  {Wang}},\ }\bibfield  {title} {\bibinfo {title} {Proving theorems by pattern
  recognition—{II}},\ }\href@noop {} {\bibfield  {journal} {\bibinfo
  {journal} {Bell System Technical J.}\ }\textbf {\bibinfo {volume} {40}},\
  \bibinfo {pages} {1} (\bibinfo {year} {1961})}\BibitemShut {NoStop}%
\bibitem [{\citenamefont {Ricci-Tersenghi}\ \emph {et~al.}(2001)\citenamefont
  {Ricci-Tersenghi}, \citenamefont {Weigt},\ and\ \citenamefont
  {Zecchina}}]{Tersenghi01}%
  \BibitemOpen
  \bibfield  {author} {\bibinfo {author} {\bibfnamefont {F.}~\bibnamefont
  {Ricci-Tersenghi}}, \bibinfo {author} {\bibfnamefont {M.}~\bibnamefont
  {Weigt}},\ and\ \bibinfo {author} {\bibfnamefont {R.}~\bibnamefont
  {Zecchina}},\ }\bibfield  {title} {\bibinfo {title} {Simplest random
  k-satisfiability problem},\ }\href
  {https://doi.org/10.1103/PhysRevE.63.026702} {\bibfield  {journal} {\bibinfo
  {journal} {Phys. Rev. E}\ }\textbf {\bibinfo {volume} {63}},\ \bibinfo
  {pages} {026702} (\bibinfo {year} {2001})}\BibitemShut {NoStop}%
\bibitem [{\citenamefont {Barthel}\ \emph {et~al.}(2002)\citenamefont
  {Barthel}, \citenamefont {Hartmann}, \citenamefont {Leone}, \citenamefont
  {Ricci-Tersenghi}, \citenamefont {Weigt},\ and\ \citenamefont
  {Zecchina}}]{Tersenghi02}%
  \BibitemOpen
  \bibfield  {author} {\bibinfo {author} {\bibfnamefont {W.}~\bibnamefont
  {Barthel}}, \bibinfo {author} {\bibfnamefont {A.~K.}\ \bibnamefont
  {Hartmann}}, \bibinfo {author} {\bibfnamefont {M.}~\bibnamefont {Leone}},
  \bibinfo {author} {\bibfnamefont {F.}~\bibnamefont {Ricci-Tersenghi}},
  \bibinfo {author} {\bibfnamefont {M.}~\bibnamefont {Weigt}},\ and\ \bibinfo
  {author} {\bibfnamefont {R.}~\bibnamefont {Zecchina}},\ }\bibfield  {title}
  {\bibinfo {title} {Hiding solutions in random satisfiability problems: A
  statistical mechanics approach},\ }\href
  {https://doi.org/10.1103/PhysRevLett.88.188701} {\bibfield  {journal}
  {\bibinfo  {journal} {Phys. Rev. Lett.}\ }\textbf {\bibinfo {volume} {88}},\
  \bibinfo {pages} {188701} (\bibinfo {year} {2002})}\BibitemShut {NoStop}%
\bibitem [{\citenamefont {Golomb}(1954)}]{Golomb54}%
  \BibitemOpen
  \bibfield  {author} {\bibinfo {author} {\bibfnamefont {S.~W.}\ \bibnamefont
  {Golomb}},\ }\bibfield  {title} {\bibinfo {title} {Checker boards and
  polyominoes},\ }\href {https://doi.org/10.1080/00029890.1954.11988548}
  {\bibfield  {journal} {\bibinfo  {journal} {American Math. Monthly}\ }\textbf
  {\bibinfo {volume} {61}},\ \bibinfo {pages} {675} (\bibinfo {year}
  {1954})}\BibitemShut {NoStop}%
\bibitem [{\citenamefont {Golomb}(1994)}]{Golomb94}%
  \BibitemOpen
  \bibfield  {author} {\bibinfo {author} {\bibfnamefont {S.~W.}\ \bibnamefont
  {Golomb}},\ }\href@noop {} {\emph {\bibinfo {title} {Polyominoes: Puzzles,
  Patterns, Problems, and Packings}}}\ (\bibinfo  {publisher} {Princeton
  Univesity Press},\ \bibinfo {address} {Princeton, NJ},\ \bibinfo {year}
  {1994})\BibitemShut {NoStop}%
\bibitem [{\citenamefont {O'Beirne}(1961)}]{OBeirne61}%
  \BibitemOpen
  \bibfield  {author} {\bibinfo {author} {\bibfnamefont {T.~H.}\ \bibnamefont
  {O'Beirne}},\ }\bibfield  {title} {\bibinfo {title} {Puzzles and paradoxes
  44: Pentominoes and hexiamonds},\ }\href@noop {} {\bibfield  {journal}
  {\bibinfo  {journal} {New Scientist}\ }\textbf {\bibinfo {volume} {12}},\
  \bibinfo {pages} {316–317} (\bibinfo {year} {1961})}\BibitemShut {NoStop}%
\bibitem [{\citenamefont {Weisstein}(2023)}]{Weisstein}%
  \BibitemOpen
  \bibfield  {author} {\bibinfo {author} {\bibfnamefont {E.}~\bibnamefont
  {Weisstein}},\ }\bibfield  {title} {\bibinfo {title} {``{H}exiamond" from
  {W}olfram {M}athworld},\ }\href
  {https://mathworld.wolfram.com/Hexiamond.html} {\bibfield  {journal}
  {\bibinfo  {journal} {mathworld.wolfram.com/Hexiamond}\ } (\bibinfo {year}
  {2023})}\BibitemShut {NoStop}%
\bibitem [{\citenamefont {Godr\`eche}(1989)}]{Godreche89}%
  \BibitemOpen
  \bibfield  {author} {\bibinfo {author} {\bibfnamefont {C.}~\bibnamefont
  {Godr\`eche}},\ }\bibfield  {title} {\bibinfo {title} {The sphinx: a
  limit-periodic tiling of the plane},\ }\href
  {https://doi.org/10.1088/0305-4470/22/24/006} {\bibfield  {journal} {\bibinfo
   {journal} {J. Phys. A: Math. Gen.}\ }\textbf {\bibinfo {volume} {22}},\
  \bibinfo {pages} {L1163} (\bibinfo {year} {1989})}\BibitemShut {NoStop}%
\bibitem [{\citenamefont {Lee}\ and\ \citenamefont {Moody}(2001)}]{LeeMoody01}%
  \BibitemOpen
  \bibfield  {author} {\bibinfo {author} {\bibfnamefont {J.-Y.}\ \bibnamefont
  {Lee}}\ and\ \bibinfo {author} {\bibfnamefont {R.~V.}\ \bibnamefont
  {Moody}},\ }\bibfield  {title} {\bibinfo {title} {Lattice substitution
  systems and model sets},\ }\href {https://doi.org/10.1007/s004540010083}
  {\bibfield  {journal} {\bibinfo  {journal} {Discrete {\&} Computational
  Geometry}\ }\textbf {\bibinfo {volume} {25}},\ \bibinfo {pages} {173}
  (\bibinfo {year} {2001})}\BibitemShut {NoStop}%
\bibitem [{\citenamefont {Goodman-Strauss}(2016)}]{GoodmanStrauss16}%
  \BibitemOpen
  \bibfield  {author} {\bibinfo {author} {\bibfnamefont {C.}~\bibnamefont
  {Goodman-Strauss}},\ }\bibfield  {title} {\bibinfo {title} {Matching rules
  for the sphinx tiling substitution},\ }\href
  {https://doi.org/10.48550/arXiv.1608.07168} {\bibfield  {journal} {\bibinfo
  {journal} {preprint arXiv:}\ }\textbf {\bibinfo {volume} {1608.07168}}
  (\bibinfo {year} {2016})}\BibitemShut {NoStop}%
\bibitem [{\citenamefont {Goodman-Strauss}(2018)}]{GoodmanStrauss18}%
  \BibitemOpen
  \bibfield  {author} {\bibinfo {author} {\bibfnamefont {C.}~\bibnamefont
  {Goodman-Strauss}},\ }\bibfield  {title} {\bibinfo {title} {Lots of aperiodic
  sets of tiles},\ }\href {https://doi.org/10.1016/j.jcta.2018.07.002}
  {\bibfield  {journal} {\bibinfo  {journal} {J. Combinatorial Theory, Series
  A}\ }\textbf {\bibinfo {volume} {160}},\ \bibinfo {pages} {409} (\bibinfo
  {year} {2018})}\BibitemShut {NoStop}%
\bibitem [{\citenamefont {Penrose}(1979)}]{Penrose79}%
  \BibitemOpen
  \bibfield  {author} {\bibinfo {author} {\bibfnamefont {R.}~\bibnamefont
  {Penrose}},\ }\bibfield  {title} {\bibinfo {title} {Pentaplexity, a class of
  non-periodic tilings of the plane},\ }\href
  {https://doi.org/10.1007/BF03024384} {\bibfield  {journal} {\bibinfo
  {journal} {Math. Intelligencer}\ }\textbf {\bibinfo {volume} {2}},\ \bibinfo
  {pages} {32} (\bibinfo {year} {1979})}\BibitemShut {NoStop}%
\bibitem [{\citenamefont {Haji-Akbari}\ \emph {et~al.}(2009)\citenamefont
  {Haji-Akbari}, \citenamefont {Engel}, \citenamefont {Keys}, \citenamefont
  {Zheng}, \citenamefont {Petschek}, \citenamefont {Palffy-Muhoray},\ and\
  \citenamefont {Glotzer}}]{HajiAkbariEtAl09}%
  \BibitemOpen
  \bibfield  {author} {\bibinfo {author} {\bibfnamefont {A.}~\bibnamefont
  {Haji-Akbari}}, \bibinfo {author} {\bibfnamefont {M.}~\bibnamefont {Engel}},
  \bibinfo {author} {\bibfnamefont {A.~S.}\ \bibnamefont {Keys}}, \bibinfo
  {author} {\bibfnamefont {X.}~\bibnamefont {Zheng}}, \bibinfo {author}
  {\bibfnamefont {R.~G.}\ \bibnamefont {Petschek}}, \bibinfo {author}
  {\bibfnamefont {P.}~\bibnamefont {Palffy-Muhoray}},\ and\ \bibinfo {author}
  {\bibfnamefont {S.~C.}\ \bibnamefont {Glotzer}},\ }\bibfield  {title}
  {\bibinfo {title} {Disordered, quasicrystalline and crystalline phases of
  densely packed tetrahedra},\ }\href {https://doi.org/10.1038/nature08641}
  {\bibfield  {journal} {\bibinfo  {journal} {Nature}\ }\textbf {\bibinfo
  {volume} {462}},\ \bibinfo {pages} {773–777} (\bibinfo {year}
  {2009})}\BibitemShut {NoStop}%
\bibitem [{\citenamefont {Senechal}(1995)}]{Senechal95}%
  \BibitemOpen
  \bibfield  {author} {\bibinfo {author} {\bibfnamefont {M.}~\bibnamefont
  {Senechal}},\ }\href@noop {} {\emph {\bibinfo {title} {Quasicrystals and
  Geometry}}}\ (\bibinfo  {publisher} {Cambridge University Press},\ \bibinfo
  {year} {1995})\BibitemShut {NoStop}%
\bibitem [{\citenamefont {Socolar}\ and\ \citenamefont
  {Taylor}(2012)}]{SocolarTaylor12}%
  \BibitemOpen
  \bibfield  {author} {\bibinfo {author} {\bibfnamefont {J.~E.~S.}\
  \bibnamefont {Socolar}}\ and\ \bibinfo {author} {\bibfnamefont {J.~M.}\
  \bibnamefont {Taylor}},\ }\bibfield  {title} {\bibinfo {title} {Forcing
  nonperiodicity with a single tile},\ }\href
  {https://doi.org/10.1007/s00283-011-9255-y} {\bibfield  {journal} {\bibinfo
  {journal} {The Mathematical Intelligencer}\ }\textbf {\bibinfo {volume}
  {34}},\ \bibinfo {pages} {18–28} (\bibinfo {year} {2012})}\BibitemShut
  {NoStop}%
\bibitem [{\citenamefont {Smith}\ \emph
  {et~al.}(2023{\natexlab{a}})\citenamefont {Smith}, \citenamefont {Myers},
  \citenamefont {Kaplan},\ and\ \citenamefont
  {Goodman-Strauss}}]{SmithEtAl1-23}%
  \BibitemOpen
  \bibfield  {author} {\bibinfo {author} {\bibfnamefont {D.}~\bibnamefont
  {Smith}}, \bibinfo {author} {\bibfnamefont {J.~S.}\ \bibnamefont {Myers}},
  \bibinfo {author} {\bibfnamefont {C.~S.}\ \bibnamefont {Kaplan}},\ and\
  \bibinfo {author} {\bibfnamefont {C.}~\bibnamefont {Goodman-Strauss}},\
  }\bibfield  {title} {\bibinfo {title} {An aperiodic monotile},\ }\href@noop
  {} {\bibfield  {journal} {\bibinfo  {journal} {preprint arXiv:}\ }\textbf
  {\bibinfo {volume} {2303.10798}} (\bibinfo {year}
  {2023}{\natexlab{a}})}\BibitemShut {NoStop}%
\bibitem [{\citenamefont {Smith}\ \emph
  {et~al.}(2023{\natexlab{b}})\citenamefont {Smith}, \citenamefont {Myers},
  \citenamefont {Kaplan},\ and\ \citenamefont
  {Goodman-Strauss}}]{SmithEtAl2-23}%
  \BibitemOpen
  \bibfield  {author} {\bibinfo {author} {\bibfnamefont {D.}~\bibnamefont
  {Smith}}, \bibinfo {author} {\bibfnamefont {J.~S.}\ \bibnamefont {Myers}},
  \bibinfo {author} {\bibfnamefont {C.~S.}\ \bibnamefont {Kaplan}},\ and\
  \bibinfo {author} {\bibfnamefont {C.}~\bibnamefont {Goodman-Strauss}},\
  }\bibfield  {title} {\bibinfo {title} {A chiral aperiodic monotile},\
  }\href@noop {} {\bibfield  {journal} {\bibinfo  {journal} {preprint arXiv:}\
  }\textbf {\bibinfo {volume} {2305.17743}} (\bibinfo {year}
  {2023}{\natexlab{b}})}\BibitemShut {NoStop}%
\bibitem [{\citenamefont {Schirmann}\ \emph {et~al.}(2023)\citenamefont
  {Schirmann}, \citenamefont {Franca}, \citenamefont {Flicker},\ and\
  \citenamefont {Grushin}}]{SchirmannFrancaFlickerGrushin23}%
  \BibitemOpen
  \bibfield  {author} {\bibinfo {author} {\bibfnamefont {J.}~\bibnamefont
  {Schirmann}}, \bibinfo {author} {\bibfnamefont {S.}~\bibnamefont {Franca}},
  \bibinfo {author} {\bibfnamefont {F.}~\bibnamefont {Flicker}},\ and\ \bibinfo
  {author} {\bibfnamefont {A.~G.}\ \bibnamefont {Grushin}},\ }\bibfield
  {title} {\bibinfo {title} {Physical properties of the {H}at aperiodic
  monotile: Graphene-like features, chirality and zero-modes},\ }\href@noop {}
  {\bibfield  {journal} {\bibinfo  {journal} {preprint arXiv:}\ }\textbf
  {\bibinfo {volume} {2307.11054}} (\bibinfo {year} {2023})}\BibitemShut
  {NoStop}%
\bibitem [{\citenamefont {Aguilar}\ \emph {et~al.}(2023)\citenamefont
  {Aguilar}, \citenamefont {Barbosa}, \citenamefont {Donangelo},\ and\
  \citenamefont {Souza}}]{AguilarBarbosaDonangeloSouza23}%
  \BibitemOpen
  \bibfield  {author} {\bibinfo {author} {\bibfnamefont {E.~J.}\ \bibnamefont
  {Aguilar}}, \bibinfo {author} {\bibfnamefont {V.~C.}\ \bibnamefont
  {Barbosa}}, \bibinfo {author} {\bibfnamefont {R.}~\bibnamefont {Donangelo}},\
  and\ \bibinfo {author} {\bibfnamefont {S.~R.}\ \bibnamefont {Souza}},\
  }\bibfield  {title} {\bibinfo {title} {Phase separation in tilings of a
  bounded region of the plane},\ }\href@noop {} {\bibfield  {journal} {\bibinfo
   {journal} {preprint arXiv:}\ }\textbf {\bibinfo {volume} {2308.03552}}
  (\bibinfo {year} {2023})}\BibitemShut {NoStop}%
\bibitem [{\citenamefont {Jung}\ \emph {et~al.}(2023)\citenamefont {Jung},
  \citenamefont {Chen},\ and\ \citenamefont {Gu}}]{JungChenGu23}%
  \BibitemOpen
  \bibfield  {author} {\bibinfo {author} {\bibfnamefont {J.}~\bibnamefont
  {Jung}}, \bibinfo {author} {\bibfnamefont {A.}~\bibnamefont {Chen}},\ and\
  \bibinfo {author} {\bibfnamefont {G.~X.}\ \bibnamefont {Gu}},\ }\bibfield
  {title} {\bibinfo {title} {Aperiodicity is all you need: Aperiodic monotiles
  for high-performance composites},\ }\href@noop {} {\bibfield  {journal}
  {\bibinfo  {journal} {preprint arXiv:}\ }\textbf {\bibinfo {volume}
  {2308.05819}} (\bibinfo {year} {2023})}\BibitemShut {NoStop}%
\bibitem [{\citenamefont {Horiyama}\ \emph {et~al.}(2015)\citenamefont
  {Horiyama}, \citenamefont {Okamoto},\ and\ \citenamefont
  {Uehara}}]{HoriyamaOkamotoUehara15}%
  \BibitemOpen
  \bibfield  {author} {\bibinfo {author} {\bibfnamefont {T.}~\bibnamefont
  {Horiyama}}, \bibinfo {author} {\bibfnamefont {Y.}~\bibnamefont {Okamoto}},\
  and\ \bibinfo {author} {\bibfnamefont {R.}~\bibnamefont {Uehara}},\
  }\bibfield  {title} {\bibinfo {title} {Ls in {L} and sphinxes in sphinx},\
  }\href@noop {} {\bibfield  {journal} {\bibinfo  {journal} {IPSJ
  (International Processing Society of Japan) SIG Technical Report (Corrected
  in Horiyama's thesis)}\ }\textbf {\bibinfo {volume} {2015-AL-154}} (\bibinfo
  {year} {2015})}\BibitemShut {NoStop}%
\bibitem [{\citenamefont {Izmailian}\ \emph {et~al.}(2019)\citenamefont
  {Izmailian}, \citenamefont {Papoyan},\ and\ \citenamefont
  {Ziff}}]{IzmailianPopoyanZiff19}%
  \BibitemOpen
  \bibfield  {author} {\bibinfo {author} {\bibfnamefont {N.~S.}\ \bibnamefont
  {Izmailian}}, \bibinfo {author} {\bibfnamefont {V.~V.}\ \bibnamefont
  {Papoyan}},\ and\ \bibinfo {author} {\bibfnamefont {R.~M.}\ \bibnamefont
  {Ziff}},\ }\bibfield  {title} {\bibinfo {title} {Exact finite-size
  corrections in the dimer model on a planar square lattice},\ }\href
  {https://doi.org/10.1088/1751-8121/ab2fed} {\bibfield  {journal} {\bibinfo
  {journal} {J. Phys. A: Math. Th.}\ }\textbf {\bibinfo {volume} {52}},\
  \bibinfo {pages} {335001} (\bibinfo {year} {2019})}\BibitemShut {NoStop}%
\bibitem [{\citenamefont {Hutchinson}\ and\ \citenamefont
  {Widom}(2015)}]{HutchinsonWidom15}%
  \BibitemOpen
  \bibfield  {author} {\bibinfo {author} {\bibfnamefont {M.}~\bibnamefont
  {Hutchinson}}\ and\ \bibinfo {author} {\bibfnamefont {M.}~\bibnamefont
  {Widom}},\ }\bibfield  {title} {\bibinfo {title} {Enumeration of octagonal
  tilings},\ }\href {https://doi.org/https://doi.org/10.1016/j.tcs.2015.03.019}
  {\bibfield  {journal} {\bibinfo  {journal} {Theoretical Computer Science}\
  }\textbf {\bibinfo {volume} {598}},\ \bibinfo {pages} {40} (\bibinfo {year}
  {2015})}\BibitemShut {NoStop}%
\bibitem [{\citenamefont {Chen}\ and\ \citenamefont
  {Spaepen}(1988)}]{ChenSpaepen88}%
  \BibitemOpen
  \bibfield  {author} {\bibinfo {author} {\bibfnamefont {L.}~\bibnamefont
  {Chen}}\ and\ \bibinfo {author} {\bibfnamefont {F.}~\bibnamefont {Spaepen}},\
  }\bibfield  {title} {\bibinfo {title} {The configurational entropy of
  two-dimensional random {P}enrose tilings},\ }\href
  {https://doi.org/https://doi.org/10.1016/0025-5416(88)90353-9} {\bibfield
  {journal} {\bibinfo  {journal} {Materials Science and Engineering}\ }\textbf
  {\bibinfo {volume} {99}},\ \bibinfo {pages} {339} (\bibinfo {year}
  {1988})}\BibitemShut {NoStop}%
\bibitem [{\citenamefont {Lieb}(1967{\natexlab{b}})}]{Lieb67b}%
  \BibitemOpen
  \bibfield  {author} {\bibinfo {author} {\bibfnamefont {E.~H.}\ \bibnamefont
  {Lieb}},\ }\bibfield  {title} {\bibinfo {title} {Residual entropy of square
  ice},\ }\href {https://doi.org/10.1103/PhysRev.162.162} {\bibfield  {journal}
  {\bibinfo  {journal} {Phys. Rev.}\ }\textbf {\bibinfo {volume} {162}},\
  \bibinfo {pages} {162} (\bibinfo {year} {1967}{\natexlab{b}})}\BibitemShut
  {NoStop}%
\bibitem [{\citenamefont {Kamio}\ \emph {et~al.}(2023)\citenamefont {Kamio},
  \citenamefont {Koizumi},\ and\ \citenamefont
  {Kakazawa}}]{KamioKoizumiKakazawa23}%
  \BibitemOpen
  \bibfield  {author} {\bibinfo {author} {\bibfnamefont {Y.}~\bibnamefont
  {Kamio}}, \bibinfo {author} {\bibfnamefont {J.}~\bibnamefont {Koizumi}},\
  and\ \bibinfo {author} {\bibfnamefont {T.}~\bibnamefont {Kakazawa}},\
  }\bibfield  {title} {\bibinfo {title} {Quadratic residues and domino
  tilints},\ }\href@noop {} {\bibfield  {journal} {\bibinfo  {journal}
  {preprint arXiv:}\ }\textbf {\bibinfo {volume} {2311.13597}} (\bibinfo {year}
  {2023})}\BibitemShut {NoStop}%
\bibitem [{\citenamefont {Luby}\ \emph {et~al.}(2001)\citenamefont {Luby},
  \citenamefont {Randall},\ and\ \citenamefont
  {Sinclair}}]{LubyRandallSinclair01}%
  \BibitemOpen
  \bibfield  {author} {\bibinfo {author} {\bibfnamefont {M.}~\bibnamefont
  {Luby}}, \bibinfo {author} {\bibfnamefont {D.}~\bibnamefont {Randall}},\ and\
  \bibinfo {author} {\bibfnamefont {A.}~\bibnamefont {Sinclair}},\ }\bibfield
  {title} {\bibinfo {title} {Markov chain algorithms for planar lattice
  structures},\ }\href {https://doi.org/10.1137/S0097539799360355} {\bibfield
  {journal} {\bibinfo  {journal} {SIAM Journal on Computing}\ }\textbf
  {\bibinfo {volume} {31}},\ \bibinfo {pages} {167} (\bibinfo {year}
  {2001})}\BibitemShut {NoStop}%
\bibitem [{\citenamefont {Kenyon}(1993)}]{Kenyon93}%
  \BibitemOpen
  \bibfield  {author} {\bibinfo {author} {\bibfnamefont {R.}~\bibnamefont
  {Kenyon}},\ }\bibfield  {title} {\bibinfo {title} {Tiling a polygon with
  parallelograms},\ }\href {https://doi.org/10.1007/BF01228510} {\bibfield
  {journal} {\bibinfo  {journal} {Algorithmica}\ }\textbf {\bibinfo {volume}
  {9}},\ \bibinfo {pages} {382} (\bibinfo {year} {1993})}\BibitemShut {NoStop}%
\bibitem [{\citenamefont {Metropolis}\ \emph {et~al.}(1953)\citenamefont
  {Metropolis}, \citenamefont {Rosenbluth}, \citenamefont {Rosenbluth},
  \citenamefont {Teller},\ and\ \citenamefont {Teller}}]{MetropolisEtAl53}%
  \BibitemOpen
  \bibfield  {author} {\bibinfo {author} {\bibfnamefont {N.}~\bibnamefont
  {Metropolis}}, \bibinfo {author} {\bibfnamefont {A.~W.}\ \bibnamefont
  {Rosenbluth}}, \bibinfo {author} {\bibfnamefont {M.~N.}\ \bibnamefont
  {Rosenbluth}}, \bibinfo {author} {\bibfnamefont {A.~H.}\ \bibnamefont
  {Teller}},\ and\ \bibinfo {author} {\bibfnamefont {E.}~\bibnamefont
  {Teller}},\ }\bibfield  {title} {\bibinfo {title} {Equation of state
  calculations by fast computing machines},\ }\href
  {https://doi.org/10.1063/1.1699114} {\bibfield  {journal} {\bibinfo
  {journal} {J. Chem. Phys.}\ }\textbf {\bibinfo {volume} {21}},\ \bibinfo
  {pages} {1087} (\bibinfo {year} {1953})}\BibitemShut {NoStop}%
\end{thebibliography}%

\end{document}